\documentclass[11pt,a4paper]{article}
\pdfoutput=1
\usepackage{lscape}
\usepackage{caption}
\usepackage{jheppub}
\usepackage[utf8]{inputenc}
\usepackage{soul,xcolor}
\usepackage{cancel}
\usepackage{makecell}
\usepackage{diagbox}
\usepackage{makecell}
\usepackage[normalem]{ulem}
\newsavebox{\foobox}

\usepackage{multirow}
\usepackage{dcolumn}
\newcommand\Tstrut{\rule{0pt}{2.6ex}}         % = `top' strut
   
\usepackage{ulem}
\usepackage{graphicx}
\usepackage{comment}
\usepackage{bm,array}
\usepackage{graphics}
\usepackage{epsf}
\usepackage{color}
\usepackage{amsmath}
\usepackage{amssymb}
\usepackage{latexsym}
\usepackage{slashed}
\usepackage{float}
\usepackage{cases}
\usepackage{multirow}
\usepackage{framed}
\def\be{\begin{equation}}
\def\ee{\end{equation}}
\def\ba{\begin{array}}
\def\ea{\end{array}}
\usepackage{amsmath}
\usepackage{amssymb}
\usepackage[utf8]{inputenc}

\def\alambda{A_\lambda}
\def\akappa{A_\kappa}
\def\mueff{\mu_\mathrm{eff}}

\def\tanb{\tan\beta}

\def\wpm{W^\pm}

\def\sQ3{\widetilde{Q}_3}
\def\sU3{\widetilde{U}_3}
\def\sD3{\widetilde{D}_3}

\def\mstonetwo{m_{\tilde{t}_{1,2}}}

\def\hsm{h_{\rm SM}}

\def\hs{h_{_S}} 
\def\as{a_{_S}}
\def\bino{\widetilde{B}}
\def\wino{\widetilde{W}}

\def\higgsinod{\widetilde{H}^0_d}
\def\higgsinou{\widetilde{H}^0_u}
\def\singlino{\widetilde{S}}
\def\ntrli{\chi_{_i}^0}

\def\ntrlone{\chi_{_1}^0}
\def\ntrltwo{\chi_{_2}^0}
\def\ntrlthree{\chi_{_3}^0}
\def\ntrlfour{\chi_{_4}^0}
\def\ntrlfive{\chi_{_5}^0}

\def\ntrltwothree{\chi_{_{2,3}}^0}
\def\ntrlthreefour{\chi_{_{3,4}}^0}

\def\charonepm{\chi_{_1}^\pm}

\def\chartwopm{\chi_{_2}^\pm}

\def\mone{M_1}
\def\mtwo{M_2}

\def\mthree{M_3}
\def\msQthree{m_{\widetilde{Q}_3}}
\def\msUthree{m_{\widetilde{U}_3}}

\def\d{\partial}

\def\msinglino{m_{_{\widetilde{S}}}}

\def\mntrlone{m_{{_{\chi}}_{_1}^0}}

\def\mntrlfive{m_{{_{\chi}}_{_5}^0}}

\def\mcharone{m_{{_{\chi}}_{_1}^\pm}}
\def\mchartwo{m_{{_{\chi}}_{_2}^\pm}}
\def\mhsm{m_{h_{\mathrm{SM}}}}

\newcommand{\Huzr}{h_u}
\newcommand{\Hdzr}{h_d}

\newcommand{\Sr}{s}

\newcommand{\vu}{v_u}
\newcommand{\vd}{v_d}
\newcommand{\vs}{v_s}
\def\vevs{{\it vevs}}
\def\vu{v_u}
\def\vd{v_d}

\def\vs{v_{\!_S}}

\def\beq{\begin{equation}}
\def\eeq{\end{equation}}
\def\beqa{\begin{eqnarray}}
\def\eeqa{\end{eqnarray}}

\newcommand{\subs}[1]{\ensuremath{_\textup{#1}}} % textual subscript
\allowdisplaybreaks
\newcommand{\fbinv}{\text{fb}$^{-1}$}

\def\etmiss{\slashed{E}_T}

\def\nmssmtools{{\tt NMSSMTools}}
\def\checkmate{{\tt CheckMATE}}
\def\smodels{{\tt SModelS}}
\def\micromegas{{\tt micrOMEGAs}}
\def\higgsbounds{{\tt HiggsBounds}}
\def\higgssignals{{\tt HiggsSignals}}

\def\madgraph{{\tt MadGraph}}
\def\pythia8{{\tt PYTHIA8}}
\def\delphes{{\tt DELPHES}}
\def\phasetracer{{\tt PhaseTracer}}
\def\cosmotransitions{{\tt CosmoTransitions}}
\def\z3nmssm{$Z_3$-NMSSM}
\newcommand{\bea}{\begin{eqnarray}}
\newcommand{\eea}{\end{eqnarray}}

\title{{\Large Electroweak Phase Transition in the $Z_3$-invariant NMSSM:} \\ \vskip 5pt {\large Implications of LHC and Dark matter searches and prospects of \\ \vspace{-0.2cm} detecting the gravitational waves}}
\author[a]{Arindam Chatterjee,}
\author[b]{AseshKrishna Datta}
\author[b,c]{and Subhojit Roy}
\affiliation[a]{Department of Physics, School of Natural Sciences, Shiv Nadar University, Gautam Budhha Nagar, Uttar Pradesh 201314, India}
\affiliation[b]{Harish-Chandra Research Institute, A CI of Homi Bhabha National
Institute, Chhatnag Road, Jhunsi, Prayagraj (Allahabad) 211019, India}
\affiliation[c]{Regional Centre for Accelerator-based Particle Physics, Harish-Chandra Research Institute, \\ Prayagraj (Allahabad) 211019, India}
\emailAdd{arindam.chatterjee@snu.edu.in, asesh@hri.res.in, subhojitroy@hri.res.in}
\preprint{HRI-RECAPP-2022-001}
\abstract{We study in detail the viability and the patterns of a strong
first-order electroweak phase transition as a prerequisite to electroweak 
baryogenesis in the framework of $Z_3$-invariant Next-to-Minimal Supersymmetric 
Standard Model (NMSSM), in the light of recent experimental results from the Higgs sector, dark matter (DM) searches and those from the searches of the lighter chargino and neutralinos at the Large Hadron Collider (LHC). For the latter, we undertake thorough recasts of the relevant, recent LHC analyses.
With the help of a few benchmark scenarios, we demonstrate that 
while the LHC has started to eliminate regions of the parameter space with relatively small $\mueff$, that favors the coveted strong first-order phase transition, rather steadily, 
there remains phenomenologically much involved and compatible regions of the same which are yet not 
sensitive to the current LHC analyses. It is further noted that such a region could also be 
compatible with all pertinent theoretical and experimental constraints. We then proceed to analyze the prospects of detecting the stochastic gravitational 
waves, which are expected to arise from such a phase transition, at various 
future/proposed experiments, within the mentioned theoretical framework and 
find them to be somewhat ambitious under the currently projected sensitivities of 
those experiments.
}
\begin{document}
\maketitle
%
%%%%%%%%%%%%%%%%%%%%%%
\section{Introduction}
\label{Introduction}
%%%%%%%%%%%%%%%%%%%%%%
%
Electroweak phase transition (EWPT), leading to electroweak symmetry breaking 
(EWSB), is central to the process of baryogenesis at the electroweak scale or Electroweak Baryogenesis (EWBG)~\cite{Kuzmin:1985mm, Shaposhnikov:1986jp, Shaposhnikov:1987tw} which can explain the observed preponderance of 
(primordial) baryons over 
antibaryons, the so-called Baryon Asymmetry in the (present-day) Universe 
(BAU). 

The customary measure of BAU, 
$Y_B$, is the ratio of the difference between baryon and antibaryon densities 
($n_B$ and $n_{\bar{B}}$, respectively) and the entropy density ($s$), i.e., 
$Y_B=(n_B-n_{\bar{B}}) / s$. Its most precise value to date
($Y_B=8.65 \pm 0.09 \times 10^{-11}$) comes from the measurement at the Planck 
experiment~\cite{Ade:2015xua} of the baryon acoustic oscillations that it gives 
rise to in the power spectrum of the cosmic microwave background (CMB).

For baryogenesis to take place, the much celebrated set of following three 
Sakharov criteria~\cite{Sakharov:1967dj} are to be necessarily met: (i) baryon 
number non-conservation ($\slashed{B}$), (ii) $C$ and $CP$ violations
($\slashed{C}$, $\cancel{CP}$) and (iii) departure from thermal equilibrium.
EWBG is no exception. However, reference~\cite{Sakharov:1967dj} was 
found to be rather prescient about scenarios based on Grand Unified Theories 
(GUTs)~\cite{Georgi:1974sy,Kolb:1990vq,Cohen:1993nk}.

BAU can also be realized
in some other motivated extensions of the Standard Model (SM) of particle physics where the same arises from a (s)lepton asymmetry (i.e., via leptogenesis)~\cite{Fukugita:1986hr, DAmbrosio:2003nfv, Pilaftsis:2003gt}
in a supersymmetric (SUSY) framework or via the Affleck-Dine mechanism~\cite{Affleck:1984fy,Dine:1995kz} or even with the help of gravitational effects \cite{Davoudiasl:2004gf}. 
Among all these, EWBG has attracted special attention as it necessarily invokes physics beyond the Standard Model (BSM) down at around the electroweak (EW) scale which is being (and will be) intensely probed at various experiments
including at the colliders.
Naturally, EWPT, as an essential trigger for EWBG, has continued to be an area of active research~\cite{Cohen:1993nk, Rubakov:1996vz, Trodden:1998ym, Riotto:1998bt, Cline:2006ts, Morrissey:2012db, White:2016nbo}.

However, at temperatures as low as around the weak scale, while
$\slashed{B}$ (as anomaly effects~\cite{tHooft:1976rip}, via finite-temperature 
`sphaleron' transitions) and $\cancel{C}$ and $\cancel{CP}$ (induced by the CKM 
phase) could be present in a scenario with electroweak interactions like 
the SM, it is difficult to find a departure from thermal
equilibrium~\cite{Cline:2006ts}.
%(when a heavy excitation with mass
%$\sim 100$ GeV requires to decay out of equilibrium, i.e., at a temperature 
%lower than its mass) as it requires exceptionally feeble couplings in the 
%theory~\cite{Cline:2006ts}.
%
EWPT can salvage the situation if it is a first-order phase transition 
(FOPT) and that also of a `strong' nature. Such a transition proceeds 
in steps starting with 
the nucleation of bubbles of the broken phase in the 
cosmological plasma of the symmetric phase, followed by their expansions and 
eventual collisions and mergers until the whole space is engulfed by the broken 
phase. The process is violent enough to trigger local departures from thermal 
equilibrium in the vicinity of the walls of the rapidly expanding bubbles
in the plasma.

Unfortunately, however, an FOPT (and hence EWBG) cannot be realized in the SM given that the observed SM-like Higgs boson is too heavy ($\mhsm \approx 125$ GeV~\cite{ATLAS:2012yve, CMS:2012qbp}) for the
purpose~\cite{Bochkarev:1987wf, Kajantie:1995kf}. This is since such a 
value of $\mhsm$ signifies a large enough Higgs quartic coupling ($\lambda_H$) 
which virtually suppresses the term cubic in the Higgs fields in the `effective' 
(higher-order) Higgs (scalar) potential. This deprives the potential of a crucial bump 
(as it varies with the field(s)) which is essential for an FOPT. Also,
$\cancel{CP}$ from the CKM phase in the SM is proven inadequate for generating enough 
chiral
asymmetries~\cite{Gavela:1993ts, Huet:1994jb, Gavela:1994dt, Kapusta:2006pm} 
for $\slashed{B}$ to occur. Hence SM, as such, cannot lead to EWBG.

It is well known that popular SUSY extensions of the SM, viz., the 
Minimal SUSY SM (MSSM) and its next-to-minimal 
incarnation (NMSSM), a priori, provide the right setup ~\cite{Espinosa:1993yi, Pietroni:1992in} for EWBG. 
In the presence of an extended Higgs sector in these scenarios and other scalar degrees of freedom (in 
particular, the top squarks) in their spectra, an effective Higgs 
potential of the right kind for an FOPT to occur can be found. Furthermore, some new Lagrangian 
parameters could now be the sources of additional $\cancel{CP}$ that triggers
$\slashed{B}$ thus facilitating the generation of BAU.
%The possibilities in this regard have been extensively studied over the past %decades.
However, since the MSSM parameter space favoring a strong first-order electroweak phase transition (SFOEWPT) has now got highly
disfavored (as it requires rather light top squarks~\cite{Liebler:2015ddv} which are constrained by the LHC searches), the NMSSM (and its 
variants) has stolen the limelight.
 
EWPT in the NMSSM is rather appealing because of the presence of a 
gauge singlet scalar field which helps generate a 
barrier between the symmetric and the broken electroweak phases of the Higgs 
potential that is required for an FOPT. Unlike in the MSSM, such a 
barrier may now arise even at the tree level and at zero-temperature thanks to the presence of cubic terms in the Higgs potential. 
Thus, the dimensionful couplings in these cubic terms in the soft 
Lagrangian involving the singlet scalar field and the doublet Higgs 
fields could play important roles in altering the barrier in favor of an
SFOEWPT~\cite{Pietroni:1992in}. In addition, thermal effects including the
so-called daisy contributions can also play an important role in the process~\cite{Menon:2004wv}.

Naturally, there has been a continued activity over the past decades
exploring myriad aspects and possibilities of EWPT in the NMSSM. In particular, some of these shed 
light on how SFOEWPT, the experimental constraints on the dark matter (DM) 
observables and the spectrum of the singlet- and/or doublet-like scalars are 
connected across the NMSSM parameter
space~\cite{Carena:2011jy, Kozaczuk:2013fga, Huang:2014ifa}, its region over 
which simultaneous compatibility of SFOEWPT and the Galactic Centre Excess 
(GCE) can be found~\cite{Bi:2015qva} while some others present 
analyses of EWBG in the presence of SFOEWPT~\cite{Kozaczuk:2014kva, Bian:2017wfv, Huber:2006wf, Balazs:2013cia, Cheung:2012pg, Huang:2014ifa}.

Further, a recent study~\cite{Athron:2019teq} has undertaken a detailed 
probe into the patterns of phase transitions, based on calculations of critical temperature, that are possible over an 
extended region of the \z3nmssm~parameter space using the package 
\phasetracer~\cite{Athron:2020sbe} which is designed specifically for the purpose. 
Subsequently, in the context of such a scenario (in the so-called `alignment 
without decoupling limit' in the Higgs sector), it has been
demonstrated~\cite{Baum:2020vfl} with the help of the package 
\cosmotransitions~\cite{Wainwright:2011kj} that ensuring a successful 
nucleation of a bubble of the broken electroweak phase is more crucial
than just confirming the presence of a critical temperature for an FOPT.
 
In both the studies mentioned above~\cite{Athron:2019teq, Baum:2020vfl}, only the Higgs-related constraints from the 
LHC and the bounds on the chargino-neutralino (electroweakinos) sectors from 
the LEP experiments are considered. Stringent bounds on the latter 
sector from the recent LHC studies and those on the DM observables, viz., the DM relic 
abundance and the DM direct detection (DMDD) rates for both the
spin-independent (SI) and the spin-dependent (SD) cases, are, however, not imposed 
in either of these works. As mentioned there, such considerations are perfectly 
justified in dedicated studies of EWPT in which neither the 
properties of these electroweakinos in general, nor those of the DM are of much 
practical concern.

Going beyond, our goal in this work is to examine the prospects of SFOEWPT in the \z3nmssm~once the latest constraints from the LHC and the DM sector are included in the analysis and their implications thereof.
Given that these constraints are already known to be intricately connected over the \z3nmssm parameter space, such an exercise takes off by throwing the physics of EWPT into the mix. Together, these are likely to shed more light on the viability of EWBG in such a framework. We would, however (as is customary in such studies), remain agnostic about the extra sources of $\slashed{C}$, $\cancel{CP}$ or how
$\slashed{B}$ is achieved with the understanding that those could always be arranged optimally.

It is perhaps straightforward to imagine~\cite{Kozaczuk:2014kva} that the issues in the DM, the LHC and the EWPT sectors are all connected via the higgsino `portal'. For, the effective higgsino mass parameter ($\mueff$) of the
scenario could affect all those sectors significantly, especially, intricately as
there are a few other model parameters that appear both in the
electroweakino and the Higgs sectors. 
As we will see, such a connection gives rise to a tantalizing possibility
that relatively light higgsinos with masses under a few hundred GeV, in the presence of an even lighter singlino and/or a bino, with or without accompanying singlet-like scalar(s), might have been, somewhat comfortably, escaping their searches at the LHC even at this matured stage
of the experiment.
In particular, we seek to explore how small
a $\mueff$ could still be viable in view of the current experimental constraints
given that it is somewhat motivated by `naturalness' and, at the same time, is preferred by SFOEWPT and hence by EWBG.

On the other side of the proceedings, the dynamics of the nucleated bubbles 
could generate gravitational waves (GW)~\cite{Apreda:2001us, Grojean:2004xa, Weir:2017wfa, Caprini:2019egz, Witten:1984rs,Hogan:1986qda, Ellis:2018mja, Alanne:2019bsm}. These would be stochastic in nature and could be detected by 
dedicated ground-based and space-borne experiments. Note that in the SM, EWPT 
is of a cross-over type. Hence it does not produce any GW in the early 
Universe. That is why the detection of such a stochastic background would 
likely to hint physics beyond the SM (BSM). In the context of the NMSSM, the production of such GW has recently been studied~\cite{Huber:2007vva, Kozaczuk:2014kva, Huber:2015znp, Bian:2017wfv}. In this work, we also present, for a chosen set of scenarios, the prospects of detection 
of such a GW background in future experiments.

The present work is organized as follows.
In section~\ref{sec:model} we briefly discuss the \z3nmssm scenario with a 
focus on its scalar and electroweakino sectors which the present study is particularly sensitive to.
Section~\ref{sec:ewbg} summarizes the generalities of EWBG by stressing SFOEWPT 
as its prerequisite. A schematic details of EWPT is then presented in the 
\z3nmssm~scenario, matched to the THDSM, in terms of the finite-temperature 
effective (scalar) potential and the target region of the parameter space for 
our present study is outlined. A brief discussion on the mechanism of GW 
production in FOPT follows where we collect its basic theoretical ingredients.
In section~\ref{sec:results} we present our results where we delineate the 
relevant region of the parameter space, choose a few benchmark scenarios for 
our purpose that meet all primary constraints, show that some of these do 
survive explicit recasts of some recent, relevant LHC analyses (first of its 
kind, in the current context) and demonstrate in some detail how SFOEWPT is 
realized in each such case which together underscores an overall preference for 
a relatively small $\mueff$. Prospects of detecting the GW at future 
experiments in these viable scenarios are then presented.
In section~\ref{sec:summary} we conclude with an outlook for the future. A 
three-part appendix outlines the key details of the implementation of our scenario in~\cosmotransitions.

\section{The theoretical framework: the \z3nmssm} 
\label{sec:model}
%%%%%%%%%%%%%%%%%
In this section we discuss the theoretical framework, i.e., the \z3nmssm with 
conserved $R$-parity, by outlining the superpotential, the soft SUSY breaking 
Lagrangian of the scenario followed by a brief description of its Higgs 
(scalar) and electroweakino sectors, that are relevant for the present work, at 
the tree-level.

The superpotential is given by~\cite{Ellwanger:2009dp}
\beq
{\cal W}= {\cal W}_\mathrm{MSSM}|_{\mu=0} + \lambda \widehat{S}
\widehat{H}_u \cdot \widehat{H}_d
        + {\kappa \over 3} \widehat{S}^3 \, ,
\label{eqn:superpot}
\eeq
where ${\cal W}_\mathrm{MSSM}|_{\mu=0}$ is the MSSM superpotential with its
higgsino mass term (the $\mu$-term) dropped, $\widehat{H}_u, \widehat{H}_d$ and 
$\widehat{S}$ are the $SU(2)$ Higgs doublet superfields and the gauge singlet 
superfield, respectively, and `$\lambda$' and `$\kappa$' are dimensionless 
parameters. The (real) scalar component of the singlet superfield $\widehat{S}$ 
assumes a non-zero vacuum expectation value ({\it vev}) $v_s$ during EWPT thus 
generating an effective $\mu$-term as $\mu_{\rm eff}=\lambda v_s/\sqrt{2}$.
Correspondingly, the soft SUSY-breaking Lagrangian is given by
\beq
-\mathcal{L}^{\rm soft}= -\mathcal{L_{\rm MSSM}^{\rm soft}}|_{B\mu=0}+ m_{S}^2
|S|^2 + (
\lambda A_{\lambda} S H_u\cdot H_d
+ \frac{\kappa}{3}  A_{\kappa} S^3 + {\rm h.c.}) \,,
\label{eqn:lagrangian}
\eeq
where $m_S$ is the soft SUSY-breaking mass of the singlet scalar field, `$S$', 
$H_u$ and $H_d$ are the doublet Higgs fields and $\alambda$ and
$\akappa$ are the NMSSM-specific trilinear soft couplings with mass dimension one. 
%
%%%%%%%%%%%%%%%%%%%%%%%%%%%%%
\subsection{The Higgs sector}
\label{subsec:higgs-sector}
%%%%%%%%%%%%%%%%%%%%%%%%%%%
The tree-level Higgs (scalar) potential of the \z3nmssm takes the following 
form: 
\beq
V_{\textrm{tree}}^{\textrm{NMSSM}} = V_F + V_D + V_\textrm{soft} \, ,
\label{Eq:NMSSMHiggspotential}
\eeq
where $V_F$, $V_D$ and $V_\textrm{soft}$ represent contributions from the $F$- 
and the $D$-terms and the soft SUSY--breaking terms, respectively and are given 
by
%%%%%%%%%%%%%
\begin{align}
V_F &= \left
|\lambda S \right|^2 (|H_u|^2 + |H_d|^2) + \left |\lambda H_u \cdot
H_d + \kappa S^2 \right |^2 , \\
V_D &= \frac18g^2(|H_u|^2 - |H_d|^2)^2 + \frac12 g_2^2|H_u^\dagger H_d|^2 \,,\\
V_\textrm{soft} &= m_{H_u}^2 |H_u|^2 + m_{H_d}^2 |H_d|^2 + m_{S}^2
|S|^2 + (\lambda A_\lambda S H_u \cdot H_d + \frac13 \kappa A_\kappa S^3 +
  \textrm{h.c.}) \, ,
\end{align}
%%%%%%%%%%%
where $g^2 = (g_1^2 + g_2^2)/2$ and $g_1$ and $g_2$ are, respectively, the 
$U(1)$ and the $SU(2)$ gauge couplings. In our present study, we consider the 
Lagrangian parameters $\lambda, \kappa, \alambda$ and $\akappa$ to be real. The complex scalar fields can be expressed as 
%%%%%%%%%%%%%
\begin{align}
H_u= \begin{pmatrix} H_u^+\\ \tfrac{1}{\sqrt{2}} \left(h_u + i a_u\right) 
\end{pmatrix}, \quad \quad
H_d= \begin{pmatrix} \tfrac{1}{\sqrt{2}}  \left(h_d + i a_d\right) \\ H_d^- 
\end{pmatrix}, \quad \quad
S= \frac{1}{\sqrt{2}} \left(s + i \sigma\right),
\label{complexfields}
\end{align}
%%%%%%%%%%%
where $\langle h_u\rangle= v_u$, $\langle h_d\rangle= v_d$ and
$\langle s\rangle= v_s$ are the \vevs~ of the real components
($CP$-even) of the neutral scalar fields that refer to the tree-level scalar 
potential at zero temperature. Note that $\sqrt{\vu^2 + \vd^2}= v \simeq 246$ 
GeV with $\tanb= \vu/\vd$ and $\mueff= \lambda \vs/\sqrt{2}$.

On electroweak symmetry breaking (EWSB), the doublet and the singlet scalars 
could mix and the physical Higgs states arise. The tree-level mass-squared 
matrices for the $CP$-even, the $CP$-odd and the charged scalars in the bases
\{$h_d, h_u, s$\}, \{$a_d, a_u, \sigma$\} and \{$H_u^+,{H_d^-}^*$\}, 
respectively, are obtained by expanding the scalar potential of
equation~\ref{Eq:NMSSMHiggspotential} around $\vd$, $\vu$ and $\vs$ (see equation~\ref{eq:NMSSMTreelevel}) and taking 
its double derivatives with respect to the scalar fields of the involved types.
Diagonalizations of these mass-squared matrices lead to three $CP$-even, two 
$CP$-odd and two charged physical Higgs states. One of the lighter $CP$-even 
states has to be the observed SM-like Higgs boson, $\hsm$. Thus, there is the 
interesting phenomenological possibility that one scalar state from each of the 
$CP$-even and the $CP$-odd sectors is light and is singlet-like ($\hs$ and
$\as$) and hence might have managed to escape detection at various collider 
experiments. Their heavier counterparts ($H$ and $A$) would then be similar to 
those found in the MSSM. Note that these Higgs masses depend on the cubic 
couplings in which the parameters $\alambda$ and $\akappa$ appear and these are 
found to play important roles in achieving FOEWPT. Also, for an FOEWPT (and 
hence for an EWBG), of particular interest is the effective scalar potential at 
finite-temperature. We discuss its salient aspects in section~\ref{sec:ewbg}.
%
%%%%%%%%%%%%%%%%%%%%%%%%%%%%%%%%%%%%%%
\subsection{The electroweakino sector}
\label{subsec:ewino-sector}
%%%%%%%%%%%%%%%%%%%%%%%%%%%
The neutralino sector of the \z3nmssm consists of five neutralinos
which are mixtures of bino ($\bino$), wino ($\wino^0_3$), two higgsinos 
($\higgsinod$, $\higgsinou$) and a singlino ($\singlino$) which is the 
fermionic component of the singlet superfield $\widehat{S}$ appearing in the 
superpotential of equation~\ref{eqn:superpot}. The symmetric, real 
$5\times 5$ neutralino mass-matrix, ${\cal M}_0$, in the gauge (weak) basis
$\psi_0 \equiv \{\widetilde{B},~\widetilde{W}^0_3, ~\widetilde{H}_d^0,~\widetilde{H}_u^0, ~\widetilde{S}\}$, is given by~\cite{Ellwanger:2009dp}
\beq
{\cal M}_0=
\left( \begin{array}{ccccc}
M_1 & 0 & -\dfrac{g_1 \vd}2 & \dfrac{g_1 \vu}2 & 0 \\[0.4cm]
\ldots & M_2 & \dfrac{g_2 \vd}2 & -\dfrac{g_2 \vu}2 & 0 \\
\ldots & \ldots & 0 & -\mu_{\rm eff} & -\dfrac{\lambda \vu}{\sqrt{2}} \\
\ldots & \ldots & \ldots & 0 & -\dfrac{\lambda \vd}{\sqrt{2}} \\
\ldots & \ldots & \ldots & \ldots & {\sqrt{2}} \kappa v_s
\end{array} \right) \, ,
\label{eqn:mneut}
\eeq
where $\mone$ ($\mtwo$) is the soft SUSY-breaking mass for the bino 
(wino). The [5,5] element of ${\cal M}_0$ is the singlino mass 
term, $\msinglino= {\sqrt{2}}\kappa \vs$. ${\cal M}_0$ can be 
diagonalized by an orthogonal $5 \times 5$ matrix `$N$', i.e.,
%%%%%
\beq
N {\cal M}_0 N^T= {\cal M}_D= {\rm diag}(m_{{_{\chi}}_{_1}^0},m_{{_{\chi}}_{_2}^0},m_{{_{\chi}}_{_3}^0},m_{{_{\chi}}_{_4}^0},m_{{_{\chi}}_{_5}^0})  \, , 
\label{eqn:diagonalise-1}
\eeq
%%%%
when the neutralino mass eigenstates, $\ntrli$, are given in terms of the weak eigenstates, $\psi_j^0$, by
%%%%
\beq
\ntrli = N_{ij} \psi_j^0 \, , \quad \text{with} \;\; i,j=1,2,3,4,5 \, ,
\label{eqn:diagN2}
\eeq
%%%%
and $\ntrli$'s are ordered in increasing mass as `$i$' increases. In this 
study, we set $\mtwo$ large. Thus, the heaviest neutralino ($\ntrlfive$) 
is almost a pure wino and is indeed heavy with a mass
$\mntrlfive \approx \mtwo$. Hence the wino practically decoupled when
${\cal M}_0$ effectively reduces to a $(4\times 4)$ matrix.
The scenario conserves $R$-parity which is odd for the SUSY excitations. Thus, 
the lightest SUSY particle (LSP) which is taken to be the lightest neutralino 
($\ntrlone$) in this work turns stable and can be a good DM candidate.

The chargino sector of the \z3nmssm is exactly the same as in the MSSM but for 
$\mu$ $\rightarrow$ $\mueff$. The $2 \times 2$ chargino mass-matrix,
${\cal M}_C$, in the gauge bases
$\psi^+ = \{ -i \widetilde{W}^+, \, \widetilde{H}_u^+ \}$ and
$\psi^- = \{ -i \widetilde{W}^-, \, \widetilde{H}_d^- \}$,
is given by~\cite{Ellwanger:2009dp}
\beq
{\cal M}_C = \left( \begin{array}{cc}
                    \mtwo   & \quad  \dfrac{g_2 \vu}{\sqrt{2}} \\
                  \dfrac{g_2 \vd}{\sqrt{2}}  & \quad \mueff 
             \end{array} \right) .
\eeq
As in the MSSM, ${\cal M}_C$ can be diagonalized by two $2 \times 2$ unitary 
matrices `$U$' and `$V$', i.e.,
\beq
U^* {\cal M}_C V^\dagger = \mathrm{diag} (\mcharone , \mchartwo) \; , \quad
\mathrm{with} \;\;  \mcharone < \mchartwo  \; ,
\label{eqn:uvmatrix}
\eeq
where, in the present work, $\charonepm$ ($\chartwopm$) is higgsino-like
(wino-like) given that we set $\mtwo$ large.

As we will find, a relatively light singlino-dominated neutralino, which,
at times, can be the LSP, has a special context in this work~\cite{Abel:1992ts}. The latter 
requires $|{\sqrt{2}}\kappa \vs| < |\mueff|, \, |M_1|$. Given
$\vs= \sqrt{2}\mueff/\lambda$, this then requires `$\kappa$' to be on the 
smaller side (with $|\kappa|< 2 \lambda$) which is just what an SFOEWPT 
prefers. Furthermore, the mutual hierarchy among these electroweakinos would 
have important implications for their phenomenologies at the LHC. 
%
%%%%%%%%%%%%%%%%%%%%%%%%%%%%%%%%%%%%%%%%%%%%%%%%%%%%%%%%%%%%%%%%%%%%%%%%
\section{EWPT in the NMSSM: a prerequisite to EWBG and its implications}
\label{sec:ewbg}
%%%%%%%%%%%%%%%%
In this section we take a quick tour into the generalities of EWBG and its 
association with (FO)EWPT followed by a brief discussion of the latter in the
\z3nmssm. Some relevant analytical details which have gone into our 
implementations of the scenario in \cosmotransitions~are deferred to the 
appendices.
%
%%%%%%%%%%%%%%%%%%%%%%%%%%%%%%%%%
\subsection{Generalities of EWBG}
\label{subsec:bg}
%%%%%%%%%%%%%%%%%
Like any successful model of baryogenesis, EWBG also requires the three 
Sakharov criteria, as mentioned in the Introduction, are to be fulfilled. As noted 
there, EWBG exploits FOEWPT which triggers electroweak symmetry breaking (EWSB) 
at the characteristic energy-scale ($\sim 100$ GeV). In the process, important 
roles are played by the radiative~\cite{Coleman:1973jx} and
finite-temperature~\cite{Kapusta:2006pm} corrections to the Higgs potential. 
For the FOEWPT, the latter ensures an optimal evolution of the potential 
as the Universe expands and cools down from an early, hot (radiation-dominated) 
epoch where the electroweak symmetry was still
intact~\cite{Dolan:1973qd,Weinberg:1974hy}.

A possible FOEWPT is envisaged when there appear (at least) two distinct 
local minima of the finite-temperature effective Higgs potential, 
separated already by a barrier when the temperature ($T$) of the Universe is 
such that $T >T_c$, where $T_c$ is the so-called `critical temperature', i.e., 
the temperature at which the two minima become degenerate, still separated by a 
barrier. One such minimum is a trivial one with a vanishing potential for null 
values of the participating scalar fields where the electroweak symmetry is 
(trivially) preserved. For $T<T_c$, the true (global) minimum emerges with a 
smaller value of the potential for finite field-values in the broken phase and 
the field(s) at the trivial (local) minimum (the false vacuum) naturally tries 
to tunnel to the true one~\cite{Linde:1977mm,Linde:1978px,Linde:1981zj}.

The tunneling process is efficiently modeled in terms of a bubble of the broken 
electroweak phase nucleated locally in the cosmological plasma (in which 
the electroweak symmetry is intact) that starts growing as the rate of 
nucleation ($\Gamma_B$, per unit volume) exceeds the same for the Hubble 
expansion. A bubble, once formed, continues to expand, collide and 
coalesce with other bubbles growing in the plasma until a giant one, formed 
this way, engulfs the whole space thus making EWSB permeate all over.
At finite-temperatures ($T$), in the semi-classical approximation,
$\Gamma_B \propto T^4 \exp(-S_3(T)/T)$~\cite{Langer:1969bc,Coleman:1977py,Affleck:1980ac}, 
where $S_3(T)$ is the effective three-dimensional Euclidean action evaluated at 
the (``bounce") solution of the classical field equation. The minimal 
requirement for a successful completion of an EWPT requires the bubble 
nucleation rate to be one per Hubble volume per Hubble time. This is met when
$\frac{S_3(T)}{T} \simeq 140$~\cite{Linde:1981zj, Mazumdar:2018dfl, Quiros:1999jp}. The corresponding nucleation temperature $T_n \,(\lesssim T_c)$ 
is the highest temperature for which $\frac{S_3(T)}{T} \lesssim 140$ is 
satisfied, as the Universe cools down. We use
\cosmotransitions~\cite{Wainwright:2011kj}
to calculate this bounce solution by employing path deformation method.

Along the way, for EWBG to take place, the three Sakharov conditions are met 
and play their roles~\cite{Kuzmin:1985mm, Cohen:1993nk} in the following 
manner.
%
%%%%%%%%%%%%%%%
\begin{itemize}
\item
The SM can give rise to $\slashed{B}$~\cite{tHooft:1976rip} thanks to the 
triangle anomaly~\cite{Adler:1969gk,Bell:1969ts}. This is described in terms of 
the vacuum configurations of the static gauge field of the unbroken $SU(2)_L$ 
gauge theory in which alternating degenerate vacua with integer-value
Chern-Simons numbers carry different baryon numbers and are separated by 
potential barriers whose constant height ($E_\text{sph}$) is given by 
static solutions that are known as
``sphalerons''~\cite{Manton:1983nd,Klinkhamer:1984di,Kunz:1992uh}.
At finite-temperatures $\slashed{B}$ occurs via sphaleron transitions (hopping of the
barriers)~\cite{Kuzmin:1985mm} from one vacuum to another. The transition rate 
(per unit volume per unit time) in the symmetric phase scales as $T^4$, while 
in the broken phase, the same is suppressed exponentially as
$\exp(\frac{-E_{\rm sph}(T)}{T})$~\cite{Arnold:1987mh,Khlebnikov:1988sr,Carson:1990jm}. The same mechanism works in the SUSY extensions of the SM, including 
the NMSSM~\cite{Moreno:1996zm,Funakubo:2005bu}.
\item 
Complementary sphaleron-induced processes would generate similar excesses in 
baryons and antibaryons thus leading to a null baryon asymmetry. When the 
underlying theory possesses $\cancel{CP}$, preferential scattering of fermions 
with a specific chirality in the symmetric phase with the expanding bubble wall 
could generate both $CP$ and $C$ asymmetries in the particle number densities 
in that phase thus biasing the sphalerons there to generate more baryons than
antibaryons~\cite{Farrar:1993sp,Farrar:1993hn}.
As noted in the Introduction, while SM does not have a strong enough source of
$\cancel{CP}$, SUSY extensions like the NMSSM have new sources of $\cancel{CP}$ in the form of phases in the extended Higgs sector and/or in the gaugino masses 
etc., which make up for the deficit.
\item
Even in the presence of $\slashed{B}$, $\slashed{C}$ and $\cancel{CP}$, the 
equilibrium average of net baryon number vanishes as a consequence of
$CPT$-invariance~\cite{Trodden:1998ym, Riotto:1998bt}. Thus, to create a 
maintainable baryon asymmetry in the front of the bubble wall, the cosmological 
plasma in its vicinity should depart from thermal equilibrium. Such a departure 
is generically realized under FOEWPT when the nucleated bubble rapidly expands 
through the plasma.
\item
Some fraction of this baryon asymmetry thus generated in the symmetric phase 
subsequently diffuses into the broken
phase~\cite{Cohen:1993nk, Bodeker:1999gx, Morrissey:2012db} thanks to the 
motion of the bubble wall. For $T < T_c$ (more precisely, for $T < T_n$),
$\frac{E_{\rm sph}(T)}{T}$ is large in the broken phase and the exponential 
suppression in the rate of sphaleron transitions, as mentioned under the first 
item above, kicks in. Quantitatively, for
$\frac{\phi_n}{T_n} \equiv \gamma_{_{\rm{EW}}} \gtrsim 1$,\footnote{ In the context of the NMSSM, $\gamma_{_{\rm{EW}}} = \frac{\phi_n}{T_n} = \frac{\Delta SU(2)}{T_n} = \frac{\sqrt{((h_d)_{\text{true}} - (h_d)_{\text{false}})^2+((h_u)_{\text{true}} - (h_u)_{\text{false}})^2}}{T_n}$.} i.e., for a 
``strong'' FOEWPT,  where $\phi_n=\langle \phi \rangle_{T_n}$ in the broken 
phase~\cite{Bochkarev:1990gb, Quiros:1999jp, Moore:1998swa}, this rate per unit 
volume falls out of equilibrium thus rendering the rate of $\slashed{B}$ too 
slow to wash out the baryon-asymmetry that has sneaked into the broken phase. 
This completes the process of successful baryogenesis.
\end{itemize}
Given that a FOEWPT is central to the process of EWBG, we briefly review the 
same in the next subsection in the context of \z3nmssm.
%
%%%%%%%%%%%%%%%%%%%%%%%%%%%%%%%%%%%%%%%%%%%%%%%%%
\subsection{Study of EWPT in the \z3nmssm}
\label{subsec:ewpt-nmssm}
%%%%%%%%%%%%%%%%%%%%%%%%%
In this section we outline the formulation of EWPT in the \z3nmssm and discuss 
the viability of SFOEWPT that facilitates EWBG over the model parameter space. 
We assume that there is no spontaneous or explicit \cancel{$CP$} in the Higgs 
sector. 
%
%%%%%%%%%%%%%%%%%%%%%%%%%%%%%%%%%%%%%%%%%%%%%%%%%%%%%%%%%%%%%%%
\subsubsection{Effective Higgs potential at finite-temperature}
\label{subsec:effpot}
%%%%%%%%%%%%%%%%%%%%%
To study the viability of SFOEWPT in the \z3nmssm, we start with the 
description of the effective potential for the Higgs sector. The
zero-temperature radiatively corrected (at one-loop) effective potential for 
the ($CP$-even) Higgs sector is given (in the $\overline{\text{MS}}$ scheme and 
in the Feynman gauge) by~\cite{Patel:2011th}
\begin{alignat}{3}
\label{eq:CW1Loop}
V_{\rm eff} & = V_{\rm tree}  && +  V_{\rm CW} && \notag \\ 
            & = V_{\rm tree}  && +  \frac{1}{64 \pi^2} && \Bigg(
  \sum_h n_h m_h^4  \left[\ln\left( \frac{m_h^2 }{Q^2}\right) - 3/2\right]
+ \sum_V n_V m_V^4 \left[\ln\left(\frac{m_V^2}{Q^2}\right) - 5/6\right] \\
& && &&- \sum_V \tfrac13 n_V m_V^4 \left[\ln\left(\frac{m_V^2}{Q^2}\right) - 3/2\right]
- \sum_f n_f m_f^4 \left[\ln\left(\frac{m_f^2}{Q^2}\right) - 3/2\right]\Bigg),
\notag
\end{alignat}
where $V_{\rm tree}$ is the tree-level potential for the $CP$-even Higgs fields 
$h_u,~ h_d$ and $s$ (see appendix \ref{NMSSMTHDMSmatch}) and $V_{\rm CW}$ is 
the well-known Coleman-Weinberg~\cite{Coleman:1973jx} one-loop correction to
$V_{\text{tree}}$ whose form is shown in the second and the third lines of the 
equation. There, $m_j$ and $n_j$ are the field-dependent
($\overline{\text{MS}}$) masses (see Appendix~\ref{field-dependent-masses}) and 
the degrees of freedom for the species `$j$', respectively, and the $n_j$'s are 
found to be as follows:
\begin{align}
& n_{h^0_i} = n_{A^0_i}  = n_{H^+_i} = n_{H^-_i} = 1 \, ,\qquad n_{W^+} =n_{W^-}=n_{Z} = 3 \, , \nonumber \\
& \hskip 40pt n_t = n_b =12,\, n_{\tau} = 4 \, , \qquad n_{\chi_i^0} = 2, 
n_{\chi_1^+}=n_{\chi_1^-} = 2 \, .
%\label{eq:field_dofs}
\end{align}
Note that the scalar states, $A^0_i$ and $H^{\pm}_i$, include the 
Goldstone bosons and that the wino-like states are taken to be decoupled
(as pointed out in section~\ref{subsec:ewino-sector}). At finite-temperatures, 
the ($CP$-even) Higgs-sector potential receives additional contributions that 
are (in the Feynman gauge) given by~\cite{Dolan:1973qd, Weinberg:1974hy, Kirzhnits:1976ts} 
{\small
\beq
\widetilde{V}_T= \frac{T^4}{2 \pi^2} \Bigg[
\sum_h n_h J_B \left(\frac{m_h^2 }{T^2}\right)
+ \sum_V n_V J_B \left(\frac{m_V^2}{T^2}\right)
-  \sum_V \frac{1}{3} n_V J_B\left(\frac{ m_V^2}{T^2}\right)
+ \sum_f n_f J_F\left(\frac{m_f^2}{T^2}\right)\Bigg] \, ,
\label{eq:thermal_one_loop}
\eeq
}
{\flushleft
{where the thermal function $J_B$ ($J_F$) captures the relevant thermal 
contribution from the bosons (fermions), and is given by}
}
\beq
J_{B/F}(y^2)=
\pm {\rm Re}
\int_0^{\infty}
		x^2 \ln
		\left(
			1 \mp \exp^{-\sqrt{x^2 + y^2}}
		\right)
{\rm d}{x} \, ,
\label{eq:jbjf}
\eeq
with the upper (lower) signs appearing for bosons (fermions). This
reveals that for $m_i^2 >> T^2$,  i.e., for large $|y^2|$, these thermal 
functions are exponentially (Boltzmann-) suppressed. Hence any massive new 
physics excitation that has been integrated out from the theory could 
never have a finite-temperature implication. In the reverse limit, i.e., at 
high temperatures with $|y^2|<<1$, $J_{B/F}$ can be approximated as
\begin{subequations}
\label{e.JBFhighT}
%\begin{gather}
\begin{align}
%\nonumber
J_B(y^2) &\approx 
J_B^{\mathrm{high}-T}(y^2) = 
-\frac{\pi^4}{45} + \frac{\pi^2}{12} y^2  - \frac{\pi}{6} y^3 - \frac{1}{32} y^4 \ln\left( \frac{y^2}{a_b}\right),
\label{eq:JBhighT}
\\
J_F(y^2) &\approx
J_F^{\mathrm{high}-T}(y^2) = 
 - \frac{7 \pi^4}{360} + \frac{\pi^2}{24} y^2 + \frac{1}{32} y^4 \ln\left(\frac{y^2}{a_f}\right), \quad
\label{eq:JFhighT}
\end{align}
%\end{gather}
\end{subequations}
where
$a_b= \pi^2 \exp(3/2 - 2 \gamma_E)$ and $a_f= 16 \pi^2 \exp(3/2 - 2 \gamma_E)$,
$\gamma_E$ being the Euler-Mascheroni constant ($\approx 0.577$). 
The term $- \frac{\pi}{6} y^3$ appearing in the high-temperature expansion of $J_B$ in equation \ref{eq:JBhighT} gives rise to a negative contribution cubic in the bosonic field in the finite-temperature effective potential $\widetilde{V}_T$. As pointed out earlier, the presence of this term can generate an energy barrier between two degenerate vacua, thus facilitating an SFOPT. Note that such a cubic term appears only for bosonic degrees of freedoms as it comes from the (Matsubara) zero mode propagator which exists only for them.\footnote{For a discussion in the context of SM, 
see, for example, reference~\cite{Anderson:1991zb} and the review
articles~\cite{Cohen:1993nk, Rubakov:1996vz, Quiros:1999jp, Riotto:1998bt,Morrissey:2012db}}

At high temperatures, the perturbative approximations at one-loop
suffer from large temperature-dependent contributions from additional
higher-order processes given by the so-called 
``daisy'' (or ``ring'') diagrams. Their dominant 
contributions to the scalar masses obtained from the resummation of these diagrams are captured in the daisy 
potential given by~\cite{Athron:2019teq, Arnold:1992rz,Parwani:1991gq,Carrington:1991hz}
\beq
V_{\rm daisy} = \frac{-T}{12 \pi} \left(
\sum_h n_h \left[ \left(M_h^2\right)^{\frac{3}{2}} - \left(m^2_h\right)^{\frac{3}{2}} \right] + \sum_V \frac{1}{3} n_V \left[\left(M_V^2\right)^{\frac{3}{2}} - \left(m^2_V\right)^{\frac{3}{2}} \right]
\right)\, ,
\label{eq:daisypot}
\eeq
where $M_h^2$ and $M_V^2 $ are the eigenvalues of the thermally improved (i.e., Debye-corrected) mass-squared matrices of the Higgs and the gauge bosons which are presented in Appendix~\ref{field-dependent-masses}~\cite{Arnold:1992rz}. Note that only the longitudinal mode of each of the gauge bosons contributes to the daisy potential of equation \ref{eq:daisypot}.
The one-loop finite-temperature effective potential thus becomes
\beq
V_T= V_{\rm eff} + \widetilde{V}_T + V_{\rm daisy} \, ,
\eeq
which is then used in the study of EWPT where one tracks its minima as a 
function of temperature. Its profile for $T \simeq T_c$ is important for the 
purpose. However, the locations of the extrema of $V_T$, as well as the ratio
$\phi_c(T_c)/T_c$, are both gauge-dependent~\cite{Dolan:1973qd,Nielsen:1975fs,Fukuda:1975di,Laine:1994zq,Baacke:1993aj,Baacke:1994ix}\footnote{
The gauge-independent quantities of the effective potential are found using the Nielsen identities~\cite{Nielsen:1975fs,Fukuda:1975di}.} see, for 
example, references~\cite{Garny:2012cg, Espinosa:2016nld, Patel:2011th, Arunasalam:2021zrs, Lofgren:2021ogg}. We have checked the minimization of $V_T$
using both Landau and Feynman gauges and have found that the gauge-dependencies 
of both $\phi_c(T_c)$ and $T_c$ are not significant for the benchmark scenarios we present.

It should be noted here that even for moderately heavy top squarks, which 
couple intensely to the doublet Higgs fields with the top quark Yukawa
coupling, $y_t$, their presence would give rise to large logarithms in
$V_\text{CW}$ (in equation~\ref{eq:CW1Loop}) because of a large enough 
hierarchy between $m_t$ and $\mstonetwo$. Such large corrections to tree-level 
potential point to significant dependence of the results on the renormalization 
scale `$Q$' and a reliable study of phase transition would thus call for 
treating the potential at higher orders.

To circumvent the problem, one could adopt the effective field theory (EFT) 
approach in which the top squarks are integrated out from the theory thus 
resulting in a scenario with two Higgs doublets, a singlet scalar, the 
electroweakinos and the entire SM spectrum. Thus, the scalar sector of this 
scenario matches with the one known in the
literature~\cite{Elliott:1993ex, Elliott:1993uc, Elliott:1993bs} as the
($Z_3$-symmetric) Two Higgs Doublet Model with a Singlet scalar (THDMS) 
extension of the SM. Hence we adopt the 
tree-level scalar potential of the THDMS in the present work to study EWPT and
make use of the relevant results obtained in
references~\cite{Elliott:1993ex, Elliott:1993uc, Elliott:1993bs, Athron:2019teq, Kozaczuk:2014kva} where
a similar consideration is made.
The model parameters of the tree-level scalar potential of the $Z_3$-symmetric 
THDMS are derived in terms of those appearing in the corresponding potential in 
the \z3nmssm at the scale $M_\text{SUSY}$ where the latter is matched onto the 
former. This correspondence is discussed in Appendix~\ref{NMSSMTHDMSmatch}. We, 
thus, adopt the following steps~\cite{Kozaczuk:2014kva} to compute the 
effective potential appropriate for our present study. 
\begin{itemize} 
\item The NMSSM model parameters are taken to be the $\overline{\rm DR}$ ones 
at the scale $M_\text{SUSY} \, (=\sqrt{\msQthree \msUthree})$ following the 
convention of the spectrum generator \nmssmtools~\cite{Ellwanger:2009dp} which 
we use for generating the particle spectrum.
\item Following references~\cite{Elliott:1993ex, Elliott:1993uc, Elliott:1993bs, Athron:2019teq, Kozaczuk:2014kva}, the relevant THDMS parameters 
appearing in $V_{\rm tree}^{\rm THDMS}$
of equation~\ref{eq:THDMS_potential} are then expressed, at the scale
$M_\text{SUSY}$, in terms of the NMSSM parameters appearing in
$V_{\rm tree}^{\rm NMSSM}$ of equation~\ref{eq:NMSSMTreelevel} by taking 
into account the relevant threshold correction that arises as the top squarks 
are integrated out (see Appendix~\ref{NMSSMTHDMSmatch}).
\item We assume that except for the additional Higgs bosons and the
higgsino-, the singlino- and the bino-like electroweakinos, all new physics 
excitations are rather heavy and hence decoupled. Thus, we use the appropriate 
set of renormalization group equations (RGEs) which now include contributions 
from all the states in the THDMS scenario, along with those from these lighter 
electroweakinos, to obtain the respective THDMS parameters at a reference 
renormalization scale $m_t$ at which the logarithmic contribution from the top 
quark to the physical minimum of the potential is minimized, and which, also 
closely resembles the energy scale for the EWSB. $V_{\rm tree}$ (at zero 
temperature) is then expressed in terms of these parameters of the THDSM at the 
scale $m_t$. To make the present work self-contained, we present the set of 
relevant RGEs~\cite{Kozaczuk:2014kva} in Appendix~\ref{RGEs}.
\item We then evaluate the zero-temperature one-loop contribution $V_{\rm CW}$ 
of equation \ref{eq:CW1Loop}. Further, we make use of a specific 
remormalization condition to ensure the dependence of $V_{\rm CW}$ on the 
renormalization scale ($Q$) is minimized~\cite{Cline:2011mm} (see appendix 
\ref{field-dependent-masses}). Finally, the finite-temperature effective 
potential, $V_T$, for the $CP$-even scalar fields, are obtained as described 
earlier.
\end{itemize}

We have used the package \cosmotransitions~\cite{Wainwright:2011kj} to track 
the evolution of the finite-temperature effective potential $V_T$ and to find
$T_c$. Further, the evolution of the potential for $T \lesssim T_c$ has also 
been studied in order to determine if successful bubble nucleation could occur 
for our benchmark scenarios. As has been done in some recent
studies~\cite{Athron:2019teq, Baum:2020vfl}, we also study in detail the 
patterns of phase transitions for some of these scenarios. These pertain to 
issues like the number of steps taken for the transition to complete, whether it is 
of a first or a second-order type and the field directions along which a 
multi-step transition occurs.
%
%%%%%%%%%%%%%%%%%%%%%%%%%%%%%%%%%%%%%%%%%%%%%%%%%%%%%%%%%%
\subsubsection{Target region of the NMSSM parameter space}
\label{subsubsec:target}
%%%%%%%%%%%%%%%%%%%%%%%%
In this section we take a brief overall look into what the possibility of an 
efficient SFOEWPT would imply for the \z3nmssm~parameter space when experimental constraints, in particular, from the observed Higgs sector and from the
DM-sector, are also factored in. This leads to our target region of the parameter space from which we choose a few benchmark scenarios to examine their viability against recently reported LHC results on searches of electroweakinos.

For the purpose, it would be instructive to take a quick look into the
tree-level NMSSM potential, $V_{\text{tree}}^{\text{NMSSM}}$, of
equation~\ref{eq:NMSSMTreelevel}. Considering only the singlet field, a 
suitable barrier in the potential profile that makes an SFOPT possible develops 
when the relative contribution from the trilinear term
$\sim \kappa \akappa s^3$ increases in comparison to the quartic term 
$\sim \kappa^2 s^4$.\footnote{Such an interplay has been reviewed in 
the context of the thermally corrected effective scalar potential of the
SM~\cite{Quiros:1999jp}.} The strength of the transition (parametrized by the 
ratio of the cubic to the quartic term) increases for a reduced `$\kappa$' 
since the latter term diminishes faster. Generically, for a given FOPT, 
increasing $\akappa$ (i.e., enhancing the cubic term above) strengthens the same. Furthermore, the term trilinear in the singlet and the doublet scalar fields ($\sim \lambda \alambda h_d h_u s$) in $V_{\text{tree}}^{\text{NMSSM}}$
could further reinforce the SFOPT (which can now take place in all field directions) for a suitable $\alambda$.
Futhermore, it has been noted~\cite{Carena:2011jy,Huang:2014ifa} that an SFOEWPT  prefers relatively light singlet- and doublet-like scalars as these enhance higher order effects in the effective potential. It is also found that thermal effects, including the daisy contributions, could turn crucial in giving rise to  coveted barriers between the involved minima.  

The upshot is the following. A smaller `$\kappa$' that an SFOPT already prefers leads to a lighter $\hs$. At the same time, this causes a singlino-like state to turn lighter which has crucial implications for the DM and the LHC phenomenologies. On the other hand, to find $\as$ on the 
lighter side, $\akappa$ needs to be so optimally small that it does not make its contribution to the cubic soft term, $\sim \kappa \akappa s^3$, insignificant. A near-parallel argument holds for the requirement on the size of $\alambda$ which
controls the masses of the doublet-like Higgs states and has a somewhat similar role to play for the potential profile via the trilinear term $\sim \lambda \alambda h_d h_u s$ as does $\akappa$ via the terms cubic in `$s$', as mentioned above.

It may, however, be noted that since 
$\hs$ could mix with $\hsm$ on 
EWSB, a light $\hs$ quickly attracts stringent bounds from the experimental 
studies of $\hsm$.
Furthermore, a light $\hs$ is also somewhat disfavored by 
the observed upper limits on the DMDD-SI rates from various DM experiments 
unless in the presence of a so-called blind spot~\cite{Badziak:2015exr, Cheung:2012qy, Cheung:2014lqa, Badziak:2015nrb, Badziak:2017uto} occurring due to a destructive interference among the diagrams with $CP$-even Higgs states appearing in their propagators. Hence settling for a lone, light $\as$ with a sizable coupling with $\hsm$ is a safer option when looking for an SFOEWPT in the \z3nmssm. Note, however, that
regions of parameter space over which $\hsm$ could have on-shell decays to $\hs$
and/or $\as$ (i.e., when $m_{\hs,\as} <\mhsm/2$) would be highly constrained by the latest LHC data on $\hsm$~\cite{Aaboud:2018fvk, Aaboud:2018gmx, Sirunyan:2020eum, ATLAS:2021hbr}. 

As we have discussed earlier, opting for smaller values of `$\kappa$' would, in 
turn, results in a light singlino-like LSP ($\msinglino=\sqrt{2}\kappa \vs$) which could be the 
viable DM candidate of the scenario. Apropos of this, as mentioned earlier, a light higgsino-triplet (comprised of a pair of neutralinos and a chargino) resulting from a smaller $\mueff=\lambda \vs/\sqrt{2}$ is very much in the context of the present work which, under circumstances, could as well provide the LSP. A larger value of $\vs$ can re-introduce the problems with large logarithms from one-loop corrections since the field-dependent masses depend on $\vs$.
Keeping this in mind, we consider $\vs \leq 2$ TeV. Thus, relatively small 
values of $\mueff \; ({\cal O}(100) \; \text{GeV})$ is achievable for reasonably large values of `$\lambda$'. This, in conjunction with
relatively small values of $\tanb$ ($< 10$), helps find $\mhsm$ in the right ballpark, even for not-so-heavy top squarks thus letting $\mhsm$ appear somewhat `natural'~\cite{Ellis:1986yg, Barbieri:1987fn, Baer:2012up,Baer:2012cf}. We, however, have not restricted ourselves very strictly
to this regime and allowed for somewhat larger values of soft masses ($\msQthree$ and $\msUthree$) for the squarks and trilinear coupling ($A_t$) from the third generation.

For smaller values of $\tanb$, on the other hand, some extra regions of the parameter space could now find
compliance with the DMDD-SI constraints by exploiting the so-called `coupling blind spot' condition $g_{\hsm \ntrlone \ntrlone}=0 \Rightarrow \mntrlone/\mueff = \pm \sin2\beta$ (`+ (-)' for singlino (bino)-like LSP)\footnote{More involved general blind spot conditions of DMDD-SI and -SD cross sections for the $4 \times 4$ neutralino (bino-higgsino-singlino) system is derived
in~\cite{Abdallah:2020yag}.} when $|\mueff|$ tends to approach the LSP mass. This allows
us to study a rather nontrivial setup within the NMSSM with a large possible
mixing of the higgsinos with the singlino or with the bino. Note that $\mone$ 
is not expected to influence the physics of the phase transitions in any drastic way since it enters the calculation of the finite-temperature effective potential via radiative corrections. Hence we have chosen its values (around the electroweak scale) to suit our purpose on DM and collider physics grounds. Thus, an involved situation might arise when all of `$\kappa$', $\mueff$ and $\mone$ are on the smaller side such that any of the
lighter electroweakinos can be dominantly of a particular type or even mixed states. As we will soon find, its implications for the phenomenology of the
electroweakinos at the LHC are rather subtle in connection to the physics of both DM and EWPT.
It must, however, be noted that since the DMDD-SD rate has the dependence $\sigma^\mathrm{SD} \propto 1/\mueff^4$, lowering $\mueff$ beyond a point would quickly attract stringent bounds from the relevant DMDD experiments.    

Furthermore, given that we, by now, find that the optimal setup prefers
smaller values of $\tanb$, we could afford to consider doublet-like heavy
Higgs bosons (`$H$' and `$A$') of the scenario to be on the lighter side and still passing the latest relevant constraints on them from the LHC
experiments in the form of bounds on the $m_{H^\pm}$--$\tanb$~\cite{ATLAS:2021upq} and $m_A$-$\tanb$~\cite{ATLAS:2020zms} planes.
This is of some importance since relatively light doublet-like Higgs bosons could potentially render the FOEWPT stronger, provided such a light `$H$' survives the DMDD-SI constraints.
The stage is now set for a brief but important discussion on the 
phenomenology of such (relatively) light electroweakinos. Dedicated LHC 
searches for these states over the past years have put stringent lower bounds 
on their masses and those are becoming even stronger with time. However, these
analyses are generally restricted to simplified MSSM scenarios in terms of the 
spectrum/hierarchy of these states and their consequent patterns of cascades
leading to the final states of interest. In general, a scenario like
\z3nmssm~could easily invalidate such assumptions in the presence of possible 
new, light states (for example, the light singlet-like scalars and the 
singlino). These could then diminish the sensitivities of various target final 
states and/or tailored signal regions to the experimental analyses thus 
weakening the lower bounds on the masses of such electroweakinos. In fact, 
there are myriad such possibilities in our current \z3nmssm~setup which could
lead to such a situation \cite{Abdallah:2020yag, Abdallah:2019znp}. On top of that, the LHC experiments mostly assume (at least, the analyses that are are relevant for the present work)  the electroweakinos produced in the hard scattering are of wino type for which the relevant cross sections are the largest.

Given the wino decouples from our analysis, for any specific mass the next largest cross section is for the higgsino-pairs which is already about half of that for a corresponding wino-like pair. This further reduces the sensitivity of various final states to the experimental analyses. As has been already pointed out, given the central role that $\mueff$ plays in the DM and EWPT sectors, the search for light higgsino-like states has now become of special significance.
The bottom line is that the published lower bound on the electroweakino masses are bound to get more relaxed for these higgsino-like states under a situation different from
what the experiments assumed for their analyses. However, it is not a
straightforward exercise to come up with the relaxed bounds for a given new situation and any such attempt requires thorough recasts of the existing analyses which we will attempt in this work.

On a conservative note, we do not consider possible situations which could result in weakened bounds on the masses of the electroweakinos when these have a compressed spectrum. In our case, for a light higgsino-triplet with a higgsino-like LSP, the lower bound on $\mueff$ could go down to a value as small as $\sim 220$ GeV, even with 139 \fbinv~of data~\cite{ATLAS:2021moa,CMS:2021edw}.

Guided by the above understanding, we lay down our strategy in
section~\ref{sec:results} for numerical exploration of the scenario 
before presenting there our results.
%
%%%%%%%%%%%%%%%%%%%%%%%%%%%%%%%%%%%%%%%%%%%%%%%%%%%%%%%%%%%%%%%
\subsection{Production of GW from first-order phase transition}\label{GW_section}
%%%%%%%%%%%%%%%%%%
Given that the NMSSM could provide us with an ideal setup for an FOPT that might have taken place in the early Universe, a study of GW 
originating from such an FOPT, in the context of our present work, is in order. 
As noted in the Introduction, GW from an FOPT would exist in the form of a
stochastic background and has been proposed to be searched for using the so-called
``cross-correlation'' method~\cite{Caprini:2015zlo, Cai:2017cbj,Caprini:2018mtu, Romano:2016dpx, Christensen:2018iqi}.
The salient mechanisms via which GW could arise from an FOPT and their corresponding contributions to the GW energy density (scaled by the critical density $\rho_c$
for the Universe with a vanishing cosmological constant $\Lambda$)
are as follows.
\begin{itemize}
\item Collisions of the expanding bubble walls release stress energy located at their walls, as well as lead to possible subsequent shocks, in the 
intervening plasma made up of relativistic particles~\cite{Kosowsky:1991ua, Kosowsky:1992vn, Kosowsky:1992rz, Kamionkowski:1993fg, Caprini:2007xq,  Huber:2008hg}. However, for a 
phase transition occurring in a thermal plasma, their  contributions to GW energy density
%, $\Omega _\phi h^2$, from the scalar field `$\phi$' itself,
are believed to be negligible~\cite{Bodeker:2017cim} and hence can be ignored.
\item Bulk motion (velocity perturbations) of the plasma generates sound 
(acoustic) waves (longitudinal modes) that propagate in the same during 
the time interval between collisions of bubbles and the expanding new phases 
dissipating their kinetic energy in the
plasma~\cite{Hindmarsh:2013xza, Giblin:2013kea, Giblin:2014qia, Hindmarsh:2015qta}.
These sound waves contribute to the GW energy density as
$\Omega_{\mathrm{sw}} h^2$, where
$h= H_0/(100 \,\mathrm{km} \,.\, \mathrm{sec}^{-1} . \, \mathrm{Mpc}^{-1})$
$\approx 0.674$~\cite{DES:2017txv}, with $H_0$ standing for the present-day
(red-shift $z=0$) value of the Hubble parameter, also known as the Hubble 
constant. Such acoustic contributions, when accumulated over the said duration, 
are expected to dominate.
\item Turbulence in the plasma of magneto-hydrodynamic (MHD) origin set up on 
collisions of the bubbles~\cite{Caprini:2006jb, Kahniashvili:2008pf, Kahniashvili:2008pe, Kahniashvili:2009mf, Caprini:2009yp, Kisslinger:2015hua} 
contributes to GW energy density as $\Omega_{\mathrm{turb}} h^2$.
\end{itemize}
The overall GW energy density can be approximated as a linear combination of the latter two contributions, i.e., 
\beq\label{GWtotal}
\Omega_{\text{GW}}h^2 \simeq \Omega_{\text{sw}}h^2+\Omega_{\text{turb}}h^2 \, .
\eeq
A few key FOPT parameters, in addition to the bubble nucleation temperature, 
$T_n$, that can be obtained from the particle physics models and which control 
these two contributions can be categorized as follows.
\begin{itemize}
\item The parameter `$\alpha$', which relates to the energy budget of the FOPT, is given by~\cite{Espinosa:2010hh}
\beq
\alpha = \frac{\rho_{\text{vac}}}{\rho^*_{\text{rad}}} = \frac{1}{\rho^*_{\text{rad}}}\left[T\frac{\d \Delta V(T)}{\d T} - \Delta V(T)\right]\Bigg|_{T_*},
\eeq
where $T_*=T|_{t_*}$ with $t_*$ being the instant of time when the FOPT 
completes. In the absence of significant effects from reheating,
$T_* \simeq T_n$. $\Delta V(T) = V_{\text{low}}(T)-V_{\text{high}}(T)$ is the 
difference between the potential energies at the false and the true minima and
$\rho^*_{\text{rad}} = g_* \pi^2 T^4/30$ where $g_*$ is the number of the 
relativistic degrees of freedom at $T=T_*$, taken here to be $\sim 100$, following recent literature.
\item The parameter `$\beta$', which gives the inverse time-duration of the 
FOPT, can be derived in terms of the effective 3-dimensional Euclidean 
action ($S_3(T)/T$) as~\cite{Kamionkowski:1993fg} 
\beq
\beta = -\frac{d S_3(T)}{dt}\Bigr|_{t_*} \simeq H_*T_*
\frac{d(S_3(T)/T)}{dT}\Bigr|_{T_*},
\eeq
%
%where $t_*$ is the instant of time when the EWPT completes, $T_*=T|_{t_*}$ and %$T_* \simeq T_n$ in the absence of significant reheating, while
where $H_* = H|_{T_*}$. For a stronger GW signal, the EWPT should occur over a larger duration of time, i.e., it should be a slow process and hence the ratio $\beta/H_*$ needs to be on the smaller side. 
\item The parameter $v_w$, which pertains to the bubble dynamics, i.e., the wall-velocity of the expanding bubble, needs to be larger for a more intense GW emission, although an optimally strong EWBG is known to be favored only for a tiny, subsonic $v_w$ instead.
\end{itemize}
The sound wave contribution to the GW energy density, $\Omega_\text{sw} h^2$, as a function of the above FOPT parameters and the frequency `$f$' of the GW, is then given by~\cite{Hindmarsh:2013xza, Hindmarsh:2016lnk, Hindmarsh:2017gnf}
\vskip -25pt
\begin{equation}\label{eq:75}
\Omega_{\text{sw}}{\rm h}^2=2.65\times 10^{-6} \; \Upsilon(\tau_\text{sw}) \left(\dfrac{\beta}{H_{\star}} \right) ^{-1} v_{w} \left(\dfrac{\kappa_{v} \alpha}{1+\alpha}\right)^2 \left(\dfrac{g_*}{100}\right)^{-\frac{1}{3}}\left(\frac{f}{f_{\text{sw}}}\right)^{3} \left[\frac{7}{4+3 \left(\frac{f}{f_{\text{sw}}}\right)^{2}}\right]^{\frac{7}{2}} \, ,
\end{equation}
where $\kappa_v$ is the fraction of the energy from the phase transition that gets converted into the bulk motion of the plasma which leads to GW and is of the form~\cite{Caprini:2019egz, Chiang:2019oms}
\begin{equation}\label{eq:76}
\kappa_v \simeq \left[ \frac{\alpha}{0.73+0.083\sqrt{\alpha}+\alpha}\right] \, ,
\end{equation}
$f_{\text{sw}}$ is the present day peak frequency for the sound wave 
contribution to GW energy density given by (with the approximation
$T_{\star} \approx T_n$)~\cite{Huber:2008hg}
\begin{equation}\label{eq:77}
f_{\text{sw}}=1.9\times10^{-5}\hspace{1mm} \text{Hz} \left( \dfrac{1}{v_{w}}\right)\left(\dfrac{\beta}{H_{\star}} \right) \left(\dfrac{T_n}{100 \hspace{1mm} \text{GeV}} \right) \left(\dfrac{g_*}{100}\right)^{\frac{1}{6}}\,\, ,
\end{equation}
$\Upsilon (\tau_\text{sw})$ is the parameter that brings in the effect of a 
finite lifetime of the sound waves which suppresses their contributions to the 
GW energy density and is given by~\cite{Guo:2020grp, Hindmarsh:2020hop}
\beq
\Upsilon (\tau_\text{sw}) = 1 - \frac{1}{\sqrt{1 + 2 \tau_{\text{sw}} H_{*}}} \, ,
\label{eq:upsilon}
\eeq
where the lifetime $\tau_{\text{sw}}$ is considered as the time scale when the 
turbulence  develops and is given by
$\tau_{\text{sw}} \approx R_{*}/\bar{U}_f$~\cite{Pen:2015qta,Hindmarsh:2017gnf},
where, in turn, $R_{*}= (8\pi)^{1/3} v_w /\beta$ is the mean bubble
separation~\cite{Hindmarsh:2019phv, Guo:2020grp} and
$\bar{U}_f \simeq \sqrt{\frac{3}{4} \frac{\kappa_v \alpha}{1 + \alpha}}$ is the root-mean-squared (RMS) fluid 
velocity obtained from a hydrodynamic
analysis~\cite{Hindmarsh:2019phv,Weir:2017wfa, Ellis:2019oqb}. Note that as
$\tau_{\text{sw}} \rightarrow \infty$, $\Upsilon \to 1$, asymptotically. On the 
other hand, for all our benchmark scenarios presented in
section~\ref{subsubsec:allowedbms}, $\tau_{\text{sw}} H_{*} < 0.1$ when $\Upsilon \to \tau_{\text{sw}} H_{*}$.
Furthermore, there is a growing realization~\cite{No:2011fi} that $v_w$ might 
not enter the calculation of the EWBG. Then, to maximize the strength of the 
GW, it is assumed that the expanding bubbles attain a relativistic terminal 
velocity in the plasma, i.e., we consider $v_w \simeq 1$.

Note that in the above calculation, the estimation of the portion of energy transferred to the fluid motion is based on the so-called bag model \cite{Espinosa:2010hh}. A recent work \cite{Giese:2020rtr} proposes a model-independent approach (by going beyond the bag model) to obtain this quantity. In that work the parameter $\alpha_{\bar{\theta}}$ quantifying the strength of the phase transition is given by
\beq\label{beyondbagmodelpseudotrace}
\alpha_{\bar\theta} \equiv \frac{D \bar\theta}{3 \omega_s(T_s)} \quad \text{with} \quad \bar\theta \equiv e - p / c^2_{s_{_b}} \;\; , 
\eeq
where the subscript $s$ ($b$) corresponds to the symmetric (broken) phase, $\bar{\theta}$ is the difference in energy ($e$) and pressure ($p$) in a given phase and known as the pseudo-trace, $D\bar\theta \equiv \bar\theta_s(T_s) - \bar\theta_b(T_s)$ is the difference in its value in the symmetric and the broken phases, $c_{s_{_b}}$ being the speed of sound in the broken phase which is defined as
\beq
c^2_{s_{_b}} \equiv \left.\frac{dp_b/dT}{de_b/dT}\right|_{T_s},
\eeq
while $\omega_s = (e + p)_s$ is the enthalpy density in the symmetric phase.

The GW power spectrum due to sound wave from beyond the bag model can then be obtained from equation~(\ref{eq:75}) by just carrying out the  following replacement:
\beq\label{beyondbagmodelrelation}
\frac{\alpha_e \kappa_\nu}{\alpha_e+1} \rightarrow \left(\frac{D\bar\theta}{4 e_s}\right)\kappa_{\bar\theta} \;\; .
\eeq
Subsequently, the GW spectra, within and beyond the bag model, are compared in  section~\ref{GWresult}.

Furthermore, the MHD turbulence contribution to the GW energy density is given by~\cite{Caprini:2015zlo}
\vskip -15pt
\beq
\Omega_{\text{turb}}h^2 =3.35\times 10^{-4} \left(\frac{\beta}{H_*}\right)^{-1} \left(\frac{\kappa_\text{turb} \alpha}{1+\alpha}\right)^{\frac{3}{2}}
\left(\frac{100}{g_*}\right)^{\frac{1}{3}}v_w\frac{(f/f_\text{turb})^3}{[1+(f/f_\text{turb})]^\frac{11}{3}(1+8\pi f/h_*)} \; ,
\eeq
where $k_\text{turb}$ is not precisely known but is expected to be in the range of 5\%--10\% of $k_v$~\cite{Hindmarsh:2015qta}. We set $k_\text{turb}=0.1 k_v$ in our calculation.
The present-day peak frequency $f_{\text{turb}}$ of the GW spectrum from the turbulence contribution is given by
\beq\label{Turbpeakfreq}
f_{\text{turb}}= 2.7 \times 10^{-5}~\text{Hz} \, \frac{1}{v_w}\left(\frac{\beta}{H_*}\right)\left(\frac{T_*}{100~\text{GeV}}\right)\left(\frac{g_*}{100}\right)^{\frac{1}{6}},
\eeq
with
$h_{*}=16.5\times10^{-6}\hspace{1mm} \text{Hz} \left(\dfrac{T_n}{100 \hspace{1mm} \text{GeV}} \right) \left(\dfrac{g_*}{100}\right)^{\frac{1}{6}}$.
%
%%%%%%%%%%%%%%%%%%%%%%%%%%%%%%%%%%
\section{Results}
\label{sec:results}

%%%%%%%%%%%%%%%%%%%
In this section we start by presenting the ranges of various input parameters 
of the scenario that we adopt to carry out a scan over the theory space. This 
is followed by a brief discussion on the pertinent constraints coming from 
relevant DM and collider experiments including the crucial ones arising from 
the studies of the observed Higgs boson to which we subject the scan. A few 
benchmark scenarios are then chosen for which SFOEWPT occurs. To pursue the 
central goal of this work, these scenarios are further classified to 
show how, in the light of what we discuss in
section~\ref{subsubsec:target}, a few of them with relatively small $\mueff$ get 
disallowed by current 
LHC searches for the electroweakinos while some others survive. The prospects of finding GW signals at future experiments are briefly discussed 
for these surviving scenarios. 

In table~\ref{tab:ranges} we present the ranges of the input parameters that we 
scan over and mention the values of the relevant ones which are kept 
fixed. The choices are broadly motivated by the discussion in
section~\ref{subsubsec:target}. 
\begin{table}[t]
\renewcommand{\arraystretch}{1.6}
\begin{center}
\begin{tabular}{|c|c|c|c|c|c|c|c|c|c|}
\hline
$\lambda$ & $|\kappa|$ & $\tanb$& \makecell{$|\mueff|$ \\ (GeV)}&  \makecell{$|\alambda|$ \\ (TeV)} &
\makecell{$|\akappa|$ \\ (GeV)}
 & \makecell{$|\mone|$ \\ (GeV)} & \makecell{$|A_t|$ \\ (TeV)} & \makecell{$\msQthree$ \\ (TeV)} & \makecell{$\msUthree$ \\ (TeV)} \\
\hline
0.2--0.7 &$\leq 0.5$&
1--20& $\leq 500$ & $\leq 2$ & $\leq 200$ & $\leq$500& $\leq 5$& 2--5& 2--5\\
\hline
\end{tabular}
\caption{Ranges of various model  parameters adopted for scanning the \z3nmssm 
parameter space. The fixed values of various soft parameters used are as 
follows:
$m_{{\widetilde{D}_{3}}}
        =m_{\tilde{L}, \widetilde{E}} = 3.5$~TeV, $A_{b,\tau}= 3$~TeV, $\mthree =3$~TeV and $\mtwo=2.5$~TeV.  }
\label{tab:ranges}
\end{center}
\end{table} 
%
%%%%%%%%%%%%%%%%%%%%%%%%%%%%%%%%%%%%%%%%%%%%%
\subsection{Constraints from various sectors}
\label{subsec:constraints}
%%%%%%%%%%%%%%%%%%%%%%%%%%
In this work, we take into account constraints from various sectors, both 
theoretical and experimental. The theoretical ones include ensuring the 
spectra to be free from tachyonic states, the scalar potential not develop an 
unphysical global minimum and the evolutions of various pertinent couplings of 
the theory with energy not encounter Landau poles, etc. The experimental
constraints include those coming from the Higgs, the DM and the flavor sectors 
and from various searches for new physics at the colliders. We further ensure 
the occurrence of SFOEWPT that facilitates EWBG and that such a transition does 
end up in the physical vacuum. To impose these constraints and for our general 
numerical analysis, we employ publicly available packages like
\nmssmtools~{\tt (v5.5.3)}~\cite{Ellwanger:2005dv,Das:2011dg}, 
\higgsbounds~{\tt (v5.8.0)}~\cite{Bechtle:2020pkv}, \higgssignals~{\tt (v2.5.0)}~\cite{Bechtle:2020uwn},
\checkmate~{\tt (v2.0.34)}~\cite{Dercks:2016npn}, 
\smodels~{\tt (v2.1.1)}~\cite{Alguero:2021dig} and
\cosmotransitions {\tt (v2.0.6)}~\cite{Wainwright:2011kj}. Below we briefly
point out some of the important constraints that are obtained from these 
packages. 
%
%%%%%%%%%%%%%%%%
\begin{table}[tph]
\renewcommand{\arraystretch}{1.3}
\vspace{0.5cm}
\resizebox{1.0\textwidth}{!}{
\small{\begin{tabular}{llccc}
\hline
Analysis (Luminosity) & Process & Final State & \texttt{SModelS} &\texttt{CheckMATE} \\
\hline\hline \\ 
		\texttt{CMS-SUS-17-004}~\cite{Sirunyan:2018ubx} (35.9 $\texttt{fb}^{\texttt{-1}}$) &$\chi_2^0\chi_1^{\pm}\rightarrow Z/\hsm \,
		\chi_1^0 \, W^{\pm}\chi_1^0$ &$(m\geq0)\ell+ (n\geq0)\tau + \etmiss$& & \hspace{0.6cm} $\checkmark$ \\\\
		\multirow{2}{*}{\texttt{CMS-SUS-16-048}~\cite{CMS:2018kag} (35.9 $\texttt{fb}^{\texttt{-1}}$)} &$\tilde{t}\tilde{t}\rightarrow b \chi_1^{\pm}b \chi_1^{\pm}$& \multirow{2}{*}{$(k\geq0)\ell + (m\geq0)b + (n\geq0)$-jet$ + \etmiss$} & & \multirow{2}{*}{\hspace{0.6cm} $\checkmark$} \\
		&$ \chi_2^0\chi_1^{\pm}\rightarrow Z^{*} \chi_1^0 \; W^{{\pm}*}\chi_1^0$&                    &                   \\\\
				\multirow{5}{*}{\texttt{CMS-SUSY-16-039}~\cite{Sirunyan:2017lae} (35.9 $\texttt{fb}^{\texttt{-1}}$)} &$\chi_2^0\chi_1^{\pm}\rightarrow \ell\tilde{\ell}\ell\tilde{\nu}$ & \multirow{5}{*}{$(n\geq0)\ell$  + $\etmiss$}  & \hspace{0.6cm}\multirow{5}{*}{\checkmark} & \hspace{0.6cm}\multirow{5}{*}{\checkmark} \\ & $\chi_2^0\chi_1^{\pm}\rightarrow\tilde{\ell}\ell\tilde{\tau}\nu$ \\ &$\chi_2^0\chi_1^{\pm}\rightarrow\tilde{\tau}\tau\tilde{\tau}\nu$ \\ & $\chi_2^0\chi_1^{\pm}\rightarrow Z \chi_1^0 \; W^{\pm}\chi_1^0$ \\ &$\chi_2^0\chi_1^{\pm}\rightarrow \hsm \chi_1^0 \; W^{\pm}\chi_1^0$ \\\\
		\multirow{2}{*}{ \texttt{CMS-SUS-17-010}~\cite{Sirunyan:2018lul} (35.9 $\texttt{fb}^{\texttt{-1}}$) }             &$\chi_1^{\pm}\chi_1^{\mp}\rightarrow W^{\pm}\chi_1^0 \; W^{\mp}\chi_1^0$ & \multirow{2}{*}{$2\ell + \etmiss$} &\hspace{0.6cm} $\checkmark$ & \\ &$\chi_1^{\pm}\chi_1^{\mp}\rightarrow \nu\tilde{\ell} \; \ell\tilde{\nu}$ \\\\
		\texttt{ CMS-SUS-16-043}~\cite{CMS:2017kyj} (35.9 $\texttt{fb}^{\texttt{-1}}$)     &            $\chi_2^0\chi_1^{\pm}\rightarrow \hsm \chi_1^0 \; W^{\pm}\chi_1^0$& 1$ \ell ~+ $ 2$b$ + $ \etmiss$                & \hspace{0.6cm} \checkmark                  & \\\\		
		\texttt{ CMS-SUS-16-045}~\cite{Sirunyan:2017eie} (35.9 $\texttt{fb}^{\texttt{-1}}$)     &            $\chi_2^0\chi_1^{\pm}\rightarrow \hsm \chi_1^0 \; W^{\pm}\chi_1^0$& 1$ \ell ~+$ 2$\gamma$ + $ \etmiss$                & \hspace{0.6cm} \checkmark                  & \\\\
		\texttt{CMS-SUS-16-034~\cite{Sirunyan:2017qaj}} (35.9 $\texttt{fb}^{\texttt{-1}}$)     &   $\chi_2^0\chi_1^{\pm}\rightarrow Z /\hsm \tilde{\chi}_1^0 \; W^{\pm}\chi_1^0$&  $(m\geq2)\ell  + (n\geq1)$-jet$+ \etmiss$               & \hspace{0.6cm} \checkmark                  & \\\\
		\multirow{2}{*}
{\texttt{ATLAS-1712-08119}~\cite{ATLAS:2017vat} (36.1 \fbinv)}
& $\tilde{\ell}\tilde{\ell}$& \multirow{2}{*}{$2\ell + (n\geq0)$-jet$+ \etmiss$}
&
& \multirow{3}{*}{\hspace{0.6cm} $\checkmark$} \\
		&$\chi_2^0\chi_1^{\pm} \rightarrow  Z^{*} \chi_1^0 \; W^{*} \chi_1^0$&                    &                    \\\\
				\multirow{3}{*}{\texttt{ATLAS-1803-02762}~\cite{Aaboud:2018jiw} (35.9 $\texttt{fb}^{\texttt{-1}}$)} &$ \chi_2^0\chi_1^{\pm}\rightarrow Z\chi_1^0 \, W^{\pm}\chi_1^0$ & \multirow{3}{*}{ $(n\geq2) \ell + \etmiss$} & \hspace{0.6cm}\multirow{3}{*}{\checkmark} & \hspace{0.6cm}\multirow{3}{*}{\checkmark} \\ &$\chi_2^0\chi_1^{\pm}\rightarrow \nu\tilde{\ell}l\tilde{\ell}$\\ &$\chi_1^{\pm}\chi_1^{\mp}\rightarrow \nu\tilde{\ell}\nu\tilde{\ell}$\\\\
\texttt{ATLAS-1812-09432}~\cite{Aaboud:2018ngk} (36.1 $\texttt{fb}^{\texttt{-1}}$)
& $\chi_2^0\chi_1^{\pm}\rightarrow \hsm \chi_1^0 \, W^{\pm} \chi_1^0$
& $(j\geq0) \ell + (k\geq 0)$-jet $+ (m\geq 0)b + (n \geq 0)\gamma + \etmiss$ & \hspace{0.6cm} \checkmark                  & \\\\
						\texttt{ATLAS-1806-02293}~\cite{Aaboud:2018sua} (36.1 $\texttt{fb}^{\texttt{-1}}$)                  & $\chi_2^0\chi_1^{\pm}\rightarrow Z\chi_1^0 \, W^{\pm}\chi_1^0$ &  $(m\geq2)\ell + (n\geq0)$-jet$ + \etmiss$ & \hspace{0.6cm}  \checkmark             & \\\\\\\\
\texttt{ATLAS-1909-09226}~\cite{Aad:2019vvf} (139 $\texttt{fb}^{\texttt{-1}}$)                  & $\chi_{2}^0\chi_1^{\pm}\rightarrow \hsm \chi_1^0 \, W^{\pm} \chi_1^0$ & $1\ell + 2b + \etmiss$                  & \hspace{0.6cm} \checkmark                  & \\\\
		\texttt{ATLAS-1912-08479}~\cite{Aad:2019vvi} (139 $\texttt{fb}^{\texttt{-1}}$)&$\chi_2^0\chi_1^{\pm}\rightarrow Z(\rightarrow \ell\ell) \, \tilde{\chi}_1^0 \;\; W(\rightarrow \ell\nu) \, \tilde{\chi}_1^0$& $3\ell + \etmiss $                           & \hspace{0.6cm} \checkmark & \hspace{0.6cm} \checkmark \\\\
\multirow{2}{*}{\texttt{ATLAS-1908-08215}~\cite{Aad:2019vnb} (139 $\texttt{fb}^{\texttt{-1}}$)} &$\tilde{\ell}\tilde{\ell}$& \multirow{3}{*}{$2\ell + \etmiss $} & \hspace{0.6cm}\multirow{3}{*}{\checkmark} & \hspace{0.6cm}\multirow{3}{*}{\checkmark} \\
		&$\chi_1^{\pm}\chi_1^{\mp}(\chi_1^{\pm}\rightarrow  W^{\pm} \chi_1^0)$&  & \\
		&$~~~~~~~~(\chi_1^{\pm}\rightarrow \tilde{\ell}\nu/\tilde{\nu}\ell)$&  &                    \\\\
\multirow{2}{*}{\texttt{ATLAS-1911-12606}~\cite{ATLAS:2019lng} (139 $\texttt{fb}^{\texttt{-1}}$)} &$\tilde{\ell}\tilde{\ell}$& \multirow{2}{*}{jets$~+ 2\ell + \etmiss $} & & \hspace{0.6cm}\multirow{2}{*}{\checkmark} \\
		&$\chi_1^{\pm}\chi_2^0 \rightarrow W^{*}(\rightarrow q q)~ \chi_1^0 ~~ Z^{*} (\rightarrow ll)~  \chi_1^0$&  &                    \\\\		
		\texttt{ATLAS-2004-10894}~\cite{ATLAS:2020qlk} (139 $\texttt{fb}^{\texttt{-1}}$)&$\chi_2^0\chi_1^{\pm}\rightarrow \hsm(\rightarrow \gamma\gamma)~\chi_1^0 \; W(\rightarrow \ell\nu)\chi_1^0$& $1\ell + 2\gamma + \etmiss $                        & \hspace{0.6cm} \checkmark & \hspace{0.6cm} \checkmark \\\\
\hline \hline
\end{tabular}
}}
\caption{Relevant experimental analyses, along with the processes and final 
states considered, in search for the electroweakinos at the $13$~TeV LHC with the data sets at $\sim 36$ \fbinv~and 139 \fbinv~hat 
are implemented in \checkmate~and/or \smodels.}
%}}}
%
\label{tab:LHCewinoanalyses}
\end{table}	
%%%%%%%%%%%
%
\begin{itemize}
\item \nmssmtools~is used to compute and constrain various relevant observables 
from the Higgs, the DM, the flavor and the collider sectors. We impose the
$2\sigma$ upper limit on the DM relic abundance, i.e., $\Omega h^2 \leq 0.131$ 
as reported by the Planck experiment~\cite{Ade:2015xua, Aghanim:2018eyx}. The 
most recent (and improved) upper bounds on the DMDD-SI~\cite{Aprile:2018dbl} 
and -SD~\cite{Aprile:2019dbj,Amole:2019fdf} rates are taken into account after 
a commensurate downward scaling of these cross-sections (as the relic abundance 
drops below the Planck-allowed band) is done. This helps the computed DMDD-SI 
and -SD rates comply with the respective stringent upper bounds. For all the 
above-mentioned DM observables, their values are obtained from a dedicated 
package like \micromegas~{\tt (v4.3)}~\cite{Belanger:2006is} as adapted in \nmssmtools. The latter also takes into
account, albeit simplistically, the constraints from the CMS analysis 
on the electroweakino searches in the $3\ell + \etmiss$  final state with
35.9~\fbinv~worth of data~\cite{Sirunyan:2018ubx}.
\item Using \higgsbounds~and \higgssignals, we retain only those parameter points which pass the thorough checks of the Higgs sector. With the help of the latter package, we allow for Higgs signal-strengths which are consistent with the experimental findings at a $2\sigma$ level. To take into account the theoretical uncertainties in the computation of $\mhsm$, we consider $\mhsm$ over the range $122 \, \mathrm{GeV} < \mhsm < 128 \, \mathrm{GeV}$.
\item A few representative (benchmark) scenarios out of the resulting set are 
then subjected to thorough recasts, via the packages \checkmate~and \smodels, 
of a multitude of relevant LHC analyses that include several recent ones with 
139~\fbinv~of data. These analyses and their availabilities in these two 
packages are indicated in table~\ref{tab:LHCewinoanalyses}. Together, these 
are expected to provide us with the most stringent lower bounds on the 
masses of the electroweakinos under diverse circumstances which are pointed out while discussing those.
In addition, there are a few more rather recent LHC analyses~\cite{ATLAS:2021moa,CMS:2021cox,CMS:2021few}
which are expected to be sensitive to the scenarios we study but yet not available in the public versions of either of these two packages. We will get back to these in section~\ref{subsubsec:allowedbms}.
\item Parameter points that pass the previous set of constraints are subjected 
to analyses via \cosmotransitions~to check for SFOEWPT that results in the 
physical EW vacuum.
\end{itemize}
We, however, do not consider the recent experimental finding on muon
($g-2$)~\cite{Muong-2:2021ojo, Muong-2:2006rrc} since the dust is yet to settle 
over its BSM implications. We, thus, have set the masses of the smuons, along 
with all the sfermions, at a multi-TeV range.

When using \checkmate, we have generated, for each such analysis, Monte 
Carlo events for the leading order productions of all pertinent pair and 
associated productions of various electroweakinos at the 13 TeV LHC, i.e., for
$p p \rightarrow \chi_{_j} \chi_{_k}$, ($\chi_{_{j,k}} \in \{\chi_i^0, \chi_1^{\pm}\}$, with $i \in \{1-4\}$), with up to two additional 
partons, using \madgraph5~\cite{Alwall:2014hca}. These events are then passed 
through \pythia8~\cite{Sjostrand:2014zea} for generating parton showers, 
hadronization and decays of the unstable particles. The additional partonic 
jets from the matrix elements are then matched to those from the parton showers 
(the so-called ME-PS matching) using the MLM
prescription~\cite{Mangano:2006rw} built-in in \madgraph5. The resulting
events are passed through \delphes~\cite{deFavereau:2013fsa} to include the detector 
effects.
For an analysis using \smodels, we just provide the package with the SLHA
file along with the \madgraph5-generated cross sections of various production
processes as mentioned earlier. To account for the significant NLO+NLL 
contributions, all production cross sections have been multiplied by a flat
$k$-factor of 1.25~\cite{Fiaschi:2018hgm}. Both the recast packages calculate a 
$r$-value for a given theory point, where $r = (S-1.64 \Delta S)/ S95$,
with `$S$', $\Delta S$ and $S95$ signifying the predicted number of signal 
events, the associated Monte Carlo error and the experimental limit on `$S$' at 
95\% confidence level, respectively. Nominally, $r < (>) 1$ indicates the 
scenario to be allowed (disallowed).
%
%%%%%%%%%%%%%%%%%%%%%%%%%%%%%%%%%%%%%%%%%%%%%%%%%%%%
\subsection{Choice and study of benchmark scenarios}
\label{subsec:choice}
\begin{figure}[t!]
\begin{center}
\includegraphics[height=5.6cm,width=0.44\linewidth]{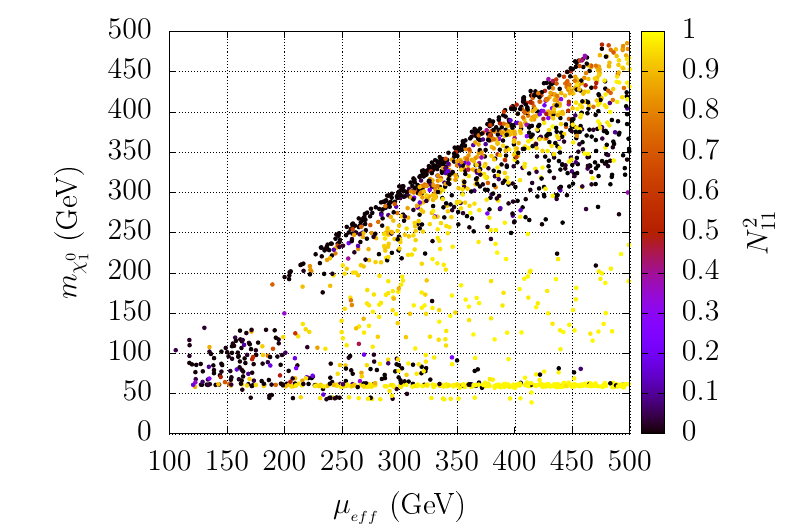}
\hskip 40pt
\includegraphics[height=5.6cm,width=0.44\linewidth]{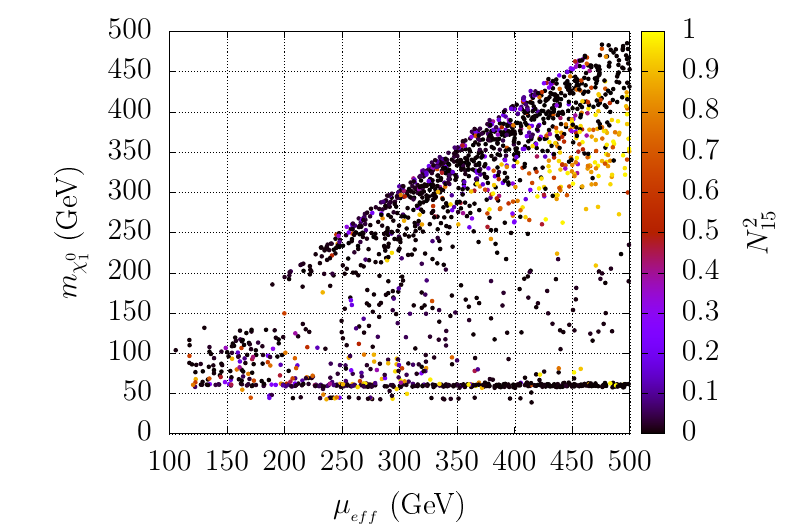}
\caption{Scatter plots in $\mueff$$-$$\mntrlone$ plane showing points that pass all relevant constraints from $\nmssmtools$ (which include various collider and DM constraints), {\tt HiggsBounds} and {\tt HiggsSignals}. Variations of the bino ($N_{11}^2$, left) and the singlino ($N_{15}^2$, right) contents in the LSP (DM) are indicated via the palettes.}
\label{fig:mueff-mlsp-n11-n15}
\end{center}
\vspace{-0.5cm}
\end{figure}

%%%%%%%%%%%%%%%%%%%%%
As pointed out earlier, we now look for a few benchmark scenarios from those 
that pass the selections of \nmssmtools, \higgsbounds~and \higgssignals.
In figure~\ref{fig:mueff-mlsp-n11-n15} we present scatter plots of parameter 
points that pass those selections in the plane of $|\mueff|-\mntrlone$. The 
choice of the said plane is motivated by the physics of the relatively light 
electroweakinos that are in the context given the recent LHC searches and 
from the viewpoint of SFOEWPT. Presenting the bino (left plot) and the 
singlino (right plot) contents of the LSP (in the palettes) further clarifies 
the situations from the involved angles.

In both plots, scenarios having a higgsino-dominated LSP arise, by construct, 
along the diagonals ($\mntrlone \approx \mueff$). Points along the two 
horizontal streaks appearing at low $\mntrlone$ correspond to a bino- or a 
singlino-dominated LSP DM that find $\hsm$ and $Z$-boson as funnels in their 
mutual annihilation. The sparse occurrence of a singlino-dominated LSP over 
these streaks points to some amount of tuning that is needed among the NMSSM 
parameters to comply simultaneously with the constraints from the Higgs and the 
DM sectors, an issue which is not of much concern for a bino-dominated LSP 
since $\mone$ could be altered practically freely without affecting the Higgs 
sector. $\mntrlone \lesssim 30$ GeV is disfavored since as a DM candidate
$\ntrlone$ would require a relatively light Higgs boson ($a_S$ or $h_S$) below 
$\sim 60$ GeV for an efficient (funnel) annihilation which, in turn, attracts 
severe constraints from the studies on $\hsm$ decays. Furthermore, the DMDD 
constraints are rather severe for such $\mntrlone$.

In each of these plots, another densely populated region appears along the edge
of the diagonal where efficient coannihilations of the DM with closely lying 
electroweakinos, backed by favorable mixing among these states, are possible.
In the rest of the (less populated) regions, compliance with the upper bound on 
the DM relic abundance is facilitated mainly by various Higgs boson funnels.
Derth of points over the region bounded roughly by $100 \, \mathrm{GeV} < \mntrlone < 200 \, \mathrm{GeV}$ and $100 \, \mathrm{GeV} < |\mueff| < 250 \, \mathrm{GeV}$ is due to the constraints derived from the CMS search for 
electrweakinos in the final state $3\ell + \etmiss$ with 35.9~\fbinv~of
data~\cite{Sirunyan:2018ubx}. A similar observation was made in
reference~\cite{Ellwanger:2018zxt} which finds further support in subsequent 
studies~\cite{Cao:2018rix, Domingo:2018ykx,Abdallah:2019znp}. A low population of points at 
higher $|\mueff|$ and for intermediate values of $\mntrlone$ is mostly since 
the DM tends to be over-abundant due to its sub-optimal conannihilation
rate and/or for a lack of suitable annihilation funnels.
%
%%%%%%%%%%%%%%%%%%
%%%%%%%%%%%%
%

In the subsequent subsections we settle for a few benchmark scenarios out of 
these allowed set which are representative of various situations of interest. 
We study their properties related to phase transitions at finite temperatures 
to ensure that an SFOEWPT (i.e., $\gamma_{_{\rm{EW}}} \gtrsim 1$) could occur.
It is important to note that while relevant LHC analyses with
$\sim 36$~\fbinv~of data would continue to constrain our scenarios, the entire 
region of the NMSSM parameter space indicated in
figure~\ref{fig:mueff-mlsp-n11-n15} can now be sensitive to some of the recent 
LHC searches for the electroweakinos with 139~\fbinv~of data which all are 
listed in table~\ref{tab:LHCewinoanalyses}. Hence we subject the benchmark 
scenarios to these analyses via their recasts using \checkmate~and \smodels. 
Furthermore, we check the future experimental sensitivity of the GW produced 
during the time of phase transition for a few of such allowed scenarios.
%
%%%%%%%%%%%%%%%%%%%%%%%%%%%%%%%%%%%%%%%%%%%%%
\subsection{Studying the benchmark scenarios}
\label{benchmarkstudy}
%%%%%%%%%%%%%%%%%%%%%%
In this subsection, we discuss how searches for the lighter electroweakinos at 
the LHC could restrict the region of parameter space which otherwise favors 
SFOEWPT and satisfy all other experimental bounds. The value of $\mueff$ 
is in direct reference since a small value of the same, while favors SFOEWPT, 
draws substantial constraint from the above-mentioned searches. Our goal is to 
first identify the lowest values of $\mueff$ (or, for that matter, the smallest values of the higgsino-like 
electroweakino masses, except when these form a triplet which contains the 
lightest of all the electroweakinos) that would be allowed under different 
circumstances, followed by a discussion of allowed scenarios with optimally 
light higgsinos. In the process, we highlight the role of different 
signal regions that play crucial roles.

For a scenario with a decoupled wino, lower bounds on the masses of the lighter 
electroweakinos to be derived from the LHC experiments would crucially depend 
on the values of two quantities, $\mueff$ and $\kappa/\lambda$. This is all the 
more so for a smaller $\kappa$ (which favors SFOEWPT) that renders the singlino 
lighter. With $\mueff$ not so large, electroweakinos, with their dominant 
contents, could exhibit altered hierarchies in their masses which result in 
contrasting patterns in their cascades. These, depending on their mutual 
mass-splits, result in their altered sensitivities to different final 
states and/or signal regions at the LHC experiments.

A smaller $M_1$ could add further intricacies to the collider
phenomenology~\cite{Domingo:2018ykx, Cao:2018rix, Abdallah:2020yag, Abdallah:2019znp} by placing the bino-like neutralino in the vicinity of the light singlino and the higgsinos while aiding compliance with various constraints from the DM sector. We also stick to small values of $\tanb$ ($\lesssim 5$) which favors SFOEWPT.
As noted earlier, we further consider relatively large values of `$\lambda$' ($\gtrsim 0.5$)  which, in conjunction with
small $\tanb$ values, aid compliance with the observed value of $\mhsm$ in a more
`natural' way. Note that we seek to allow for 
relatively small values of $m_H$ as well since those are what is preferred by 
SFOEWPT. For small values of $\tanb$ ($1 \lesssim \tanb \lesssim 5$) that we  would like to restrict ourselves to, stringent lower 
bounds on $m_{H^\pm, H,A}$ come dominantly from their searches in the $tb$ (for $H^\pm$) and $\tau\tau$ (for $H,A$) final states \cite{ATLAS:2021upq,ATLAS:2020zms}. While in the general scan of the parameter space, these constraints have eliminated some scenarios, the benchmark scenarios that we work with happen to lie outside the constrained regions. However, for the latter, as and when these become sensitive to similar future analyses, the presence of light 
electroweakinos in the spectrum could help evade those if the Higgs states could also decay to these electroweakinos. In this regard, searches for
$H^\pm$ is expected to be of immediate relevance and hence we mention its
altered branching fraction to $tb$ final state for  our benchmark scenarios.

Also given that the patterns of vacuum transitions get to be rather involved, we 
adopt the following convention to describe those in the upcoming discussions.
The total number of steps involved in a given phase transition process is
denoted by the roman numerals (i.e., I, II, etc.) while the type of the phase 
transition, i.e., whether it is of a first or a second-order kind, is denoted 
by the arabic numerals (i.e., 1, 2, etc.). On the other hand, for multi-step 
phase transitions, the various field directions along which the phase transitions 
occur are indicated by `S', for the singlet direction and `D', for the $SU(2)$ 
field directions. For example, a direct, i.e., a one-step, FOPT along all three 
directions is denoted by `I(1)', whereas a two-step FOPTs in which the first 
transition takes place along the singlet direction and the subsequent one along the 
$SU(2)$ field directions is labeled as `II-S(1)-D(1)'.
%
%%%%%%%%%%%%%%%%%%%%%%%%%%%%%%%%%%%%%%%%%%%%%%%%%%%%%%
\subsubsection{Disallowed scenarios with low $\mueff$}
%%%%%%%%%%%%%%%%%%%%%%%%%%%%%%%%%%%%%%%%%%%%%%%%%%%%%%
Benchmark points presented in table~\ref{tab:BPs_set1} are chosen with the 
following considerations. We seek to get an idea of how large a value of 
$\mueff$ which is still consistent with SFOEWPT but is expected to be ruled out 
by the electroweakino searches at the LHC. We employ \checkmate~for the 
purpose by putting all its currently implemented set of LHC analyses in action. 
In case we find some such benchmark scenarios to be barely allowed, we subject 
the same further to \smodels~(see table~\ref{tab:LHCewinoanalyses}) in which a 
host of relevant LHC analyses (with 139~\fbinv~of data) are incorporated to 
check if that is still the case.

In BP-D1, the LSP is singlino-dominated with $\mntrlone \sim 60$ GeV. The 
higgsino-like electroweakinos are the immediately heavier states with their 
masses governed by $|\mueff|$ ($\sim 275$ GeV) and range over 280 GeV -- 310 
GeV. We set $\mone\sim480$ GeV such that the bino-dominated neutralino
effectively decouples on both collider and cosmology considerations. The DM 
relic is under-abundant thanks to the presence of the $\hsm$-funnel which,
`$\lambda$' being large (=0.68), is rather efficient. This effectively
scales down the 
reported upper limits of the DMDD-SI and -SD rates (as a function of
$\mntrlone$) thus aiding compliance of the scenario with these constraints. 

The calculation for $T_c$ in BP-D1 suggests the possibility of a direct
(type-I(1)) SFOEWPT ($\Delta_{SU(2)}/T_C=1.14$) from the trivial false minimum 
at \{$h_d,h_u,h_s\} \equiv \{0,0,0\}$ to the broken, true (global) minimum at
\{$h_d,h_u,h_s\} \equiv \{25.5,145.6,-474.4\}$ GeV at $T_c=129.5$ GeV. Note, 
however, that this is the only benchmark point that we present for which the 
phase with the true minimum does not nucleate successfully and hence the system 
would remain trapped at the metastable false minimum ($\{0,0,0\}$). This could 
render much of the parameter space (that otherwise favors SFOEWPT) 
cosmologically nonviable~\cite{Baum:2020vfl}. Nonetheless, we retain this point
as a benchmark to demonstrate some characteristic collider-aspects, as 
discussed below, which could be equally instrumental in a scenario that does 
not have this shortcoming.

A \checkmate~analysis rules out BP-D1 (with $r$=$1.12$) via a CMS 
analysis~\cite{Sirunyan:2017lae} of 35.6~fb$^{-1}$ worth data in the
$3\ell +\etmiss$ final state where an opposite-sign, same-flavor (OS-SF) 
lepton ($e$ or $\mu$)-pair originates in the decay of an on-shell $Z$-boson 
coming from the decay of a heavier neutralino. Such a scenario, with  $\mueff$ 
as small as 275 GeV, would anyway be excluded more convincingly (i.e., with a 
larger $r$-value) by the recent LHC analyses in references~\cite{ATLAS:2021moa, CMS:2021cox} for the same final state which exploit 139~\fbinv~of data.

The benchmark point BP-D2 is somewhat similar to that in BP-D1 in terms of the
phenomenological features that are relevant for our discussion, i.e, the LSP is
still singlino-dominated with a very similar mass as before ($\approx 60$ GeV), 
the higgsino-like states are again the next heavier excitations with masses not 
very different (though on a little higher side) from those in BP-D1. Like BP-
D1, BP-D2 also possesses relatively light singlet-like scalars. The DM 
phenomenologies, in both qualitative and quantitative terms, are rather 
similar in these two cases.

However, in contrast to that in BP-D1, in BP-D2, it is a two-step phase 
transition (type-II-S(1)-D(2)) as is suggested by the calculations of $T_c$. 
First, a broken phase ($\{0,0,539.9\}$ GeV) appears only along the 
singlet direction at $T_c =$ 151.5 GeV with a possibility of a first-order 
phase transition. This is followed by the appearance of another configuration 
at $T_c=112.7$ GeV for which $SU(2)$ is now broken in the true minimum. This 
triggers the possibility of a second-order phase transition in which the 
scalar field could move from the evolved false minimum to the said true 
minimum. On the other hand, the calculation for $T_n$ now suggests that the 
tunneling process corresponding to $T_c=151.5$ GeV is so slow that what 
takes place instead is a strong ($\Delta_{SU(2)}/{T_n} =$ 2.2),
one-step first-order transition along all three directions 
simultaneously (type-I(1)) from the trivial to the physical phase
($\{67.0,197.8,774.8\}$ GeV) at $T_n = 96.2$~GeV.
%

%
%%%%%%%%%%%%%%%%
\begin{table}[tp]
\renewcommand{\arraystretch}{1.11}
%\begin{table}[H]
\begin{center}
{\tiny\fontsize{6.0}{6.0}\selectfont{
%\resizebox{!}{0.3\paperheight}{%
\begin{tabular}{|c|@{\hspace{0.08cm}}c@{\hspace{0.08cm}}|@{\hspace{0.08cm}}c@{\hspace{0.08cm}}|c|@{\hspace{0.08cm}}c@{\hspace{0.08cm}}|c|}
\hline
\Tstrut
Inputs/Observables & \makecell{BP-D1} & \makecell{BP-D2} & \makecell{BP-D3}  \\
\hline
\Tstrut
$\lambda, \, \kappa$  &  $0.683, \, 0.060$ 
                      &  $0.547, \, 0.044$
                      &  $0.565, \, 0.071$ \\
%$\kappa$   &  $0.060$ &  $0.044$ &  $0.071$  \\
$A_\lambda, \, \akappa$~(GeV)  &  $-1352.3, \, 134.5$ 
                               &  $978.4,   \, -110.0$
                               &  $963.5, \, -112.5$ \\
%$A_\kappa$~(GeV)   &  134.5    &  $-110.0$ & $-112.5$ \\
$\mueff$~~(GeV)       &   $-274.4$    &  308.0 &  308.0  \\
$\tan\beta$        &  4.77    &  2.87 &  2.87\\
$M_1$~(GeV)        &  478.8     &  460.3  & $-57.2$ \\
$m_{\widetilde{Q}_3}, \, m_{\widetilde{U}_3}$~(GeV)  
   &  2956.7, 3378.3
   &  3710.8, 3562.8
   &  3710.8, 3562.8 \\
%   $M_{\widetilde{U}_3}$~(GeV)        &  3378.3     &  3562.8 &  3562.8\\
$A_{t}$~(GeV)        &  -1019.7   &  2204.0 &  2204.0 \\ [0.05cm]
\hline
\Tstrut
%$m_{\chi_1^0}$~(GeV)    &  60.9      &  60.6  &  $-59.6$ \\
$m_{\chi_{1,2,3,4}^0}$~(GeV)  &  60.9, -304.3, 307.9, 479.4 
                              &  60.6, 312.7, $-338.3$, 468.1 
                              &  $-59.6$, 91.1, 327.2, $-338.4$ \\
%$m_{\chi_2^0}$~(GeV)    &  -304.3     &  312.7 & $91.1$  \\
%$m_{\chi_2^0}$~(GeV)    &  -304.3     &  312.7 & $91.1$  \\
%$m_{\chi_3^0}$~(GeV)    &  307.9    &  $-338.3$ & $327.2$   \\
%$m_{\chi_4^0}$~(GeV)    &  479.4    &  468.1 & $-338.4$ \\
%%$m_{\chi_5^0}$~(GeV)    &  2535.2    &  2539.5 &  2537.0 \\
%
$m_{\chi_1^\pm}$~(GeV)  &  -284.1    &  316.3 &  316.0 \\
$m_{h_1}, \, m_{h_2}, \, m_{a_1}$~(GeV)  &  79.2, 124.4, 126.6 
                                         &  78.1, 122.2, 109.5
                                         &  86.9, 123.0, 142.6  \\
%$m_{h_2}$~(GeV)       & 124.4  &  122.2 &  123.0 \\
%%$m_{h_3}$~(GeV)        &  1336.6   &  980.3 &  980.6 \\
%$m_{a_1}$~(GeV)         &  126.6      &  109.5 &  142.6 \\
%%$m_{a_2}$~(GeV)         &  1336.7    &  981.1 &  981.1 \\
%
% & & & \\
$m_{H^{\pm}}$~(GeV)        & 1359.0     &  963.8  &  963.6  \\[0.05cm]
\hline
\Tstrut
 $N_{11}$, $N_{21}, N_{31}$, $N_{41}$  & $-0.03,~~0.04, -0.13, 0.99$    & $0.03,-0.25, -0.02, ~~0.97$ & $0.99,-0.07, -0.1, ~~0.06 $\\
$N_{13}$, $N_{23}, N_{33}$, $N_{43}$  & $\!\!\!\!~~0.01, -0.71, ~~0.70, ~~0.06$   & $-0.04,~~0.70, ~~0.70,-0.16 $ & $0.11,-0.03, ~~0.71, -0.70 $ \\
$N_{14}$, $N_{24}, N_{34}$, $N_{44}$  & $\!\!\!\!~~0.38,~~0.65, ~~0.65, ~~0.12$    & $\!\!\!\!~~-0.27,-0.65, ~~0.68, ~0.19$ & $\!\!\!\!-0.04,-0.27, -0.68, -0.68$  \\
$N_{15}$, $N_{25}, N_{35}$, $N_{45}$  & $0.93,-0.26, -0.27, -0.02$   & $~~0.96,-0.16,~~0.22,~~0.02 $ & $0.07,-0.96, -0.18, -0.21 $ \\[0.05cm]
\hline
\Tstrut
BR($\chi^\pm_1\to\chi_1^0 W^\pm $)  &   1.00    &  1.00 &  0.19 \\[0.05cm]
BR($\chi^\pm_1\to\chi_2^0 W^\pm $)  &  0.00  & 0.00 &  0.81  \\[0.05cm]
\hline
\Tstrut
BR($\chi^0_2\to\chi_1^0 Z $)    &  0.52    &  0.58 & off-shell  \\[0.05cm]
%BR($\chi^0_2\to\chi_1^0 h_1 $)  &  0.01   &  0.05 & off-shell \\[0.05cm]
BR($\chi^0_2\to\chi_1^0 h_2 $)  &  0.37    &  0.33 & off-shell  \\[0.05cm]
%BR($\chi^0_2\to\chi_1^0 a_1 $)  &  0.02   &  0.007 & off-shell  \\[0.05cm]
BR($\chi^0_2\to\chi_1^0 \gamma $)  & 0.00   &  0.00 & 0.15  \\[0.05cm]
\hline
\Tstrut
BR($\chi^0_3\to\chi_1^0 Z $)    &  0.43   &  0.36 &  0.03 \\[0.05cm]
BR($\chi^0_3\to\chi_2^0 Z $)    &  0.00   &  0.00 &  0.53 \\[0.05cm]
%BR($\chi^0_3\to\chi_1^0 h_1 $)  &  0.05   &  0.02 &  0.00 \\[0.05cm]
%BR($\chi^0_3\to\chi_2^0 h_1 $)  &  0.00  &  0.00 &  0.02\\[0.05cm]
BR($\chi^0_3\to\chi_1^0 h_2 $)  &  0.34   &  0.42 &  0.10\\[0.05cm]
BR($\chi^0_3\to\chi_2^0 h_2 $)  &  0.00   & 0.00 & 0.27 \\[0.05cm]
BR($\chi^0_3\to\chi_1^0 a_1 $)  &  0.19  &  0.16 &  0.01 \\[0.05cm]
\hline
\Tstrut
BR($\chi^0_4\to\chi_1^0 Z $)    &  0.16    &  0.12 &  0.18 \\[0.05cm]
BR($\chi^0_4\to\chi_2^0 Z $)    &  0.12    &  $\sim 0$ &  0.31 \\[0.05cm]
BR($\chi^0_4\to\chi_3^0 Z $)    &  0.02   &  0.09 &  0.00 \\[0.05cm]
%BR($\chi^0_4\to\chi_1^0 h_1 $)  &  0.001   &  0.004 &  0.005 \\[0.05cm]
BR($\chi^0_4\to\chi_2^0 h_1 $)  &  0.00   &  0.11 &  0.02 \\[0.05cm]
BR($\chi^0_4\to\chi_1^0 h_2 $)  &  0.01    &  $\sim 0$ & 0.02  \\[0.05cm]
BR($\chi^0_4\to\chi_2^0 h_2 $)  &  0.01  &  0.21  & 0.30  \\[0.05cm]
BR($\chi^0_4\to\chi_2^0 a_1 $)  &  0.01    &  $\sim 0$ & 0.15 \\[0.05cm]
BR($\chi^0_4\to\chi_1^{\pm} W^{\mp} $)  &  0.48    &  0.46 & 0.00 \\[0.05cm]
\hline
\Tstrut
BR($H^{+}\to t \bar{b}$)  &  0.12   &  0.39 &  0.37 \\[0.05cm]
\hline
\Tstrut
$\Omega h^2$  &  $4.9\times10^{-4}$  &  $4.4\times10^{-4}$ & $4.8 \times10^{-3}$ \\[0.10cm]
$\sigma^{\rm SI}_{\chi^0_1-p(n)}\times \xi$~(cm$^2$)  &  $4.5(4.6)\times 10^{-47}$  &  $2.4(2.5)\times 10^{-47}$ &  $2.5(2.6)\times 10^{-47}$\\[0.30cm]
$\sigma^{\rm SD}_{\chi^0_1-p(n)}\times \xi$~(cm$^2$)  &  $3.5(3.2)\times 10^{-42}$   &  $7.6(5.8)\times 10^{-43}$ &  $1.9(1.5)\times 10^{-43}$\\[0.15cm]
\hline
\Tstrut
%\texttt{First $T_c$-based transition}  &  1st-order    & 1st-order & 1st-order \\
\texttt{First $T_c$ {\tt(GeV)} / Transition type}  &  129.4 / 1st-order    & 151.5 / 1st-order & 165.7 / 1st-order \\
\texttt{$\{h_d,h_u,s\}_{_\text{False\_vac.}}$} {\tt (GeV)} & $\{0,~0,~0\}$  &  $\{0,~0,~0\}$ &  $\{0,~0,~0\}$  \\
\texttt{$\{h_d,h_u,s\}_{_\text{True\_vac.}}$} {\tt (GeV)}  & $\{25.5,145.6,-474.4\}$   & $\{0,0,539.9\}$ & $\{0,0,557.5.9\}$ 
%
%\texttt{$T_c$} {\tt (GeV)} &  $129.4$  &  151.5 &  165.7
\\
 & & & \\ 
\texttt{Second $T_c$ {\tt (GeV)} / Transition type}  &  $-$  & 112.7 / 2nd-order & 105.6 / 1st-order\\
\texttt{$\{h_d,h_u,s\}_{_\text{False\_vac.}}$} {\tt (GeV)}  & $-$    & $\{0,~0,661.7\}$ & $\{0,~0,662.3\}$  \\
\texttt{$\{h_d,h_u,s\}_{_\text{True\_vac.}}$} {\tt (GeV)}  & $-$    & $\{9.5,31.5,668.2\}$ & $\{12.8,41.6,669.0\}$ \\
%\texttt{$T_c$} {\tt (GeV)} &  $-$     & 112.7 & 105.6\\\hline
 & & & \\
\texttt{$T_n$ {\tt (GeV)} / (Nucleation) Transition type}  &  $-$  & 96.2 / 1st-order & 55.9 / 1st-order  \\
\texttt{$\{h_d,h_u,s\}_{_\text{False\_vac.}}$} {\tt (GeV)}  & $-$     & $\{0,~0,~0\}$ & $\{0,~0,~0\}$ \\
\texttt{$\{h_d,h_u,s\}_{_\text{True\_vac.}}$} {\tt (GeV)} & $-$ & $\{67.0,197.8,774.8\}$ & $\{68.1, 199.2, 759.2\}$ \\
 & & & \\
%\texttt{$T_n$} {\tt (GeV)} &  $-$  & 96.2 & 55.9 \\ \hline \\
%
\texttt{$\gamma_{_{\rm{EW}}} = \Delta_{SU(2)}/T_n$}  & $-$   & $2.2$ &  $3.8$ \\ \hline
 & & & \\
\texttt{CheckMATE} result  &  Excluded    & Excluded &  Excluded \\
$r$-value              &  1.12      & 1.01 & 2.13 \\
Analysis ID         &  CMS$\_$SUS$\_$16$\_$039~\cite{Sirunyan:2017lae} & CMS$\_$SUS$\_$16$\_$039~\cite{Sirunyan:2017lae}  & CMS$\_$SUS$\_$16$\_$039~ \cite{Sirunyan:2017lae}\\
Signal region ID &  SR$\_$A30  & SR$\_$A30 &  SR$\_$G05\\[0.05cm]\hline
\end{tabular}
%}
}}
%\vskip 5pt
\caption{Benchmark scenarios allowed by all relevant theoretical and experimental constraints (see text for details) except for those from the LHC searches for the electroweakinos. Shown are the various relevant masses, mixings, branching fractions along with the values of DM observables and details of the EWPT. The most sensitive LHC signal regions that rule out these scenarios, along with the LHC analyses they belong to, are also presented. Other fixed parameters are as indicated in the caption of table~\ref{tab:ranges}.  The parameter $\xi \, (= \frac{\Omega h^2}{0.1187})$ is used to scale (down) the DD rates.}
\label{tab:BPs_set1}
\end{center}
\end{table}

On the collider front, BP-D2 yields somewhat smaller production cross 
sections for the higgsino-like states than what BP-D1 gives because these 
states are a little heavier in BP-D1. A \checkmate~analysis indicates that in the CMS analysis in reference~\cite{Sirunyan:2017lae} that uses 35.9~\fbinv~of data, the same final state 
($3\ell +\etmiss$) with an identical signal region as in BP-D1
becomes the most sensitive of the searches while, this time, 
barely disallowing ($r=1.01$) the parameter point. A subsequent 
\smodels~study indicates that the analyses in
references~\cite{Aad:2019vvi, Aad:2019vnb, Aad:2019vvf}, all involving
139~\fbinv~of data, are even less sensitive. As for BP-D1, BP-D2 is also likely 
to be ruled out convincingly by the analyses of $3\ell +\etmiss$ final state 
with 139~\fbinv~of data presented in
references~\cite{ATLAS:2021moa, CMS:2021cox}. The resulting $r$-values would 
hint at how big a $\mueff$ could thus be excluded in such a setup.

The point BP-D3 contains a somewhat heavier ($\sim 91$ GeV) singlino-like neutralino state 
where SFOEWPT ($\Delta_{SU(2)}/{T_n}=$ 3.8) remains viable, `$\kappa$' being 
still small ($\sim 0.071$).
The relic for such a singlino as a DM candidate is bound to be
over-abundant in the absence of a suitable funnel, as is the case with BP-D3. The possibility of an 
efficient coannihilation, say with a bino-like state, requires a small
mass-split between them which then tends to make the DD rates way too large to 
be acceptable. Instead, a bino-like LSP having a smaller mass and possessing a 
suitable annihilation funnel via $Z$- or $\hsm$ could qualify as a DM.  
This is what we find in BP-D3 (with an
$\hsm$ funnel, with $|\mntrlone| \approx 60$ GeV). Spectrum-wise, this 
constitutes its basic difference from BP-D2. The pattern of phase-transition in 
BP-D3, as obtained from the critical temperature calculation, is also of a two-step kind (type-II-S(1)-D(1)) but is slightly different from what occurs in 
BP-D2, as can be seen in table~\ref{tab:BPs_set1}. Although the bubble nucleation calculation indicates that both benchmark points have one-step SFOPTs in all three directions (type-I(1)).

However, the hierarchy among the lighter neutralinos now triggers important 
effects with major implications for the searches of the electroweakinos at the 
LHC. The higgsino-like states predominantly decay to the singlino-like NLSP and 
the $Z$-boson and/or $\hsm$ thanks to an enhanced higgsino-singlino mixing for 
a value of `$\lambda$' which is on the larger side ($\sim 0.57$) \cite{Abdallah:2020yag}. Subsequently, 
the NLSP neutralino would undergo dominant decays to off-shell $Z$-boson or
$\hsm$. Such cascades result in strengthened multi-lepton (more than three 
leptons) final states which now become far more sensitive to the recent LHC 
analyses when compared to the trilepton final states. The reason behind this is 
a much suppressed SM background for the former~\cite{Sirunyan:2017lae}. Indeed, 
a dedicated \checkmate~analysis confirms this effect and rules out BP-D3 rather 
emphatically ($r=2.13$) by getting sensitive to the right (dedicated for 
finals states with more than 3 leptons) signal region (``G05") of the CMS 
analysis in reference~\cite{Sirunyan:2017lae} which considers data worth
35.9~\fbinv~only. This needs to be contrasted with the verdict on higgsinos of 
very similar masses in BP-D2 in which those masses appear to be barely 
disallowed ($r=1.01$) with the same set of data. Further, given the 
heightened sensitivity of the analysis to the multi-lepton finals states, it 
could eventually rule out even heavier higgsino-like states in such a setup.

The exercise undertaken in this section indicates how different 
types of spectrum for the light higgsino-like electroweakinos (i.e., smaller
$\mueff$), which otherwise comply with all relevant bounds including those from the DM sector and which allow for SFOEWPT, get ruled out by the LHC analyses with $\sim 36$ \fbinv~of data even when the latter's sensitivities to the targeted final states deteriorate significantly. The benchmark scenarios are so chosen that we end up with $r \gtrsim 1$. Such a value nominally reflects how light such electroweakinos could get before they start attracting bounds from the LHC analyses. Of course, more recent LHC analyses~\cite{ATLAS:2021moa, CMS:2021cox} with 139 \fbinv~of data are expected
to push these mass-bounds (and hence $\mueff$) upwards but these are yet to be
implemented in a recast package.
%
%%%%%%%%%%%%%%%%%%%%%%%%%%%%%%%%%%%%%%%%%%%%%%%%%%%%%%%%%%%%%%%%%%%%%%
\subsubsection{Allowed benchmark scenarios with successful nucleation}\label{subsubsec:allowedbms}
%%%%%%%%%%%%%%%%%%%%%%%%%%%%
In this section we present a few benchmark scenarios that have all the 
good qualities of those listed in table~\ref{tab:BPs_set1} but now also 
pass the lower bounds on the electroweakino masses coming from some 
of the recent LHC analyses. Naively, this pushes $\mueff$ up which impedes
an efficient SFOEWPT with successful nucleation. The SFOEWPT now tends 
to proceed in two steps the details of which are presented in
table~\ref{tab:Cosmo_PT} for our benchmark points. This is somewhat 
typical when the trivial and the global minima have a large separation between 
them in the field space~\cite{Baum:2020vfl}. This is since a larger $\mueff$ corresponds to a larger $\vs$ at zero temperature for a given $\lambda$, a feature that governs the field-separation at $T_c$.

The benchmark points in table~\ref{tab:BPs_set2} are picked up keeping in mind 
the following issues. While our goal is to find compatible points with smaller 
values of $\mueff$, we like to see the resulting scenarios have LSPs with
different dominant admixtures. Allowing for this has a considerable 
bearing on both the DM phenomenology and searches for the electroweakinos at 
the LHC. Furthermore, these benchmarks possess a light singlet-like scalar 
below 100 GeV. This is since we set both `$\kappa$' and $\akappa$ small which 
is preferred by SFOEWPT. Note that such a light singlet state inevitably 
affects both DM and collider phenomenologies, more so since the nature of the 
lighter electroweakinos, including the LSP, could get altered, simultaneously. 
As has been noted in section \ref{subsubsec:target}, to facilitate SFOEWPT we 
look for relatively light doublet-like Higgs bosons (by choosing $\alambda$ suitably) which are still allowed by the LHC Higgs searches.

In BP-A1 we have a higgsino-like lightest triplet with masses in the range
$\sim 400 - 430$ GeV with $\mueff \sim$ 422 GeV. Thus, the LSP and the NLSP are 
both higgsino-like (with their higgsino contents at $70 \%$ and $98 \%$, 
respectively) while the lighter chargino is a nearly pure higgsino. As far as 
the DM sector is concerned, such a higgsino-like LSP DM is naturally under-
abundant ($\Omega h^2 = 3.78 \times 10^{-4}$). This, in turn, generically helps 
satisfy the DMDD-SI and -SD constraints via downward scaling of the respective 
cross-sections.

As for the pattern of EWPT in BP-A1, the calculation for $T_c$ suggests 
that this is a two-step process of the type II-S(1)-D(1) 
as indicated in table~\ref{tab:Cosmo_PT} where first, at $T_c = 946.7$ GeV, a 
broken phase appears in the singlet-direction followed by another in the
$SU(2)$ field directions at $T_c = 91.1$ GeV. Subsequently, successful 
nucleations take place closely below the respective $T_c$'s, down at 
$T_n =$ 946.6 GeV and 90.2 GeV. Note that in this particular case, calculations 
for both $T_c$ and $T_n$ suggest that the first phase transition (in the 
singlet-only direction) is just of a first-order kind while the second one, in 
the all-important $SU(2)$ field directions that breaks the electroweak 
symmtery, is of a `strong' first-order type ($\gamma_{_{EW}} = 1.1$) which is a 
crucial requirement for EWBG. 
%
%
%%%%%%%%%%%%%%%%
\begin{table}[tph]
\renewcommand{\arraystretch}{1.13}
\begin{center}
{\tiny\fontsize{7.4}{7.3}\selectfont{
\begin{tabular}{|c|@{\hspace{0.06cm}}c@{\hspace{0.06cm}}|@{\hspace{0.06cm}}c@{\hspace{0.06cm}}|@{\hspace{0.06cm}}c@{\hspace{0.06cm}}|c|@{\hspace{0.06cm}}c@{\hspace{0.06cm}}|}
\hline
\Tstrut
Input/Observables & \makecell{BP-A1} & \makecell{BP-A2} & \makecell{BP-A3} & \makecell{BP-A4}  \\
\hline
\Tstrut
$\lambda$   &  $0.609$  &  $0.609$         &  $0.633$      & $0.523$   \\
$\kappa$   &  $0.326$  &  $0.326$ & $0.216$  & $0.041$   \\
$\tan\beta$        &  1.98    &  1.98   & 1.79    & 3.65 \\
$A_\lambda$~(GeV)  &  $477.0$    &  $477.0$    & $-558.7$    & $-1253.9$ \\
$A_\kappa$~(GeV)   &  38.7   &  37.8   & $-46.3$  & $138.1$   \\
$\mueff$~~(GeV)       &  $421.8$   &   $421.8$  & $-398.7$   &  $-334.5$   \\
$M_1$~(GeV)        &  480.1    &  $-365.1$   & 286.3   & $-143.8$ \\
$M_{\widetilde{Q}_3}$~(GeV)        &  4262.7   &  4262.7  & 3950.3   & 2292.0  \\ 
$M_{\widetilde{U}_3}$~(GeV)        &  3450.4   &   3450.4  & 3544.4  & 3435.8  \\
$A_{t}$~(GeV)        &  -639.2   &  -639.2  & 1372   & 3862.4  \\ [0.05cm]
\hline
\Tstrut
$m_{\chi_1^0}$~(GeV)    &  395.9   &  -360.9   & 284.5  & $-61.3$   \\
$m_{\chi_2^0}$~(GeV)    &  $-445.6$   & 415.1   & $-289.5$  & $-139.2$  \\
$m_{\chi_3^0}$~(GeV)    &  476.8  &  $-447.5$  & $-421.8$  & $-359.3$   \\
$m_{\chi_4^0}$~(GeV)    &  509.5  &  493.2  & $-426.9$  & 359.7  \\
$m_{\chi_5^0}$~(GeV)    &  2538.7  &  2538.7  & 2542.1 & 2534.2  \\

$m_{\chi_1^\pm}$~(GeV)  &  431.5  &  431.5  & $-412.1$ & $-345.3$  \\
$m_{h_1}$~(GeV)         &  122.6    &  122.7   & 126.9  & 74.0  \\
$m_{h_2}$~(GeV)         &  449.2  &  449.0  & 288.5  & 124.7  \\
$m_{h_3}$~(GeV)         &  822.8  &  824.8  & 806.4  & 1296.6  \\
$m_{a_1}$~(GeV)         &  75.01  &  79.0   & 84.8   & 121.0  \\
$m_{a_2}$~(GeV)         &  819.4  &  821.4   & 805.4   & 1296.6  \\
$m_{H^{\pm}}$~(GeV)         &  816.5   & 818.4   & 800.9  & 1293.3  \\[0.05cm]
\hline
\Tstrut
{\tiny$N_{11}$, $N_{21}, N_{31}$}, $N_{41}$  & {\tiny$-0.43,~-0.01, -0.62, ~0.66$}   & {\tiny$0.995,~~0.05, -0.15, ~~0.02$}  & {\tiny$-0.99, -0.02, -0.06, ~~0.08$}   &  {\tiny$~~0.12, ~~0.98, ~-0.17, ~~0.04$}  \\
{\tiny$N_{13}$, $N_{23}, N_{33}$, $N_{43}$}  & {\tiny$-0.56,~-0.70, -0.08, -0.43$}   & {\tiny$0.15,-0.58, ~~0.69, -0.41$}  & {\tiny$-0.10, -0.06, -0.71, ~~0.70$}     &  {\tiny$0.01,-0.15, ~-0.70, ~~0.70$} \\
{\tiny$N_{14}$, $N_{24}, N_{34}$, $N_{44}$}  & {\tiny$\!\!\!\!~~0.61,~~0.70, ~~-0.06, ~~0.36$}   & {\tiny$0.07,~~0.66, ~~0.70, ~~0.27$}   & {\tiny$-0.01, ~~0.27, ~~0.67, ~~0.70$}   &  {\tiny$0.25,~~0.06, ~~0.68, ~~0.69$}  \\
{\tiny$N_{15}$, $N_{25},N_{35}$, $N_{45}$}  & {\tiny$\!\!\!-0.35,~~0.11, ~~0.80, ~~0.50$}   & {\tiny$0.02,-0.48, ~~0.11, ~~0.87$}  & {\tiny$~~0.02,~~0.96, -0.23, -0.15$}   &  {\tiny$0.96,~~-0.14, -0.15, -0.19$}  \\[0.05cm]
\hline
\Tstrut
BR($\chi^\pm_1\to\chi_1^0 W^\pm $)  &  off-shell  &  off-shell  & 0.14  & 0.81  \\[0.05cm]
BR($\chi^\pm_1\to\chi_2^0 W^\pm $)  &  off-shell  &  off-shell & 0.86  & 0.19  \\[0.05cm]
\hline
\Tstrut
%BR($\chi^0_2\to\chi_1^0 Z $)    &  off-shell  &  off-shell  & 0.0  & off-shell   \\[0.05cm]
BR($\chi^0_2\to\chi_1^0 h_1 $)  &  off-shell  &  off-shell  & $\sim0$   & 0.98  \\[0.05cm]
%BR($\chi^0_2\to\chi_1^0 h_2 $)  &  off-shell  & off-shell  & 0.0   & off-shell   \\[0.05cm]
%BR($\chi^0_2\to\chi_1^0 a_1 $)  &  off-shell  &  off-shell  & 0.0  & off-shell   \\[0.05cm]
BR($\chi^0_2\to\chi_1^0 \gamma $)  &  $\sim {0.01}$  &  0.001  & 0.98  &  $\sim0$  \\[0.05cm]
\hline
\Tstrut
BR($\chi^0_3\to\chi_1^0 Z $)    &  off-shell  &  off-shell  & 0.06  &  0.47  \\[0.05cm]
BR($\chi^0_3\to\chi_2^0 Z $)    &  off-shell  &  off-shell  & 0.58  & 0.05  \\[0.05cm]
%BR($\chi^0_3\to\chi_1^0 h_1 $)  &  off-shell  &  off-shell  & 0.005  & 0.04  \\[0.05cm]
BR($\chi^0_3\to\chi_2^0 h_1 $)  &  off-shell &  off-shell  & 0.33  & 0.03  \\[0.05cm]
BR($\chi^0_3\to\chi_1^0 h_2 $)  &  off-shell  &  off-shell  & $\sim0$  & 0.27  \\[0.05cm]
BR($\chi^0_3\to\chi_2^0 h_2 $)  &  off-shell  &  off-shell  & $\sim0$ &  0.12  \\[0.05cm]
BR($\chi^0_3\to\chi_1^0 a_1 $)  &  0.23  &  0.13  & 0.01  &  0.01  \\[0.05cm]
%BR($\chi^0_3\to\chi_1^0 \gamma $)  &  0.00  &  0.00  & 0.00 & 0.00  \\[0.05cm]
\hline
\Tstrut
BR($\chi^0_4\to\chi_1^0 Z $)    &  0.14  &  0.90  & 0.07  & 0.37  \\[0.05cm]
BR($\chi^0_4\to\chi_2^0 Z $)    &  off-shell  &  $\sim0$ & 0.30  & 0.17  \\[0.05cm]
%BR($\chi^0_4\to\chi_3^0 Z $)    &  off-shell  &  0.00  & 0.00  & 0.00  \\[0.05cm]
%BR($\chi^0_4\to\chi_1^0 h_1 $)  &  off-shell  &  0.0  & 0.03  & 0.02  \\[0.05cm]
%BR($\chi^0_4\to\chi_2^0 h_1 $)  &  off-shell  &  0.0  & 0.02  & 0.002 \\[0.05cm]
BR($\chi^0_4\to\chi_1^0 h_2 $)  &  off-shell  &  $\sim0$  & 0 & 0.32  \\[0.05cm]
%BR($\chi^0_4\to\chi_2^0 h_2 $)  &  off-shell  &  0.00  & 0.0  & 0.02  \\[0.05cm]
BR($\chi^0_4\to\chi_1^0 a_1 $)  &  0.14  &  0.01  & $\sim0$   & 0.08  \\[0.05cm]
BR($\chi^0_4\to\chi_2^0 a_1 $)  &  off-shell  &  $\sim0$  & 0.58  & 0.01  \\[0.05cm]
%BR($\chi^0_4\to\chi_1^{\pm} W^{\mp} $)  &  0.00  &  0.00  & 0.00 & 0.00  \\[0.05cm]
\hline
\Tstrut
BR($H^{+}\to t \bar{b}$)  &  0.93  &  0.92  & 0.84 & 0.27  \\[0.05cm]
\hline
\Tstrut
$\Omega h^2$  &  $3.78 \times 10^{-4}$  &  0.107  & 0.119 & $1.96 \times 10^{-3}$  \\[0.10cm]
$\sigma^{\rm SI}_{\chi^0_1-p(n)}\times \xi$~(cm$^2$)  &    $1.2(1.3)\times 10^{-46}$&  $7.2(7.6)\times 10^{-48}$  & $1.2(1.2)\times 10^{-46}$ & $4.1(4.3)\times 10^{-47}$   \\[0.30cm]
$\sigma^{\rm SD}_{\chi^0_1-p(n)}\times \xi$~(cm$^2$)  &    $4.6(4.5)\times 10^{-44}$&  $9.4(7.3)\times 10^{-42}$  & $3.5(2.8)\times 10^{-42}$  & $1.1(0.8)\times 10^{-41}$   \\[0.15cm]
\hline
\Tstrut
\texttt{CheckMATE} result  &    Allowed  &  Allowed  &  Allowed  & Allowed  \\
$r$-value               &   0.03 &  0.08     &  0.14  & 0.55  \\
Analysis ID          &   CMS$\_$SUS$\_$16$\_$039~\cite{Sirunyan:2017lae} &  CMS$\_$SUS$\_$16$\_$039~\cite{Sirunyan:2017lae}  &   CMS$\_$SUS$\_$16$\_$039~\cite{Sirunyan:2017lae}  & CMS$\_$SUS$\_$16$\_$039~\cite{Sirunyan:2017lae}\\
Signal region ID &    SR$\_$A01 &  SR$\_$A08  & SR$\_$A28  & SR$\_$A31\\[0.05cm]\hline
\end{tabular}
}}
%\vskip 5pt
\caption{ Same as in table~\ref{tab:BPs_set1} except for showing benchmark scenarios (with successful nucleation) allowed by all relevant theoretical and experimental constraints including the recent ones from the LHC electroweakino searches. The details of the EWPT are indicated separately in
tables~\ref{tab:Cosmo_PT} and \ref{tab:Table_GW_data}.}
\label{tab:BPs_set2}
\end{center}
\end{table}
%
%%%%%%%%%%%%%%%%%%%%%%%%%%%%%%%%%%
\hspace{-2cm}
\begin{table}[t]
\renewcommand{\arraystretch}{1.3}
%   \centering
%   \setlength{\extrarowheight}{5pt}
\small{\begin{tabular}{|c||c|c|c|c|c|c|}
\hline
  BM&\multicolumn{2}{c|}{$T_i$ (GeV)} & ${\{h_d, h_u, h_s\}}_{\text{false}}$  & \multirow{2}{*}{$\xrightarrow[\text{type}]{\text{Transition}}$} & ${\{h_d, h_u, h_s\}}_{\text{true}}$  & $\gamma_{_{\rm{EW}}}$ \\
 No. & \multicolumn{2}{c|} {\tt \footnotesize (Transition pattern)}  & (GeV) &  & (GeV) &$=\frac{\Delta_{SU(2)}}{T_n}$\\
\hline 
\hline
% \multirow{3}{*}{BP6} & 
\multirow{4}{*}{BP-A1} & $T_c$ & 946.7 & \{0, 0, 0\} & FO & \{0, 0, 63.2\} & \\
%                        &       &             &                     &  &\\ 
\cline{3-6}
 & \tt{II-S(1)-D(1)} & 91.1 & \{0, 0, 1000.9\} & ,, & \{40.4, 79.6, 1000.7\} &\\   
\cline{2-6} 
 & $T_n$ & 946.6 & \{0, 0, 0\} & ,, & \{0, 0, 64.9\} &\\ 
\cline{3-6}
 & \tt{II-S(1)-D(1)} & 90.2 & \{0, 0, 1000.9\} & ,, & \{44.2, 86.9, 1000.6\} & 1.08 \\ 
\hline
\hline
% \multirow{3}{*}{BP6} & 
\multirow{4}{*}{BP-A2} & $T_c$ & 946.0 & \{0, 0, 0\} & ,, & \{0, 0, 64.4\} & \\
%                        &       &             &                     &  &\\ 
\cline{3-6}
 & \tt{II-S(1)-D(1)} & 91.3 & \{0, 0, 1000.9\} & ,, & \{39.9, 78.6, 1000.6\} &\\   
\cline{2-6} 
 & $T_n$ & 945.6 & \{0, 0, 0\} & ,, & \{0, 0, 66.2\} &\\ 
\cline{3-6}
 & \tt{II-S(1)-D(1)} & 86.2 & \{0, 0, 1000.8\} & ,, & \{57.1, 112.5, 1000.3\} & 1.46 \\ 
\hline
\hline
% \multirow{3}{*}{BP6} & 
\multirow{4}{*}{BP-A3} & $T_c$ & 644.4 & \{0, 0, 0\} & ,, & \{0, 0, $-100.0$\} & \\
%                        &       &             &                     &  &\\ 
\cline{3-6}
 & \tt{II-S(1)-D(1)} & 95.8 & \{0, 0, $-916.3$\} & ,, & \{41.4, 72.9, $-915.3$\} &\\   
\cline{2-6} 
 & $T_n$ & 644.3 & \{0, 0, 0\} & ,, & \{0, 0, $-104.8$\} &\\ 
\cline{3-6} 
 & \tt{II-S(1)-D(1)} & 94.5 & \{0, 0, $-914.9$\} &,, & \{48.5, 85.6, $-914.8$\} & 1.04\\ 
\hline
\hline
% \multirow{3}{*}{BP6} & 
\multirow{3}{*}{BP-A4} & $T_c$ & 185.0 & \{0, 0, 0\} & ,, & \{0, 0, $-668.9$\} & \\
%                        &       &             &                     &  &\\ 
\cline{3-6}
 & \tt{II-S(1)-D(2)} & 136.5 & \{0, 0, $-846.6$\} & SO & \{2.3, 9.1,  $-846.7$\} &\\   
\cline{2-6} 
 & $T_n$ & \multirow{2}{*}{116.9} & \multirow{2}{*}{\{0, 0, 0\}} & \multirow{2}{*}{FO} & \multirow{2}{*}{\{30.3, 113.8, $-877.4$\}} & \multirow{2}{*}{1.01}\\ 
  &  \tt{I-(1)} &  &  &  &  & \\ 
\hline
\end{tabular}}
   \caption{Phase transition characteristics of the benchmark points presented in table~\ref{tab:BPs_set2}. For each benchmark point, presented are the $T_c$'s and $T_n$'s, the corresponding field values, the transition types (`FO' for first-order and `SO' for second-order) and the strengths of the phase transition along the $SU(2)$-direction ($\gamma_{_{\rm{EW}}}$). See text for details.}
  \label{tab:Cosmo_PT}
\end{table}

Searches for the lighter electroweakinos in BP-A1 effectively amounts to those 
for the higgsinos only where these states appear as the lightest triplet of
electroweakinos which includes the LSP. This is since the heavier neutralinos, $\ntrlthree$ and 
$\ntrlfour$, are singlino- and bino-like, respectively, whose productions are 
coupling-suppressed. In contrast to scenarios in which the higgsinos do not 
form the lightest triplet, here one loses out on the cascade of one of the 
neutralinos (which is the LSP in the present case). This restricts their abilities to contribute to diverse final 
states. On top of that, $\charonepm$ and $\ntrltwo$, once produced in such a 
scenario, decays to the LSP, which is not far away in mass, via off-shell gauge 
and Higgs bosons (as indicated in table~\ref{tab:BPs_set2}) thus resulting in 
associated leptons/jets to be generically soft. Both these issues have negative 
impacts on the experimental sensitivities of such a scenario. This is clearly 
reflected in the LHC analyses of such
scenarios~\cite{ATLAS:2021moa, CMS:2021edw} which report much relaxed lower 
bounds (down to $\sim 220$ GeV, conservatively) on the masses of such
higgsino-like electroweakinos as a function of their mass-split with the LSP.

A \checkmate~analysis that includes all readily available analyses in its 
repository results in a `$r$' value far below 1 for the point BP-A1 thus
marking its total insensitivity to the LHC searches and hence allowed by the 
same. The relevant analysis and the most significant signal region therein are 
also indicated. Note that the higgsino masses for this benchmark point 
are way above their current lower bounds for such a scenario as mentioned 
above. In passing, we note that in the future runs of the LHC such a 
scenario would likely attract bounds from the searches of the doublet-like 
heavy Higgs bosons sooner than from the direct searches for such 
electroweakinos.

Benchmark point BP-A2 is almost the same as BP-A1 except for $M_1$ now being 
brought down below $\mueff$. Thus, the LSP DM is now highly
bino-dominated and its relic abundance ($\Omega h^2 = 0.107$) now falls within 
the Planck-observed band. Towards this, the required depletion in the relic is 
again facilitated by the coannihilation of the bino-like DM with the
higgsino-like chargino and neutralinos. Note that the sign on $M_1$ (with 
respect to that of $\mueff$) ensures compliance with the experimentally observed latest 
upper bound on the DMDD-SI cross-section by setting up a so-called `coupling 
blind spot' as discussed in section~\ref{subsubsec:target}.

As in BP-A1, EWPT in BP-A2 is also of the type
II-S(1)-D(1). The only notable 
difference that is found with respect to BP-A1 is in the delayed nucleation for 
the crucial phase transition in the $SU(2)$ field directions ($T_n = 86.2$ GeV, 
as opposed to 90.2 GeV in BP-A1) as shown in table~\ref{tab:Cosmo_PT}. This 
results in a stronger FOEWPT ($\gamma_{_{EW}} = 1.5 $, compared to
$\gamma_{_{EW}} = 1.1 $ in BP-A1). Its implications for the GW physics will be discussed in section \ref{GWresults}.
The delayed nucleation can be explained by the altered $\mone$ which 
modifies the thermal correction to the effective potential via terms 
that are only quadratic and quartic in $m(\phi)/T$ given that the bino is a 
fermion (see equation~\ref{eq:JFhighT}). For our present benchmark scenario,  
successful nucleation would then require some appropriate modification in the 
term cubic in $m(\phi)/T$ which we achieve by a minor tweaking of $\akappa$. In 
the process, the potential barrier gets modified in a way that 
leads to delayed nucleation compared to BP-A1.

On the LHC front, unlike in BP-A1, in BP-A2 cascades of both higgsino-
like neutralinos ($\ntrltwothree$) will be important for the relevant final 
states. Although, just as in BP-A1, $\ntrltwothree$ and the higgsino-like
$\charonepm$ would undergo off-shell decays to $Z$, $\hsm$ and $W^{\pm}$, the 
corresponding branching fractions for $\ntrltwothree$ get suppressed in the 
presence of their significant on-shell branchings to a photon (for $\ntrltwo$) 
and to a light $\as$ (for $\ntrlthree$). The relevant lower bound from the LHC 
on the masses of lighter electroweakinos with such mass-splits is 
presented in reference~\cite{ATLAS:2021moa} for a wino(NLSP)-bino(LSP) 
system which can be conservatively taken as $\sim$ 300 GeV. In a scenario 
like BP-A2, such a bound would get weakened not only because of the suppressed 
off-shell branching fractions of the neutralinos as mentioned above but 
also, as described in section \ref{subsubsec:target}, since the collective production cross-sections for the higgsino-like 
electroweakinos are known to be smaller than if they were wino-like, for any 
given mass. This is corroborated by our \checkmate~analysis which indeed 
allows BP-A2. Compressed scenarios like BP-A1 and BP-A2 would, however, be 
sensitive to the HL-LHC. Also, as for BP-A1, BP-A2 is likely to be probed 
first in the searches for doublet-like heavy Higgs bosons at future LHC runs.

The benchmark point BP-A3, to start with, differs from BP-A2 in having a light 
singlino-like (NLSP, $\ntrltwo$) state in-between the bino-like LSP
($\ntrlone$) and the higgsino-like chargino ($\charonepm$) and neutralinos
$\ntrlthreefour$. This is achieved by lowering the ratio $\kappa/\lambda$. The 
split between $\ntrltwo$ and $\ntrlone$ is tailored to be rather small 
($\sim$ 5 GeV). Expectedly, the abundance of the highly bino-dominated LSP DM 
depletes via its coannihilation with the singlino-dominated NLSP. The DM relic 
abundance is found to lie within the Planck-observed band. Note that the 
proximity in the masses of these two states could, apriori, infuse a 
significant singlino component within the LSP thus pushing up the DMDD-SI cross 
sections dangerously. For the current benchmark scenario, such contamination 
has been tamed by requiring a relative sign between
$M_1$ and $\msinglino (= 2 \kappa \mueff/\lambda)$~\cite{Abdallah:2020yag}. 
Achieving the coveted 
relative sign between these two quantities through a relative sign between 
$M_1$ and $\mueff$ has an additional advantage since, as in BP-A2, this 
further helps restrict the DMDD-SI cross-section below its 
experimentally observed upper limit. The pattern of EWPT in
BP-A3 is pretty similar to those in BP-A1 and BP-A2, i.e., this is a two-
step process of type II-S(1)-D(1). However, the first 
transition along the singlet direction occurs somewhat later in time at 
around 644 GeV (in place of 945 GeV, as in BP-A2).

On the collider front, the higgsino-like  $\ntrlthreefour$ ($\charonepm$) 
preferentially decay to singlino-like NLSP, $\ntrltwo$ (thanks to their 
enhanced coupling given `$\lambda$' is reasonably
large~\cite{Abdallah:2020yag}) along with an on-shell $Z$-boson and 
Higgs bosons ($W^{\pm}$ boson). In turn, it is found that $\ntrltwo$ dominantly 
decays to $\ntrlone \gamma$ ($\sim$ 98$\%$) as its decays to off-shell 
$Z$- and Higgs bosons are much suppressed due to a small mass-split between
$\ntrltwo$ and the LSP. Conservatively, when such photons go undetected 
due to their softness, cascades of the higgsino-like states via $\ntrltwo$ 
would be effectively equivalent to their direct decays to LSP thus resulting in 
canonically sensitive final states like
$3\ell+\etmiss$ and $1\ell+2b$-$jets+\etmiss$. Hence the reported bounds on the
masses of the wino-like electroweakinos from such final states, after 
correcting (relaxing) for the higgsino-like ones, would hold straightaway. Our  
\checkmate~analysis shows that BP-A3 survives this bound and is expected 
to be probed at the HL-LHC via the above standard searches for the 
electroweakinos as well as in the hunt for doublet-like heavier Higgs 
bosons.

The benchmark point BP-A4 differs from BP-A3 in the flipping of the nature of 
the LSP and NLSP, i.e., the LSP (NLSP) becomes singlino-dominated (bino-
dominated). Furthermore, this is the only benchmark point where we find the
$CP$-even singlet Higgs boson to be the lightest of the scalars ($\hs \sim 74$ 
GeV). Also, BP-A4 contains the smallest $|\mueff|$ ($\sim 335$ GeV) among all 
four benchmark points presented in this table. The DM is found to be 
underabundant in the presence of multiple funnels ($\as$ and
$\hsm$). Hence the DMDD bounds are again satisfied thanks to the downward 
scaling of the DD cross sections.

The EWPT still takes place in two steps but is of 
type II-S(1)-D(2) as is suggested by the calculations of $T_c$. The first of these is of the strong first-order type occurring 
along the singlet-direction at $T_c = 165$ GeV. The subsequent transition 
occurs along the $SU(2)$ direction at $T_c = 136.5$ GeV and second-order in 
nature. However, the nucleation calculation indicates that the tunneling  rate 
corresponding to the first transition is too small. Consequently, the actual 
nucleation from the trivial phase to the physical phase takes place directly 
(type I-(1)) at a later time at $T_n = 116.9$ GeV. The possibility of such kind 
of a phase transition has already been pointed out in
reference~\cite{Baum:2020vfl}.
%
%
%%%%%%%%%%%%%%%%%
\begin{figure}[t]
\begin{center}
\includegraphics[height=5.6cm,width=0.45\linewidth]{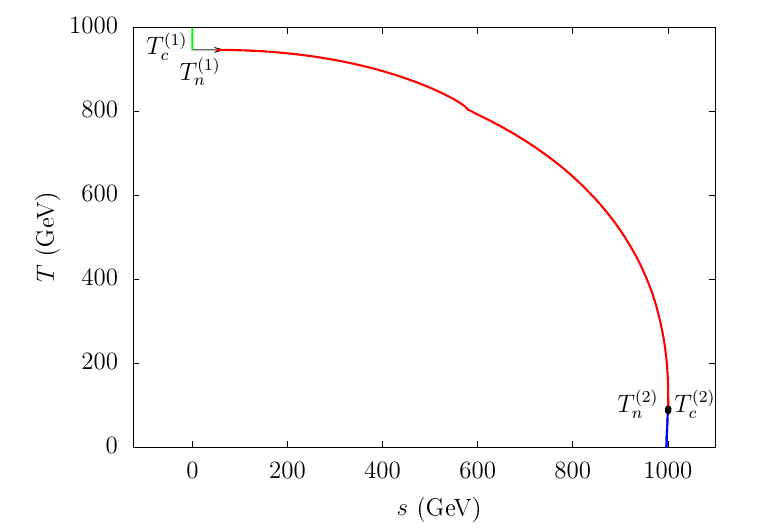}
\hskip 40pt
\includegraphics[height=5.6cm,width=0.44\linewidth]{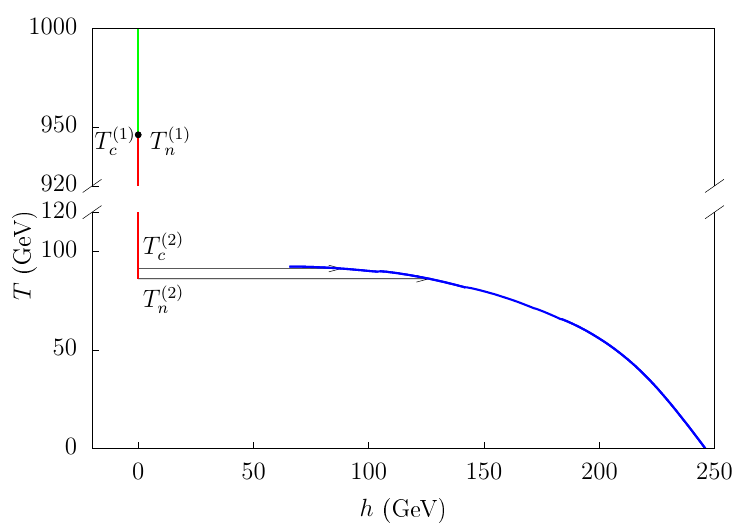}
\hskip 40pt
\includegraphics[height=5.6cm,width=0.44\linewidth]{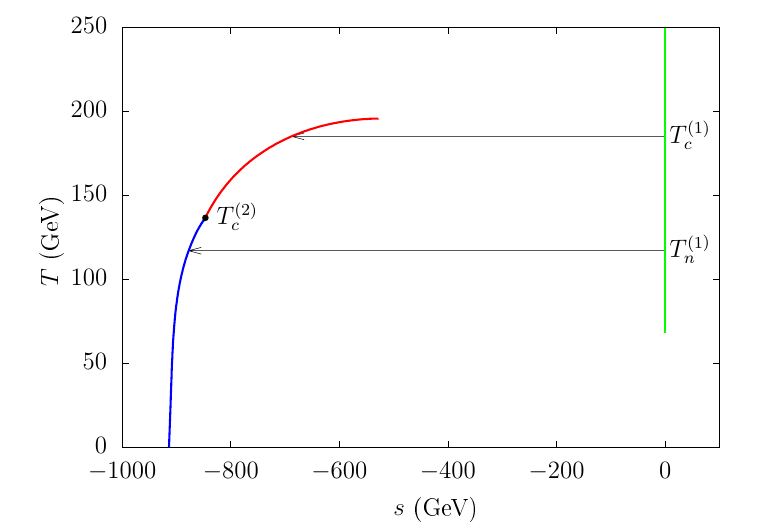}
\hskip 40pt
\includegraphics[height=5.6cm,width=0.44\linewidth]{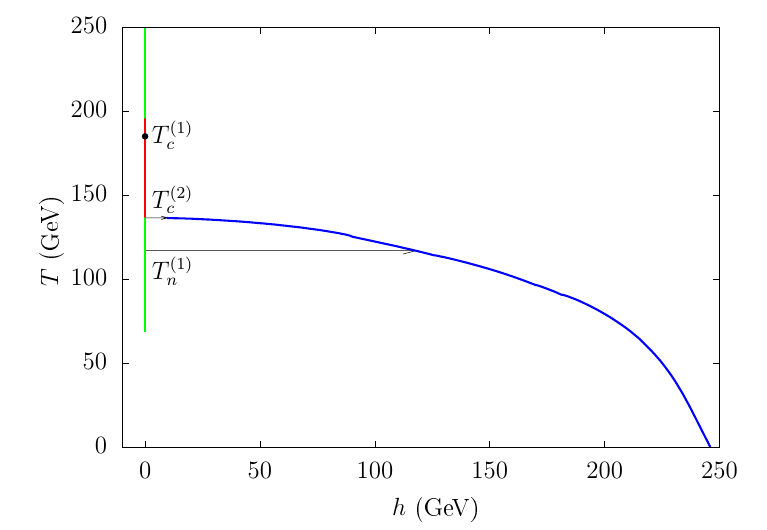}~~
\caption{Phase flows in benchmark scenarios
BP-A2 and BP-A4. Each color stands for a particular minimum of the 
potential (phase) while the individual lines represent the evolution of the 
phase in different field directions (along the singlet direction $s$ (left) and 
along the $SU(2)$-direction $h = \sqrt{h_d^2+h_u^2}$ (right) as a function
of temperature. For each phase transition denoted are the $T_c$ and $T_n$. The 
arrows represent the directions of transition from the false to the true vacuum 
as obtained from the calculations at $T_c$ and $T_n$ in the corresponding field 
space whereas a bullet in black denotes that along this transition the corresponding field value does not alter too much. $T_{c,n}^{(i)} \, (i \in \{1,2\})$ stands for $i$-th transition from the calculation of $T_{c,n}$.}

\label{fig:PT-plot}
\end{center}
\vspace{-0.6cm}
\end{figure}
%%%%%%%%%%%%
%

As in BP-A3, heavier higgsinos, $\ntrlthreefour$, decay to a
bino-dominated NLSP ($\ntrltwo$) and a singlino-dominate LSP ($\ntrlone$) 
accompanied by an on-shell gauge or a Higgs boson. 
Here also, $\ntrlthreefour$'s decays to the singlino-dominated state 
(the LSP in this case) are favored as `$\lambda$' is on the larger side. On the 
other hand, the bino-dominated $\ntrltwo$ now undergoes a dominant decay to the 
singlet-like $CP$-even Higgs boson $\hs$ and the singlino-dominated $\ntrlone$. 
This can play a crucial role in relaxing the relevant collider
bounds~\cite{Abdallah:2020yag, Abdallah:2019znp} when the heavier higgsinos 
first decay to  $\ntrltwo$ which all have branching fractions around $20\%$ 
for the present benchmark point (see table~\ref{tab:BPs_set2}).

Among the implemented analyses in the \checkmate~package an older CMS one 
(with 35.9~\fbinv~of data)~\cite{Sirunyan:2017lae} and another from the ATLAS (with 
139~\fbinv~of data)~\cite{ATLAS:2020qlk} show maximal sensitivities in the 
final states with $3\ell+\etmiss$ and $1\ell+2\gamma+\etmiss$, respectively. 
The corresponding `$r$' values are found to be 0.55 and 0.53 which signify that 
the BP-A4 is still allowed by a wide margin by the electroweakino 
searches at the LHC. This may not be unexpected given that the heavier 
higgsinos do not always undergo one-step decays to the LSP as is 
assumed by the experimental collaborations.
A subsequent analysis with \smodels~that incorporates very recent ATLAS studies 
for the final states like
$1\ell + \hsm(\bm \to bb) + \etmiss$~\cite{Aad:2019vvf},
$2\ell + \etmiss$~\cite{Aad:2019vnb} and $3\ell + \etmiss$~\cite{Aad:2019vvi} 
with 139~\fbinv~of data keeps this benchmark point alive.

In this regard, recent analyses by the ATLAS and the CMS collaborations 
of the final state $3\ell+\etmiss$, with and without extra jets, at
139~\fbinv~and 137~\fbinv~of data,
respectively~\cite{ATLAS:2021moa,CMS:2021cox}, are expected to have heightened 
sensitivities to the present benchmark scenario but are yet to be implemented in the 
recast packages. However, we have managed to check the constraints from the 
ATLAS analysis~\cite{ATLAS:2021moa} for this benchmark scenario and we 
find~\cite{private} BP-A4 to be still allowed. The scenario is 
expected to get probed at the future LHC runs in the 
electroweakino searches first rather than in the searches for the heavy Higgs 
bosons. This is since these doublet-like heavy Higgs bosons are heavier in the 
present case ($\sim 1.3$ TeV). 

As we have just discussed, benchmark points BP-A1, BP-A2, and BP-A3 have 
similar phase transition patterns while BP-A4 has one of a different kind. 
Hence, in figure~\ref{fig:PT-plot}, we choose to show the relevant phase 
diagrams for only BP-A2 (having an SFOEWPT in two steps with a palpable 
split between $T_c$ and $T_n$) and BP-A4 (one-step phase transition with a reasonably large 
$T_c$ and $T_n$ for the SFO) which may serve as the representative 
scenarios for the purpose.
\hspace{-5cm}
%%%%%%%%%%%%%%%%
\begin{table}[t]
\renewcommand{\arraystretch}{1.25}
   \centering
   \setlength{\extrarowheight}{5.5pt}
\small{\begin{tabular}{|c|c|c|c|}
\hline   
 \multirow{1}{*}{BP No.}& \multirow{1}{*}{$T_n$ (GeV)} & \multirow{1}{*}{$\alpha$} & \multirow{1}{*}{$\beta/H_n$} \\
\hline 
 \multirow{2}{*}{BP-A1} & 946.7  & $2.04\times10^{-5}$ & $1.31\times10^7$\\& 90.2 & $2.34\times10^{-2}$ & $2.53\times10^4$ \\
 \hline
  \multirow{2}{*}{BP-A2} & 945.9  & $2.15\times10^{-5}$ & $1.19\times10^7$\\& 86.2 &  $4.33\times10^{-2}$ & $1.21\times10^3$\\
 \hline
   \multirow{2}{*}{BP-A3} & 644.3  & $1.12\times10^{-4}$ & $2.06\times10^6$\\& 94.5 &  $1.82\times10^{-2}$ & $3.71 \times10^4$\\
 \hline
    \multirow{1}{*}{BP-A4} & \multirow{1}{*}{116.9}  & \multirow{1}{*}{$8.63\times10^{-2}$} & \multirow{1}{*}{$2.22 \times10^2$}\\
\hline
\end{tabular}}
   \caption{Values of the parameters $T_n$, $\alpha$ and $\beta/H_n$ (that control the GW intensity) for the benchmark points presented in
   table~\ref{tab:BPs_set2}.}
   \label{tab:Table_GW_data}
   \vspace{-0.3cm}
\end{table}
%
%%%%%%%%%%%%%%%%%%%%%%%%%%%%%%%%%%%%%%%%%%%%%%%%%%%%%%%%%%
\subsubsection{Prospects of GW detection}\label{GWresults}
\label{GWresult}
%%%%%%%%%%%%%%%%%%
%
\begin{figure}[t!]
\begin{center}
\includegraphics[height=5.6cm,width=0.43\linewidth]{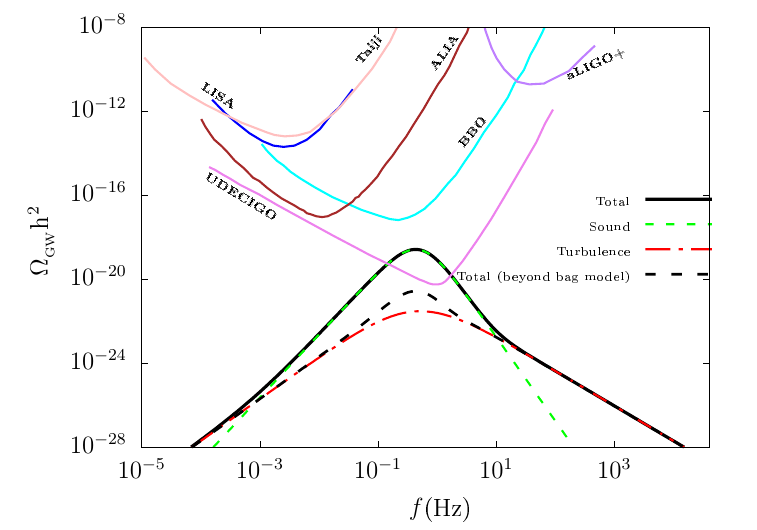}
\hskip 40pt
\includegraphics[height=5.6cm,width=0.44\linewidth]{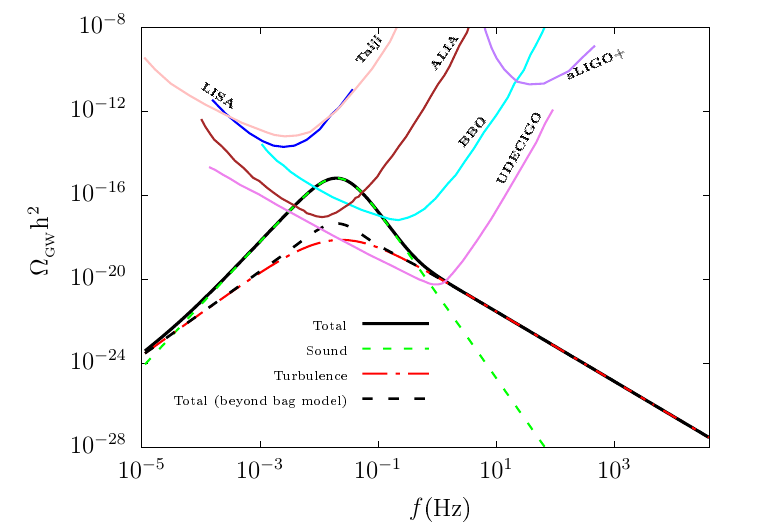}
\hskip 40pt
\includegraphics[height=5.6cm,width=0.44\linewidth]{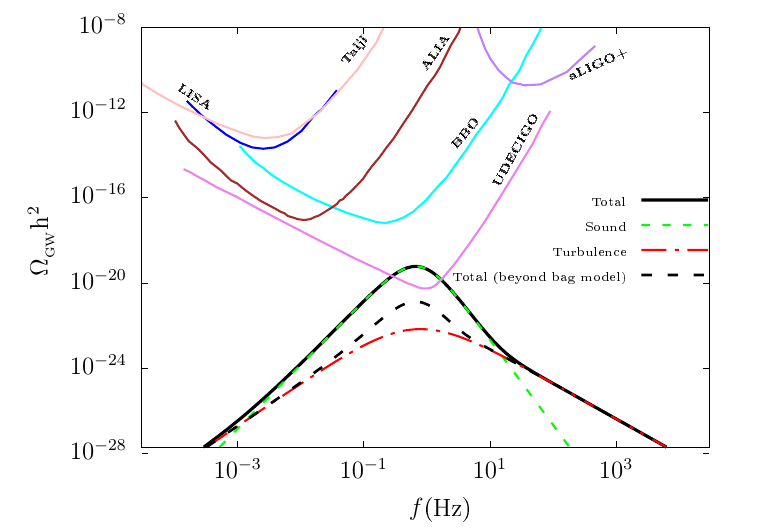}
\hskip 40pt
\includegraphics[height=5.6cm,width=0.44\linewidth]{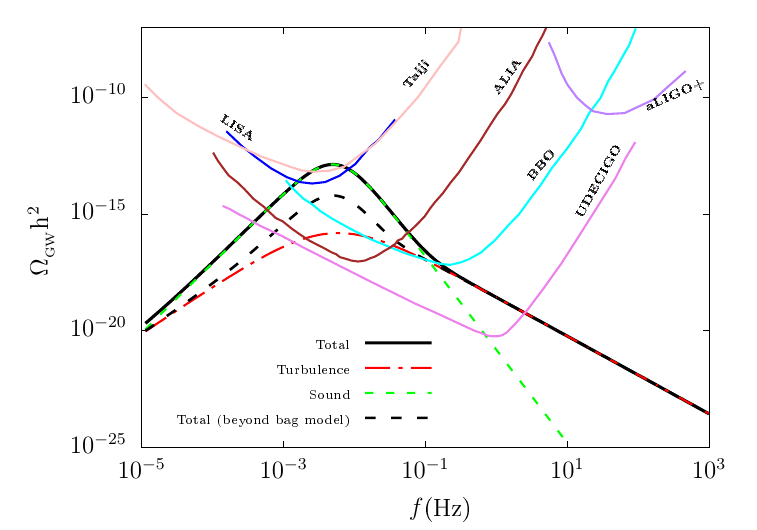}~~
\caption{GW energy density spectrum with respect to frequency for the four 
benchmark scenarios BP-A1 (top, left), BP-A2 (top, right), BP-A3 (bottom, left) 
and BP-A4 (bottom, right) illustrated against the experimental sensitivity 
curves of some GW detectors like LISA, Taiji, TianQin, aLigo+, BBO and
U-DECIGO. In each plot, the solid black line denotes the total GW energy density within the bag model
whereas the broken lines in 
green and red represent the individual contributions from sound waves (within the bag model) and turbulence, respectively. The broken black lines denote the total GW spectrum beyond the bag model.}
\label{fig:GW-freq-plot}
\end{center}
\vspace{-0.5cm}
\end{figure}
%%%%%%%%%%%%
%
Values of various key parameters ($T_n$, $\alpha$, $\beta/H_n$) pertaining to 
the GW spectra arising from the FOPTs for the benchmark scenarios BP-A1 to
BP-A4 are shown in table~\ref{tab:Table_GW_data}. The corresponding GW 
(frequency) spectra are calculated using
equations~\ref{GWtotal}--\ref{Turbpeakfreq} and are shown in 
figure~\ref{fig:GW-freq-plot}. These are further compared with the sensitivity 
of some space- and ground-based gravitational wave detectors, viz.,
LISA~\cite{LISA:2017pwj}, Taiji~\cite{Gong:2014mca}, TianQin~\cite{TianQin:2015yph}, aLigo+~\cite{Harry:2010zz}, Big Bang Observer (BBO)~\cite{Corbin:2005ny} and  Ultimate(U)-DECIGO~\cite{Kudoh:2005as}.
Note that for each of the benchmark points BP-A1, BP-A2 and BP-A3 we observe a two-step first order phase transition. The phase transitions along the singlet-direction for these benchmark points happen at relatively larger temperatures (happens to be at larger $\beta/H_n$ and for smaller $\alpha$). Thus, the contributions  of the first FOPT (along the singlet direction) to the GW spectrum are relatively much smaller compared to the ones from the second FOPT which occur along the $SU(2)$ field directions. This is why a spectral peak due to the first FOPT along the singlet direction for these benchmark points does not appear in figure~\ref{fig:GW-freq-plot}.\footnote{An interesting situation with a GW spectrum having multiple peaks (from a two-step phase transition in the NMSSM) could be observed in these experiments if the 
transition along the singlet direction also takes place at $T_n \sim$ 100 
GeV (which enhances the corresponding `$\alpha$' value) and with a relatively smaller value of $\beta/H_n$. We reserve this study for a future 
work.} In all four plots, the individual contributions from sound waves  (within the bag model) and turbulence  are shown with broken lines, in green and red colors, respectively. The total GW spectra, within the bag model, are denoted by solid, black lines. On the other hand, the same beyond the bag model (taking into account equations~\ref{beyondbagmodelpseudotrace}--\ref{beyondbagmodelrelation}) are indicated in these plots by broken black lines.

As can be found from these plots, the peak region of the
GW spectrum obtained in the bag model for BP-A1 and BP-A3 lie only within the sensitivity of U-DECIGO. However, beyond the bag model, calculations suggest that the peak GW intensities for these two points fall short of the sensitivity of U-DECIGO. As for BP-A2, the peak of GW spectrum in the bag model lies within the sensitivity of ALIA, BBO and U-DECIGO while beyond the bag model it only falls within the sensitivity of U-DECIGO. Also, note that the peaks of these spectra are higher for BP-A2 when compared to BP-A1. Given that these two benchmarks have otherwise very similar types of phase transitions, the reason 
behind this can be traced back to the fact that in BP-A2 there occurs a 
stronger FOPT in the $SU(2)$ field directions due to a relatively late-time 
nucleation and the associated values taken by `$\alpha$' and $\beta/H_n$ are 
such that the GW peak intensities shoot up thus increasing the prospects of observing the same. In contrast, for 
BP-A4, an extended section of the GW spectrum, around its peak, in the bag model, falls within the reaches of multiple 
experiments like LISA, Taiji, ALIA, BBO and U-DECIGO and, beyond the bag model, the same falls within the sensitivities of ALIA, BBO and U-DECIGO.

The quantity signal-to-noise ratio (SNR) is used to measure the
detectability of the GW signal at the experiments. SNR is defined
as~\cite{Caprini:2015zlo}
\beq\small{
\text{SNR} = \sqrt{\delta\times \mathcal{T}\int_{f_{min}}^{f_{max}}df\bigg[\frac{h^2\Omega_{\text{GW}}(f)}{h^2\Omega_{\text{exp}}(f)}\bigg]^2},}
\eeq
where $\mathcal{T}$ is the duration of the experimental mission in years,
$\delta$ stands for the number of independent channels employed by an 
experiment to exploit cross-correlations (required to pin down the stochastic 
origin of the GW) and $\Omega_{\text{exp}}(f) \, h^2$ denotes the effective 
power spectral density of strain noise of the experiment. Here, we consider
$\delta=2$ for BBO and U-DECIGO and $\delta=1$ for LISA while for 
all of them we take $\mathcal{T}=5$. The SNR values for the benchmark points 
are found to be way below 1, with the exception of BP-A4, for which it 
is comparatively large but still $< 1$. These are to be compared with 
the reference minimum threshold value of SNR for the detection of GW 
which is taken to be 10~\cite{Caprini:2015zlo}. Thus, none 
of the benchmark scenarios meets this detectability criterion. It 
is, however, expected that there is a region of parameter space in the NMSSM 
which might give rise to stronger FOPTs (larger $\alpha$) and the corresponding 
GW spectra lie deeper within the regions of experimental sensitivity (depending 
upon $\beta/H_n$) thus yielding an SNR value larger than
10.\footnote{The exploration, however, demands a more detailed study which we 
will take up in a future work.}
%

%%%%%%%%%%%%%%%%%%%%%%%%%%%%%
\section{Summary and outlook}
\label{sec:summary}
%%%%%%%%%%%%%%%%%%%
Inspired by the prospects the \z3nmssm~scenario holds in explaining the baryon 
asymmetry of the Universe via EWBG, which in turn requires SFOEWPT, we have 
sought to figure out how accommodating the scenario appears in the face 
of recent LHC results, in particular, the ones pertaining to the searches of 
the lighter electroweakinos which might happen to be higgsino-like and 
are favored by SFOEWPT. Various pertinent theoretical requirements, constraints 
on the Higgs sector (including the observed properties of the SM-like Higgs 
boson) from the LHC, flavor-constraints and bounds on various DM observables 
obtained from a host of dedicated experiments do already play their parts in 
delineating the allowed region of the NMSSM parameter space that still 
remains compatible with SFOEWPT. We further look into the prospects of 
detecting the (stochastic) GW arising from an SFOEWPT at various future 
experiments.

The backdrop of our present study has been the looming tension between the 
physics of SFOEWPT and the recent LHC results from the electroweakino searches.
While the former prefers $\mueff$ in the range of a few hundreds of a GeV, the 
latter are tending to push $\mueff$ steadily above such a ballpark. The general 
goal of such a study could then be to check if there is a meeting ground 
somewhere in the middle where both constraints are simultaneously complied 
with. A further observation is that EWPT is somewhat stubborn in its need for
relatively small $\mueff$. A middle ground can thus only be found if the 
reported constraints from the LHC could be evaded under circumstances that 
have not been considered explicitly by the LHC experiments. In this work, we 
exploit such caveats to our advantage via recasts of the relevant LHC analyses 
using popular packages like \checkmate~and \smodels.

Thus, the region of the \z3nmssm parameter space that concerns us in this work 
is characterized by reasonably small $\mueff$ that yields
relatively light higgsino-like states. Also, SFOEWPT prefers a
relatively light 
$CP$-even singlet-like scalar, $\hs$,
thus requiring `$\kappa$' to be small. This leads
to a relatively light singlino in the spectrum which can be the DM particle while the light singlet scalars play crucial roles in the DM phenomenology. Also, for our benchmark scenarios, we choose relatively large values of
`$\lambda$' ($\gtrsim 0.5$) and smaller values of $\tanb$ ($\lesssim 5$). 
Together, these yields $\mhsm$ in the right ballpark ($\sim 125$ GeV) without requiring too large SUSY radiative corrections. 
Furthermore, a smaller $\tanb$ could allow the heavier doublet Higgs bosons to 
remain relatively light (reminiscent of the ``alignment  without decoupling''
scenario) which might aid SFOEWPT. With `$\lambda$' on the larger side, the
mixing among the higgsinos and the singlino, all of 
which can be relatively light thus being the candidates for the LSP DM and the 
NLSP, can be sizable. Hence, with such choices of theory parameters, the 
physics of the EWPT (and hence EWBG) becomes intricately connected to the
DM and collider (LHC) phenomenologies.

Results of a scan over the parameter space are presented depicting first how
relatively small $\mueff \, (\lesssim 500 \, \text{GeV})$ fairs against  various theoretical and basic experimental bounds including those from the observed Higgs sector and the ones from the DM 
experiments.
Two sets of benchmark scenarios are then presented to demonstrate the
SFOEWPT--DM--LHC connection. These scenarios are checked to give rise to SFOEWPT by using the package \cosmotransitions~in which we implemented the framework of \z3nmssm, matched to THDMS, 
as has been a pretty standard practice for the purpose.  

With one set of benchmark scenarios, we have sought to find out up to what a 
ballpark maximum value of not so large a $\mueff$ can still be ruled out by 
recent LHC analyses, in particular, when one departs from the 
simplified assumptions on the decays and branching fractions of the cascading 
electroweakinos which is expected to relax the reported bounds on the 
electroweakino masses. Subjecting this set of otherwise highly motivated 
scenarios to thorough recasts of some pertinent LHC analyses (with both
36~\fbinv~and 139~\fbinv~of data) with the help of \checkmate~and
\smodels~reveals that $\mueff \lesssim 300$ GeV, with low values of `$\kappa$'
($\lesssim 0.1$) and larger `$\lambda$' ($\gtrsim 0.5$), is mostly ruled out. It should be noted that
smaller values of  $\mueff$ already attract severe constraints from the DM direct detection experiments. Thus, in this regard, the LHC searches might not always yield a robust improvement over the DM bounds. An even smaller $\mueff$ could, however, survive the LHC bounds for a compressed electroweakino spectrum. In this work, being conservative, we do not consider this possibility.

With the other set of benchmark scenarios, we have demonstrated how low a
$\mueff$ could still be allowed instead. A similar exercise shows that,
under favorable circumstances, upwards of $\mueff \sim 335$ GeV could survive the LHC onslaught. This is rather encouraging since we find that EWPT could still 
remain to be of strong, first-order type even for $\mueff$ as large as
$\sim 425$ GeV which is the case for a couple of benchmark scenarios that we have 
presented. These also show that a viable LSP DM can be bino- or singlino-like or even a mixture of bino, singlino and higgsino states. We have 
thoroughly studied the properties of EWPT in these scenarios with the help of 
\cosmotransitions~and have found that for $\mueff$ on the larger side,
a two-step phase transition is a more likely phenomenon with the first 
transition taking place in the singlet field direction followed by the other
in the $SU(2)$ field directions.

For these latter set of scenarios, we have thoroughly studied the stochastic 
GW (background) spectra that might carry the imprints of FOPT from new physics 
beyond the SM. We find that the signal intensities lie inside the sensitivity 
limits of one or more of the future/proposed experiments like the LISA, BBO, 
UDECIGO, Taiji, Alia, etc. However, the SNR values, as such, are not found to be healthy 
enough to guarantee a positive detection.

In summary, the present work corroborates the basic findings reported in the 
literature pertaining to SFOEWPT, in particular, and EWPT, in general, in the 
framework of the \z3nmssm. We broadly concur with various reported 
patterns and features of EWPT in such a scenario and the different conditions 
under which those manifest. We then go beyond to shed light on what the recent 
searches of the electroweakinos at the LHC have to say about the viability of 
SFOEWPT in the current framework while compatibility with the constraints 
from various pertinent theoretical and experimental sectors including the DM 
sector is ensured all through. Furthermore, it appears that the GW signals resulting from the strong FOPTs in these scenarios are likely to remain too 
weak to be detected at future dedicated experiments. 

As for an outlook, new LHC studies with data from the recently terminated LHC 
Run~2 and those that would arrive soon from high luminosity LHC (HL-LHC) are 
likely to shed a more unambiguous light on the broad viability of EWBG 
within the \z3nmssm while these continue to explore electroweakinos 
at larger masses and in difficult scenarios like the compressed ones. Furthermore, improvements are possible in the theoretical 
calculations of several key EWBG objects, viz., the bubble wall profile, wall 
velocity and $CP$-violation and in the dealing of the transport equations which 
could lend a more accurate estimate of the relation between the NMSSM 
parameters and EWBG. With these, a reassessment of the detectability of such GW signals may be warranted which might prove the latter's role as complementary to the LHC searches. A synergy like this between LHC and GW physics is likely to be rather intriguing and we reserve such a study for a future work.
%
%%%%%%%%%%%%%%%%%%%%%%%%%
\section{Acknowledgments}
%%%%%%%%%%%%%%%%%%%%%%%%%
AC thanks Harish-Chandra Research Institute (HRI) for hosting him during the
course of this collaborative work. AC also acknowledges partial support from the Department of Science and Technology, India, through INSPIRE faculty fellowship (grant no: IFA 15 PH-130, DST/INSPIRE/04/2015/000110). SR is supported by the funding available 
from the Department of Atomic Energy (DAE), Government of India for the 
Regional Centre for Accelerator-based Particle Physics (RECAPP) at HRI. SR 
would like to thank Peter Athron, Sebastian Baum, Junji Cao, Ulrich Ellwanger, 
Andrew Fowlie, Tathagata Ghosh, Thomas Konstandin, Sabine Kraml, Indrani Pal, Avik Paul, Krzysztof 
Rolbiecki, Tim Stefaniak, Di Zhang, Yang Zhang  for helpful discussions and 
communications. SR also acknowledges the use of cluster computing available at 
the High Performance Scientific Computing facility at HRI and thanks Rajiv 
Kumar for his technical help at this facility.
%
%%%%%%%%%%%
%\vskip 30pt
\appendix
\section*{Appendices}
\addcontentsline{toc}{section}{\protect\numberline{}Appendices}%
%
%%%%%%%%%%%%%%%%%%%%%%%%%%%%%%%%%%%%%%%%%%%%%%%%%%%%%%%%%%%%%%%%%%%%%%%%%%%%%%%
\section{Matching the NMSSM parameters to those in the THDMS potential}
\label{NMSSMTHDMSmatch}
%%%%%%%%%%%%%%%%%%%%%%%
In terms of the $CP$-even Higgs fields $h_d$, $h_u$ and $s$, the tree-level
\z3nmssm potential of equation~\ref{Eq:NMSSMHiggspotential}, which is relevant 
for the study of phase transitions, can be written as~\cite{Ellwanger:2009dp}
\begin{multline}
%\begin{align}
\hskip -8pt
V_{\rm tree}^{\rm NMSSM}(h_d, h_u, s) =
\frac{1}{32}(g_1^2 + g_2^2) \left( h_d^2 - h_u^2 \right)^2 + \frac{1}{4} \kappa^2 s^{4} - \frac{1}{2} \lambda \kappa s^2 h_d h_u + \frac{1}{4} \lambda^2 \left[h_d^2 h_u^2 + s^2 \left(h_d^2 + h_u^2\right)\right] \\
 - \frac{1}{\sqrt{2}} \lambda A_{\lambda}   s h_d h_u +  \frac{1}{3\sqrt{2}}  \kappa  A_{\kappa}  s^3
+ \frac{1}{2} m_{H_d}^2 h_d^2  + \frac{1}{2} m_{H_u}^2 h_u^2  +  \frac{1}{2}  m_S^2  s^2.
\label{eq:NMSSMTreelevel}
%\end{align}
\end{multline}
On the other hand, the tree-level $Z_3$-symmetric THDMS  potential is given
by~\cite{Elliott:1993ex, Elliott:1993uc, Elliott:1993bs, Athron:2019teq}
\begin{align}
V_{\rm tree}^{\rm THDMS}  = & \frac{1}{2} \lambda_1 \left|  H_d\right|^4  +  \frac{1}{2}\lambda_2 \left|H_u\right|^4 + \left(\lambda_3 + \lambda_4\right) \left|H_d\right|^2\left|H_u\right|^2 - \lambda_4 \left|H_u^{\dagger} H_d\right|^2 + \lambda_5 \left|H_d\right|^2\left|S\right|^2 \nonumber \\
&+\lambda_6 \left|H_u\right|^2\left|S\right|^2  + \lambda_7\left(S^{*2}H_d\cdot H_u  + {\rm h.c.} \right) + \lambda_8 \left|S\right|^4  +  m_1^2\left|H_d\right|^2 + m_2^2\left|H_u\right|^2 + m_3^2 \left|S\right|^2 \nonumber\\
&-m_4 \left(H_d\cdot H_u S + {\rm h.c.}\right) - \frac{1}{3} m_5 \left(S^3  +  {\rm h.c.}\right).
\label{eq:THDMS_potential}
\end{align}
All parameters in equation~\ref{eq:NMSSMTreelevel} and~\ref{eq:THDMS_potential} 
are taken to be real as we do not consider any $CP$-violation in the Higgs 
sector in this work. We use ${\tt NMSSMTools}$ to obtain the particle spectrum 
of the \z3nmssm at the scale $M_{\rm SUSY}$. Except for the electroweakinos 
with masses around a few hundred GeV, we consider all other SUSY 
excitations to be much heavier such that those may be considered
effectively decoupled from the physics of phase transitions. However, to {avoid 
large logarithmic corrections from appearing in} $V_{\rm CW}^{\rm NMSSM}$ due 
to the top squarks, those are integrated out at the scale $M_{\rm SUSY}$ 
in an EFT approach. Below this scale, $V_{\rm tree}^{\rm NMSSM}$ can be mapped 
onto $V_{\rm tree}^{\rm THDMS}$. Comparing equations~\ref{eq:NMSSMTreelevel} 
and~\ref{eq:THDMS_potential}, after expanding the latter in terms of the 
component fields $h_d, h_u$ and `$s$', the matched conditions among the model 
parameters of these two scenarios, at the scale $M_{\rm SUSY}$, are given
by~\cite{Elliott:1993ex, Elliott:1993uc, Elliott:1993bs, Athron:2019teq}
\begin{align}
\label{eq:Matchconditions}\small{
\begin{gathered}
\lambda_1  =  \lambda_2  = \frac{1}{4} \left({g_1}^2 + g_2^2 \right),
\quad
\lambda_3 = \frac{1}{4}  \left( g_2^2 - g_1^2 \right),
\\
\lambda_4 = \frac{1}{2}  \left(2 |\lambda|^2 - g_2^2 \right),
\quad
\lambda_5 = \lambda_6 = |\lambda|^2,
\quad
\lambda_7 = -\lambda  \kappa,
\quad
\lambda_8 = |\kappa|^2,
\\
m_1^2 = m_{H_d}^2,
\quad
m_2^2 =  m_{H_u}^2,
\quad
m_3^2 =  m_{S}^2,
\quad
m_4 = A_\lambda  \lambda,
\quad
m_5 = - A_\kappa  \kappa.
\end{gathered}}
\end{align}
At one-loop, the only relevant threshold correction that arises (as the top 
squarks are integrated out at the scale $M_{\rm SUSY}$) is to $\lambda_2$ and 
is given
by~\cite{Ellis:1991zd,Haber:1993an, Casas:1994us, Carena:1995bx}
\begin{equation}
\label{eq:OneLoopThreshold}\small{
\Delta \lambda_2
=
 \frac{3 y_t^4  A_t^2}{8 \pi^2 M\subs{SUSY}^2}
 \left(1 - \frac{A_t^2}{12 M\subs{SUSY}^2}  \right),}
\end{equation}
where $A_t$ is the soft-SUSY-breaking top squark-Higgs trilinear coupling in 
the scalar potential and $y_t$ is the top quark Yukawa coupling, both defined
at the scale $M_{\rm SUSY}$. Note that all the NMSSM parameters are also 
provided at the scale $M_{\rm SUSY}$ and in the $\overline{\text{DR}}$ scheme. 
Thus, after matching, all the THDMS parameters also get defined at the same 
scale and in the same renormalization scheme.
%
%%%%%%%%%%%%%%%%%%%%%%%%%%%%%%%%%%%%%%%
\section{RGEs in the THDMS}\label{RGEs}
%%%%%%%%%%%%%%%%%%%%%%%%%%%%%%%%%%%%%%%%
We borrow the set of relevant RGEs from reference~\cite{Kozaczuk:2014kva} which 
we use to run the THDMS model parameters from the scale $M_{\rm SUSY}$ to the 
scale $m_t$. Contributions to the $\beta$-functions from the SM gauge bosons, 
the Higgs bosons, the top quark, the higgsinos and the singlino are included. 
As for the gauginos, the contribution from the bino is known to be small (even when 
$\mone$ does not get to be too large, as is the case in our present analysis) 
while $\mtwo$ is set at a rather large value. Hence we ignore their effects.
With these, the one-loop $\beta$-functions for the model 
parameters $q_i$ are given by
\begin{equation}
\label{eq:beta}\small{
\beta_{q_i}  =  \frac{1}{16\pi^2} \frac{\partial}{\partial  \ln  \Lambda}  q_i  \, ,}
\end{equation}
where `$\Lambda$' is the energy scale.
The one-loop RGEs for the quartic couplings
$\lambda_i, i \in \{1,2,3, ..., 8\}$, the mass parameters ($m_{4,5}$) and
$v_s$ appearing in $V_{\rm tree}^{\rm THDMS}$ of
equation~\ref{eq:THDMS_potential} are as
follows \cite{Kozaczuk:2014kva, Ellwanger:2009dp, Elliott:1993ex}:
\begin{equation}
\label{eq:RGEs}
\footnotesize{
\begin{aligned}
\beta_{\lambda_1} = &12 \lambda_1^2+4\lambda_3^2+4\lambda_3\lambda_4+2\lambda_4^2+2\lambda_5^2-\lambda_1(3g_1^2+9g_2^2) +\frac{3}{4} g_1^4+\frac{9}{4}g_2^4+\frac{3}{2}g_1^2g_2^2
-4\lambda^4+4\lambda^2\lambda_1,\\
\beta_{\lambda_2} = &12\lambda_2^2+4\lambda_3^2+4\lambda_3\lambda_4+2\lambda_4^2+2\lambda_6^2-\lambda_2(3g_1^2+9g_2^2)+\frac{3}{4}g_1^4+\frac{9}{4}g_2^4+\frac{3}{2}g_1^2g_2^2 +12y_t^2\lambda_2-12y_t^4\\
&-4\lambda^4+4\lambda^2\lambda_2, \\
\beta_{\lambda_3} = &(\lambda_1+\lambda_2)(6\lambda_3+2\lambda_4)+4\lambda_3^2+2\lambda_4^2+2\lambda_5\lambda_6-\lambda_3(3g_1^2+9g_2^2)+\frac{3}{4} g_1^4+\frac{9}{4}g_2^4-
\frac{3}{2}g_1^2g_2^2+6y_t^2\lambda_3\\
&-4\lambda^4+8\lambda^2\lambda_4+8\lambda^2\lambda_3,\\
\beta_{\lambda_4} = &2\lambda_4(\lambda_1+\lambda_2+4\lambda_3+2\lambda_4)+4\lambda_7^2-\lambda_4(3g_1^2+9g_2^2)+3g_1^2g_2^2+6y_t^2\lambda_4+4\lambda^4-4\lambda^2\lambda_4,\\
\beta_{\lambda_5} = &\lambda_5(6\lambda_1+4\lambda_5+8\lambda_8)+\lambda_6(4\lambda_3+2\lambda_4)+8\lambda_7^2-\frac{1}{2}\lambda_5(3g_1^2+9g_2^2)-12\kappa^2\lambda^2-4\lambda^4+4\kappa^2\lambda_5+6\lambda^2\lambda_6,\\
\beta_{\lambda_6} = &\lambda_5(4\lambda_3+2\lambda_4)+\lambda_6(6\lambda_2+4\lambda_6+8\lambda_8)+8\lambda_7^2-\frac{1}{2}\lambda_6(3g_1^2+9g_2^2)+6y_t^2\lambda_6-16\kappa^2\lambda^2-4\lambda^4\\
&+4\kappa^2\lambda_6+6\lambda^2\lambda_6,\\
\beta_{\lambda_7} = &\lambda_7(2\lambda_3+4\lambda_4+4\lambda_5+4\lambda_6+4\lambda_8)-\frac{1}{2}\lambda_7(3g_1^2+9g_2^2)+3y_t^2\lambda_7+8\kappa\lambda^3+4\kappa^2\lambda_7+6\lambda^2\lambda_7, \\
\beta_{\lambda_8} = &2\lambda_5^2+2\lambda_6^2+4\lambda_7^2+20\lambda_8^2+8(\kappa^2+\lambda^2)\lambda_8-16\kappa^2-4\lambda^4,\\
\beta_{m_4} = &(2\lambda_3+4\lambda_4+2\lambda_5+2\lambda_6+4\lambda^2+2\kappa^2 - \frac{9}{2}g_2^2-\frac{3}{2}g_1^2+3y_t^2)m_4+4\lambda_7 m_5,\\
\beta_{m_5} = &(12\lambda_8+6\lambda^2+6\kappa^2)m_5+12\lambda_7 m_4,\\
\beta_{v_s}= &-2v_s(\kappa^2+\lambda^2). \quad 
\end{aligned}}
\end{equation}
The individual THDMS parameters at the scale $m_t$ are then calculated using 
the expression\footnote{Note that the NMSSM parameters are in the
$\overline{\rm DR}$ scheme~\cite{Ellwanger:2009dp} whereas, for the 
calculations of phase transitions, the THDMS parameters are provided in the
$\overline{\rm MS}$ scheme~\cite{Kozaczuk:2014kva}. We ignore the effect of 
this shift in the scheme as this would modify the quartic couplings only mildly 
due to small threshold corrections. To convince ourselves, we have 
compared the mass-eigenvalues of the $CP$-even Higgs mass-squared matrix and 
the related mixing matrix that are obtained from \cosmotransitions~to 
the corresponding ones obtained using \nmssmtools~and their agreements are 
found to be within the level of a few percent.}
(see reference~\cite{Kozaczuk:2014kva} for details) 
\begin{equation}\label{eq:fixed_order}\small{
\begin{aligned}
&q_i(m_t) \simeq q_i(M_{\rm SUSY})-\beta_{q_i} \ln \frac{M_{\rm SUSY}}{m_t}\, .
\end{aligned}}
\end{equation}  
%
%%%%%%%%%%%%%%%%%%%%%%%%%%%%%%%%%%%%%%%%%%%%%%%%%%%%%%%%%%
\section{Field-dependent masses and the daisy corrections}
\label{field-dependent-masses}
%%%%%%%%%%%%%%%%%%%%%%%%%%%%%%
Here we present the field-dependent mass-squared matrices for the scalar sector 
which are derived from
equation~\ref{eq:THDMS_potential}~\cite{Elliott:1993ex, Athron:2019teq}.
The $3\times3$, symmetric matrix (${\cal M}_{H}^2$) for the $CP$-even scalars, 
in the basis $\{h_d,h_u,s\}$, is given by
\begin{equation}
\label{CpEvenHiggsMat}
\tiny{
\begin{pmatrix}m_1^2
 + \tfrac32 \lambda_1 \Hdzr^2
 + \tfrac12 \lambda_5 \Sr^2
 + \tfrac12 (\lambda_3 + \lambda_4) \Huzr^2 &  - \tfrac1{\sqrt{2}} m_4 {\Sr}
 + \tfrac12 \lambda_7 \Sr^2
 + (\lambda_3 + \lambda_4) \Huzr \Hdzr &  - \tfrac1{\sqrt{2}} m_4 \Huzr
 +  \lambda_5 \Hdzr {\Sr}
 +  \lambda_7 \Huzr {\Sr} \\
                                                ...  & m_2^2
 + \tfrac32 \lambda_2 \Huzr^2
 + \tfrac12 \lambda_6 \Sr^2
 + \tfrac12 (\lambda_3 + \lambda_4) \Hdzr^2 &   - \tfrac1{\sqrt{2}} m_4 \Hdzr
 + \lambda_7 \Hdzr {\Sr}
 + \lambda_6 \Huzr {\Sr}\\
                                                ...    &   ... & m_{3}^2
 - \sqrt{2} m_5 {\Sr}
 + \tfrac12 \lambda_5 \Hdzr^2
 + \lambda_7 \Huzr \Hdzr
 + \tfrac12 \lambda_6 \Huzr^2
 + 3 \lambda_8 \Sr^2
                                    \end{pmatrix} \; ,}
\end{equation}
whereas, the corresponding one for the $CP$-odd scalars (${\cal M}_A^2$), in 
the same basis as above, can be written as
\begin{equation}
\label{CPoddmassmatrix}
\tiny{
\begin{pmatrix}m_1^2
 + \tfrac12 \lambda_1 \Hdzr^2
 + \tfrac12 \lambda_5 \Sr^2
 + \tfrac12 (\lambda_3 + \lambda_4) \Huzr^2 & \tfrac1{\sqrt{2}} m_4 {\Sr}
 - \tfrac12 \lambda_7 \Sr^2 & \tfrac1{\sqrt{2}} m_4 \Huzr
 + \lambda_7 \Huzr {\Sr}  \\
                                              ...     &     m_2^2
 + \tfrac12 \lambda_2 \Huzr^2
 + \tfrac12 \lambda_6 \Sr^2
 + \tfrac12 (\lambda_3 + \lambda_4) \Hdzr^2 & \tfrac1{\sqrt{2}} m_4 \Hdzr
 + \lambda_7 \Hdzr {\Sr}\\
                                              ...     &    ...   &   m_3^2
 + \sqrt{2} m_5 {\Sr}
 + \tfrac12 \lambda_5 \Hdzr^2
 - \lambda_7 \Huzr \Hdzr
 + \tfrac12 \lambda_6 \Huzr^2
 + \lambda_8 \Sr^2
                                       \end{pmatrix} \;.}
\end{equation}
On the other hand, the field-dependent $2\times2$, symmetric mass-squared 
matrix for the charged Higgs sector (${\cal M}_{H^\pm}^2$) in the 
basis $\{h_d,h_u\}$ is given by
\begin{equation} \label{ChargedHiggsmassmatrix}
\small{
{\cal M}_{H^\pm}^2 = \begin{pmatrix}m_1^2
 + \tfrac12 \lambda_5 \Sr^2
 + \tfrac12 \lambda_1 \Hdzr^2
 + \tfrac12 \lambda_3 \Huzr^2 & \tfrac1{\sqrt{2}}  m_4 \Sr
 - \tfrac12 \lambda_7 \Sr^2
 - \tfrac12 \lambda_4 \Hdzr \Huzr\\
                                              ...    &    m_2^2
 + \tfrac12 \lambda_6 \Sr^2
 + \tfrac12 \lambda_3 \Hdzr^2
 + \tfrac12 \lambda_2 \Huzr^2
                                       \end{pmatrix} \;.}
\end{equation}
The mass parameters $m_1^2$, $m_2^2$ and $m_{3}^2$ are determined via the 
minimization conditions (the tadpoles) of 
$V_{\rm tree}^{\rm THDMS}$ of equation~\ref{eq:THDMS_potential}
and are given by 
\begin{equation}
\label{eq:TreeLvlSymmBrkCond}
\footnotesize{
\begin{aligned}
m_1^2 &=
 - \tfrac12 (\lambda_3 + \lambda_4) \vu^2
 - \tfrac12 \lambda_1 \vd^2
 - \tfrac12 \lambda_5 \vs^2
 - \tfrac12 \lambda_7 \frac{\vu \vs^2}{\vd}
 + \tfrac1{\sqrt{2}} m_4 \frac{\vu \vs}{\vd},
\\
m_2^2 &=
 - \tfrac12 \lambda_2  \vu^2
 - \tfrac12 (\lambda_3 + \lambda_4) \vd^2
 - \tfrac12 \lambda_6  \vs^2
 - \tfrac12 \lambda_7  \frac{\vd \vs^2}{\vu}
 + \tfrac1{\sqrt{2}} m_4  \frac{\vd \vs}{\vu},
\\
m_{3}^2 &=
- \tfrac12 \lambda_6 \vu^2
- \tfrac12 \lambda_5 \vd^2
- \lambda_8 \vs^2
- \lambda_7 \vd \vu
+ \tfrac1{\sqrt{2}} m_4\frac{\vu \vd}{\vs}
+ \tfrac1{\sqrt{2}} m_5 \vs .
\end{aligned}}
\end{equation}
Diagonalization of ${\cal M}_A^2$ in equation~\ref{CPoddmassmatrix} leads to 
two neutral $CP$-odd scalars and a Goldstone boson. Similarly, diagonalization 
of ${\cal M}_{H^\pm}^2$ in equation~\ref{ChargedHiggsmassmatrix} results in a charged Higgs boson and a charged Goldstone boson. The masses of these neutral 
and charged Goldstone bosons are zero at the electroweak minima where the $CP$-even fields acquire values $\{\vd, \vu, \vs \}$. However, the gauge-fixing 
terms in the Lagrangian alter the tree-level mass matrices. In the Feynman 
gauge that we opt for this work, the mass matrices get modified and the 
Goldstone bosons no longer remain massless at the electroweak minima. We include 
these gauge-dependent contributions to ${\cal M}_A^2$ and ${\cal M}_{H^\pm}^2$ 
which are listed in~reference~\cite{Athron:2019teq}.

Note that the one-loop CW potential (of equation~\ref{eq:CW1Loop}) shifts the 
location of the electroweak minimum from where it was appearing in the 
field space for the tree-level potential. In the $\overline{\rm MS}$ 
renormalization scheme that we adopt, one can find suitable counter-terms (as 
described in reference~\cite{Cline:2011mm}) that modify the quadratic terms 
($m_1^2 h_d^2 + m_2^2 h_u^2+m_3^2 s^2$) in the potential (of
equation~\ref{eq:THDMS_potential}) thus ensuring the minimum of the effective potential coincides with that of the tree-level potential. The accompanying shifts in 
the mass parameters of the potential (obtained from
equation~\ref{eq:TreeLvlSymmBrkCond}) are given by
\begin{equation}
\small{
m_1^2  \rightarrow m_1^2  -  \frac{1}{v_d}\frac{\partial V_\text{{CW}}}{\partial h_d}\Bigr|_{\substack{h_d=\vd\\h_u=\vu\\s=\vs}},
\quad m_2^2  \rightarrow m_2^2  -  \frac{1}{v_u}\frac{\partial V_\text{CW}}{\partial h_u}\Bigr|_{\substack{h_d=\vd\\h_u=\vu\\s=\vs}},
\quad m_3^2  \rightarrow m_3^2  -  \frac{1}{v_s}\frac{\partial V_\text{CW}}{\partial s}\Bigr|_{\substack{h_d=\vd\\h_u=\vu\\s=\vs}} \, ,}
\end{equation}
Note that in the tree-level field-dependent mass-squared matrices 
(see equations~\ref{CpEvenHiggsMat}, \ref{CPoddmassmatrix} and 
\ref{ChargedHiggsmassmatrix}) the values of $m_1^2$, $m_2^2$ and $m_3^2$ are 
without this modification since the latter are solutions of the 
corresponding tree-level tadpole equations as presented in
equation~\ref{eq:TreeLvlSymmBrkCond}. 

In the fermionic sector, we consider the top quark, the bottom quark, the tau 
lepton along with the four neutralinos ($\chi_{1,2,3,4}^0$) and 
the one chargino ($\charonepm$), since the wino-like states are taken to 
be much heavier and hence are decoupled from the physics of phase 
transitions. The field-dependent masses of the top quark, the bottom quark and 
the tau lepton are given by~\cite{Ellwanger:2009dp}
\begin{equation}
\label{eq:FermionMasses}
\begin{gathered}
m_t  =  \tfrac{1}{\sqrt{2}}  y_t \Huzr \, , \quad
m_b  =  \tfrac{1}{\sqrt{2}}  y_b \Hdzr \, , \quad
m_{\tau}  =  \tfrac{1}{\sqrt{2}}  y_{\tau} \Hdzr \, .
\end{gathered}
\end{equation}
The field-dependent $4\times4$, symmetric neutralino mass matrix in the basis
$\{\widetilde{B}, ~\widetilde{H}_d^0,~\widetilde{H}_u^0, ~\widetilde{S}\}$ is
given by
\begin{equation} \label{eq:mneuhat}
\small{
  {\cal M}_{\chi^0} = \begin{pmatrix} M_1 & -\frac{g_1 h_d}{2} & \frac{g_1 h_u}{2}  & 0 \\
                                                    ...  & 0 & -\frac{\lambda s}{\sqrt{2}} & -\frac{\lambda h_u}{\sqrt{2}}\\
                                                    ...  & ... & 0 & -\frac{\lambda h_d}{\sqrt{2}}\\
                                                    ...  & ... & ... & \sqrt{2}\kappa s 
                                    \end{pmatrix} \; .}
\end{equation}
On the other hand, the field-dependent mass of the higgsino-like chargino is 
given approximately by $\mcharone \simeq \frac{\lambda s}{\sqrt{2}}$. Note that 
the masses of these electroweakinos are given in terms of the NMSSM model 
parameters which are defined at the scale $M_{\rm SUSY}$. This is acceptable 
for our purpose since these masses do not appear in the tree-level
potential.\footnote{The electroweakino masses, however, contribute to
higher-order (starting at one-loop) corrections to the tree-level potential. Thus, the 
consideration of running of these masses amounts to having an even 
higher-order correction to the potential. Hence we ignore such a running.}

The field-dependent masses of the gauge bosons, $\wpm$ and $Z$, are given by
\begin{equation}
\label{eq:gaugebosonmasses}
\small{
\begin{gathered}
m_{W^\pm}^2  =  \tfrac{1}{4} g_2^2 \left(\Huzr^2  +  \Hdzr^2 \right), \quad
m_Z^2 =  \tfrac{1}{4}  \left({g_1}^2  +  g_2^2 \right) \left( \Huzr^2  +  \Hdzr^2 \right).
\end{gathered}}
\end{equation}

Note that in the daisy potential of equation~\ref{eq:daisypot}, $M_h^2$ and 
$M_V^2$ are the eigenvalues of the thermally improved (i.e., Debye-corrected) 
mass-squared matrices for the Higgs and the gauge bosons, respectively, i.e., 
generically, $M^2 =$ eigenvalues$[{\cal M}^2 + \Delta (T^2)]$ where
$\Delta (T^2) = c_{ij} T^2$ and $c_{ij}$'s are the so-called daisy 
coefficients. From the high temperature expansion of the thermal one-loop 
potential $\widetilde{V}_T$ (of equation~\ref{eq:thermal_one_loop}) using
equation~\ref{e.JBFhighT}, the daisy coefficients can be found from the 
following relation:
\begin{equation}\small{
c_{ij}  =   \left.\frac{1}{T^2}\frac{\partial^2   \widetilde{V}_T}{\partial  \phi_i  \partial  \phi_j}\right|_{T^2  \gg m^2}.}
\end{equation}
For an FOPT, the daisy correction is especially important since it has an 
impact on the all very critical cubic term of the potential at a
finite temperature. Effects of only the scalars and the longitudinal 
modes of the vectors are included in this contribution. Thermal contributions 
to the transverse modes are suppressed due to gauge
symmetry~\cite{Espinosa:1992kf}. With these in mind, the various daisy 
coefficients (neglecting the electroweakino contributions) are as follows~\cite{Comelli:1996vm, Basler:2018cwe, Carrington:1991hz, Athron:2019teq}:
\begin{subequations}
\label{eq:daisy-coeff}
\small{\begin{align}
c^{H}_{11} = c^{A}_{11} = c^{H^{\pm}}_{11} &= \tfrac{1}{24} \left(6 \lambda_2 + 4 \lambda_3 + 2 \lambda_4 + 2 \lambda_6 + 6 y_t^2 + \frac{3}{2} {g_1}^2+ \frac{9}{2} g_2^2 \right ),\label{Eq:DaisyCoeffs_Hu}\\
c^{H}_{22} = c^{A}_{22} = c^{H^{\pm}}_{22} &= \tfrac{1}{24} \left(6 \lambda_1 + 4 \lambda_3 + 2 \lambda_4 + 2 \lambda_5 + + 6 y_b^2 + 2 y_\tau^2 + \frac{3}{2} {g_1}^2+ \frac{9}{2} g_2^2 \right),\label{Eq:DaisyCoeffs_Hd}\\
c^{H}_{33} = c^{A}_{33} &= \tfrac{1}{24} \left( 4 \lambda_5 + 4 \lambda_6 + 8 \lambda_8 \right) ,\label{Eq:DaisyCoeffs_S}
\end{align}}
\end{subequations}
where the subscripts $\{1,2,3\}$ refer to the fields $\{h_d, h_u, s\}$. Note 
that the gauge symmetries plus the discrete $Z_3$ symmetry of the model set the 
off-diagonal terms of the $\Delta(T^2)$ matrix to zero (i.e. $\Delta(T^2)$ is a 
diagonal matrix). The longitudinal components of the gauge bosons receive 
thermal corrections. For  $\wpm$ bosons the correction is
$c^{\wpm}_{L} = 2 g_2^2 T^2$. Thus, the thermally improved mass of the 
longitudinally polarized $\wpm$ bosons is given by
\begin{align}\small{
M_{W^{\pm}_L}^2  =  {1 \over 4}  g_2^2  (h^2_d + h^2_u)  +  2 g_2^2  T^2.}
\end{align} 
Similarly, the longitudinal components of the $Z$-boson and the photon ($A$) 
fields also receive thermal corrections. Their masses can be determined by 
diagonalizing the following matrix:
\beq
\small{
\frac{1}{4} (h^2_d + h^2_u)
\begin{pmatrix}
g_2^2 &  -g_2 g_1 \\
-g_1 g_2 & g_1^2
\end{pmatrix}
+
\begin{pmatrix}
2 g_2^2 T^2 & 0 \\
0 & 2 g_1^2 T^2 
\end{pmatrix}.}
\eeq
The thermally improved masses of the longitudinally polarized $Z$-boson 
and the photon are given by
\beq
\small{
M_{Z_L, \gamma_L}^2 = \frac{1}{8} (g_2^2 + g_1^2) (h_d^2 + h_u^2) + (g_2^2 + g_1^2)T^2 \pm \delta,} 
\eeq
where 
\beq
\small{
\delta  = \sqrt{\frac{1}{64}  (g_2^2 + g_1^2 )^2(h_{d}^2 + h_{u}^2+8T^2)^2
- g_2^2 g_1^2 T^2 ( h_{d}^2 + h_{u}^2 + 4 T^2)}. }
\eeq
%
%%%%%%%%%%%%%%%%%%%%%%%%%%%

\end{document}